%% file: main.tex
\documentclass[onecolumn]{aastex63}
\usepackage{times}
\usepackage{epsfig}
\usepackage{amsmath, amsthm, amssymb}
\usepackage{color}
\usepackage{microtype}
\usepackage{float}
\usepackage{stfloats}
\usepackage{natbib}
\usepackage{verbatim}
\graphicspath{{./figures/}}
\usepackage{wrapfig}
\hypersetup{backref,breaklinks,colorlinks,citecolor=blue}
\usepackage[all]{hypcap}
\usepackage{definitions}
\usepackage{subfiles}
\usepackage{booktabs}

\usepackage[linesnumbered]{algorithm2e}
\usepackage{enumerate}

\newcommand{\figref}[1]{Figure~\ref{#1}}

\usepackage{placeins}

\newcommand{\TVD}{\textnormal{\tiny\textsc{Tvd}}}
\newcommand{\TCI}{\textnormal{\tiny\textsc{TCI}}}
\newcommand{\ye}{Y_{\mathrm{e}}}
\newcommand{\Lo}{\mbox{\tiny\textsc{L}}}
\newcommand{\Hi}{\mbox{\tiny\textsc{H}}}
\newcommand{\Center}{\mbox{\tiny\textsc{C}}}

\shortauthors{Pochik et al.}
\received{\today}
\shorttitle{\thornado-hydro: DG methods with nuclear EoS}

\begin{document}

\title{\thornado-hydro: a discontinuous Galerkin method for supernova hydrodynamics with nuclear equations of state\footnote{This manuscript has been authored in part by UT-Battelle, LLC, under contract DE-AC05-00OR22725 with the US Department of Energy (DOE). The US government retains and the publisher, by accepting the article for publication, acknowledges that the US government retains a nonexclusive, paid-up, irrevocable, worldwide license to publish or reproduce the published form of this manuscript, or allow others to do so, for US government purposes. DOE will provide public access to these results of federally sponsored research in accordance with the DOE Public Access Plan (http://energy.gov/downloads/doe-public-access-plan).}}

\author[0000-0002-8310-0271]{David Pochik}
\affiliation{Department of Physics and Astronomy, University of Tennessee Knoxville, TN 37996}

\author[0000-0002-8825-0893]{Brandon L. Barker}
\altaffiliation{NSF Graduate Research Fellow}
\affiliation{Department of Physics and Astronomy, Michigan State University, East Lansing, MI 48824, USA}

\author[0000-0003-1251-9507]{Eirik Endeve}
\affiliation{Computer Science and Mathematics Division, Oak Ridge National Laboratory, TN 37831}
\affiliation{Department of Physics and Astronomy, University of Tennessee Knoxville, TN 37996}

\author[0000-0002-7712-5835]{Jesse Buffaloe}
\affiliation{Department of Physics and Astronomy, University of Tennessee Knoxville, TN 37996}

\author[0000-0003-4008-6438]{Samuel J. Dunham}
\affiliation{Department of Physics and Astronomy, Vanderbilt University, Nashville, TN 37212}
\affiliation{Department of Physics and Astronomy, University of Tennessee Knoxville, TN 37996}

\author[0000-0003-1891-8402]{Nick Roberts}
\affiliation{Department of Physics and Astronomy, University of Tennessee Knoxville, TN 37996}

\author[0000-0001-9816-9741]{Anthony Mezzacappa}
\affiliation{Department of Physics and Astronomy, University of Tennessee Knoxville, TN 37996}

\correspondingauthor{Eirik Endeve}
\email{endevee@ornl.gov}

\begin{abstract}
This paper describes algorithms for non-relativistic hydrodynamics in the \textbf{t}oolkit for \textbf{h}igh-\textbf{or}der \textbf{n}eutrino r\textbf{ad}iation hydr\textbf{o}dynamics (\thornado), which is being developed for multiphysics simulations of core-collapse supernovae (CCSNe) and related problems with Runge--Kutta discontinuous Galerkin (RKDG) methods.  
More specifically, \thornado\ employs a spectral type nodal collocation approximation, and we have extended limiters --- a slope limiter to prevent non-physical oscillations and a bound-enforcing limiter to prevent non-physical states --- from the standard RKDG framework to be able to accommodate a tabulated nuclear equation of state (EoS).  
To demonstrate the efficacy of the algorithms with a nuclear EoS, we first present numerical results from basic test problems in idealized settings in one and two spatial dimensions, employing Cartesian, spherical-polar, and cylindrical coordinates.  
Then, we apply the RKDG method to the problem of adiabatic collapse, shock formation, and shock propagation in spherical symmetry, initiated with a 15~$M_\odot$ progenitor.  
We find that the extended limiters improve the fidelity and robustness of the RKDG method in idealized settings.  
The bound-enforcing limiter improves robustness of the RKDG method in the adiabatic collapse application, while we find that slope limiting in characteristic fields is vulnerable to structures in the EoS --- more specifically, in the phase transition from nuclei and nucleons to bulk nuclear matter.  
The success of these applications marks an important step toward applying RKDG methods to more realistic CCSN simulations with \thornado\ in the future.  
\end{abstract}

\keywords{Computational methods (1965), Core-collapse supernovae (304), Hydrodynamical simulations (767), Nuclear astrophysics (1129)}


\section{Introduction}
\label{sec:Intro}


Stars with zero-age main sequence (ZAMS) masses $M_{\mathrm{ZAMS}} \gtrsim 8M_{\odot}$ end their lives as spectacular explosions known as core-collapse supernovae (CCSNe).  
These explosions are at the heart of some of the most important questions in astrophysics.
They are the primary catalysts of galactic chemical evolution, producing and dispersing many of the elements heavier than hydrogen and helium, and provide feedback into the interstellar medium.
They may even be a source of the lighter first peak r-process elements \citep{pinedo:2014}, though neutron star mergers are likely the primary production site for the r-process \citep{kasen:2017}.  
Their cores are the foundries for compact objects including those recently detected by Advanced LIGO and Virgo \citep{abbott:2016, abbott:2017, abbott:2017a, abbott:2020}.
Through their observables and the compact objects left behind, we may even begin to probe the nature of nuclear matter \citep{schneider:2019}.

Throughout their lives, these massive stars undergo successive cycles of nuclear fusion, forging heavier elements in their cores.  
At the end of a star's lifetime, fusion processes build up a degenerate iron core that is unable to undergo nuclear fusion itself.  
This iron core, supported thus far by electron degeneracy pressure, grows to the effective Chandrasekhar mass \citep{baron:1990} and, no longer able to balance gravity, subsequently collapses.
During collapse, runaway electron capture processes accelerate the collapse and produce vast
numbers of neutrinos, while photodissociation of iron group nuclei robs the core of more energy.  
Eventually the core reaches nuclear density and the nuclear strong force becomes repulsive, effectively stiffening the Equation of State (EoS) tremendously, and collapse is halted in the inner core.  
The collapse rebounds and produces a strong shock that is driven through the outer core.
Ultimately, through a combination of neutrino cooling and dissociation of iron group nuclei, the shock runs out of energy and stalls before escaping the core, becoming an accretion shock.  
Meanwhile, the inner core regains equilibrium in the form of a newborn proto-neutron star (PNS).  

Providing a mechanism to revive the stalled shock and drive the explosion is among the forefront questions in the study of CCSNe.  
Of the proposed mechanisms, the most favored has been the delayed neutrino-driven mechanism \citep{bethe:1985}.  
Neutrinos emitted from the surface of the cooling PNS, aided by hydrodynamic and magnetohydrodynamic instabilities, deposit energy below the stalled shock and reinvigorate the explosion.  
Of the other proposed mechanisms, the magneto-rotational mechanism -- wherein a rapidly rotating PNS supplies energy to power the shock \citep{akiyama:2003} -- has potential, but likely doesn't account for most CCSNe. 
A key characteristic of magneto-rotationally driven SNe is the formation of collimated jets, which are not seen in the vast majority of supernova remnants \citep[e.g., see][]{soderberg:2010}.  
Additionally, for this mechanism to be effective the stellar core must be very rapidly rotating, beyond the rotation rates commonly achieved through stellar evolution \citep{heger:2005}.  
Ultimately, any successful mechanism must not only revive the shock but also explain the observations of supernovae (e.g., light curves and spectra).

For several decades this was the state of the field.  
These mechanisms saw little success until relatively recently: spherically symmetric (spatially one-dimensional [1D]) simulations of CCSNe consistently failed to produce explosions.  
It wasn't until computing resources allowed for axisymmetric (spatially two-dimensional [2D]), and eventually full-physics three-dimensional (3D), simulations that successful explosions could be consistently produced without modified or parametrized physics. 
Ultimately, the reason for this is 1D fails to capture the fundamentally non-spherical nature of CCSNe and hydrodynamic instabilities are unable to develop. 
The CCSN explosion mechanism has been the subject of decades of work and still remains incompletely described \citep[for in-depth reviews, see, e.g., ][]{bethe:1990, mezzacappa:2001, mezzacappa:2005, janka:2012a, janka:2016, burrows:2013, hix:2014, muller:2016, couch:2017}.


Hydrodynamics, along with gravity and neutrino transport, plays a key role in the dynamics of CCSNe.  
This starts with the progenitors, which in nature are multi-dimensional and likely involve a complicated mixing of elements in the convectively burning shells (see, e.g., \citet{ArnettMeakin2011}).  
Further, it has been shown that asphericities in progenitors can mean the difference between a model that explodes, and a model that doesn't \citep{Couch2013}.  
However, regardless of the progenitor, after the core rebounds it is known that the shocked fluid develops instabilities.

Once the bounce-shock stalls and the neutrino hearing (or gain) region is established below the shock, at least two hydrodynamical instabilities may contribute to the evolution of the shock: neutrino-driven convection \citep{Herant1992} and the Standing Accretion Shock Instability (SASI; \citet{Blondin2003}).  
Both of these instabilities create turbulence in the post-shock flow, and that turbulence contributes ram pressure that enlarges the extent of the gain region \citep{murphy:2013}, thus
increasing the efficacy of neutrino heating, thus aiding the explosion (see \citet{couch:2015a}, and references therein).  
Which effect is more dynamically important, however, may depend on the progenitor mass \citep{muller:2012,hanke:2013,summa:2016,Vartanyan2019}.  
Regardless of which effect is dominant, simulations should be able to satisfactorily quantify the turbulence, and in particular should be able to capture the turbulent energy cascade from the energy carrying scale through the inertial scale, down to the (numerical) dissipation scale.  
However, a consensus has not yet been reached as to what, in terms of angular resolution, is required to adequately capture the turbulent energy cascade.
In particular, \citet{Radice2015}, \citet{Abdikamalov2015}, and \citet{Casanova2020} suggest that resolutions much lower than $1^{\circ{}}$ may be necessary (due to the
numerical dissipation of the scheme, which creates a ``bottleneck'' for energy transfer at a scale set by the scheme), but recently \citet{Melson2020} argued that $1^{\circ{}}$ resolution is sufficient to obtain a clear distinction between the inertial and dissipation scales.  
Additionally, \cite{endeve:2012a} showed that turbulence from the SASI can amplify magnetic fields, and more recently, \citet{MullerMHD2020} found that turbulently amplified magnetic fields can aid neutrino-driven explosions, even in slowly-rotating progenitors.  
See \citet{Radice2018} for a recent review of turbulence in CCSNe.  

In addition to the hydrodynamic instabilities occurring in the shocked mantle, the PNS undergoes convection and potentially other instabilities due to entropy and electron fraction gradients \citep{bruenn:2004}, which has an effect on the luminosity of heavy flavored neutrinos as well as the mean energies of all neutrino flavors \citep{Buras2006}. This may not directly affect the shock dynamics, but it does give rise to the recently discovered Lepton Number Emission Self-Sustained Asymmetry (LESA; \citet{Tamborra2014}), which may hold implications for the composition of the ejecta.  
For more detailed discussions on the role of hydrodynamic instabilities in CCSNe, we refer to the recent review by \citet{Muller2020}.  


Insight into hydrodynamic phenomena can often be gained by treating the fluid as polytropic \citep[in the CCSN context, see, e.g.,][]{yahil:1983,Blondin2003}; i.e. the fluid pressure $p$ is assumed to be proportional to a power law of the mass density $\rho$, which gives rise to the polytropic EoS, $p \propto \rho ^{\Gamma}$, where $\Gamma=\big(\pderiv{\ln p}{\ln \rho}\big)$ is the adiabatic index.\footnote{Contrary to a realistic model, the adiabatic index for a polytropic model remains constant through space and time.}  
However, relating the state variables by this expression neglects the nuclear interactions and compositions in stellar collapse; e.g. the polytropic EoS fails to capture the response in pressure due to the thermal or compositional changes that are typical in a stellar environment.  
For the conditions prevalent in stellar interiors, particularly in the high-density regimes of stellar collapse, a simple analytic form for the EoS likely does not exist.
Instead, an EoS for this case is often created by minimizing a thermodynamic potential --- e.g. the Helmholtz free energy --- for a system of particles under stellar conditions \citep[see, e.g.,][]{swesty:1996,fryxell:2000,timmes:2000}.  
Once the free energy is known, other relevant quantities, such as pressure, internal energy, and entropy, can easily be obtained.  

The task of developing an equation of state for realistic CCSN simulations has remained a pertinent objective for several decades.  
Important contributions toward this effort include the \cite{lattimer:1991} (LS) and \cite{shen:1998} (STOS) EoSs.  
The LS EoS used a compressible liquid-drop model \citep[see, e.g.,][]{lattimer:1985}, while STOS used a relativistic mean field (RMF) model with the TM1 parameter set \citep[see, e.g.][]{Sugahara:1994}.  
However, due to the importance of including light nuclei in CCSN simulations, a notable drawback for both the LS and STOS EoS was their exclusion of all light nuclei other than alpha particles \citep[][]{hempel:2012,steiner:2013a}.  
Further advances include the hadronic EoSs from G.~Shen \citep[][]{shen:2011b,shen:2011a}, which build upon the NL3 \citep[][]{lalazissis:1997} and FSUgold \citep[][]{Todd-Rutel:2005} parameter sets.  
Additionally, unlike the LS and STOS EoSs, the statistical model of \cite{hempel:2012} (HS) 
\citep[see also][]{steiner:2013a} does not use the single-nucleus approximation for heavy nuclei, but includes a more realistic compositional distribution of nuclei.

Moreover, recent neutron star observations \citep[see, e.g.,][]{Greif:2020,steiner:2013b} and observations of other astronomical phenomena \citep[see, e.g,][and references therein]{Greif:2020}, experiments in nuclear physics \citep[see, e.g.,][]{Greif:2020}, and experiments in relativistic heavy-ion collisions \citep[see, e.g.,][and references therein]{oertel:2017}, have led to the development of multiple EoSs for dense nuclear matter that are applicable to CCSN simulations \citep[see, e.g.,][]{steiner:2013b,steiner:2013a}.  
These equations of state provide thermodynamic quantities as functions of density, temperature, and electron fraction.  
The SHFo/SFHx EoSs from \cite{steiner:2013b,steiner:2013a} build upon the statistical model used in HS and constrain properties of nucleonic matter with an RMF model \citep[see, e.g.,][]{shen:1998,shen:2011b,shen:2011a}.  
The most probable mass-radius relationship derived from neutron star (NS) observations was used to build the ``optimal" SFHo EoS, while the ``extreme" SFHx EoS is built around a minimized radius model for low-mass NSs \citep[][]{steiner:2010,steiner:2013b}.  
For our purposes, the importance of these equations of state lies in their ability to resolve various physical regimes in CCSNe, including the phase transition from nuclei and nucleons to bulk nuclear matter at high densities ($\rho\sim10^{14}$~g~cm$^{-3}$) \citep[][]{steiner:2013a}, and the high-density rebound of the core, which determines the initial strength of the shock \citep[][]{shen:1998}.  
We note that these EoSs do not include lower density/temperature regimes; i.e., they do not describe matter out of nuclear statistical equilibrium (NSE); but see, e.g., \citet{bruenn:2020} for treatment of non-NSE regions in CCSN models.  



Clearly, multidimensional, multiphysics models of CCSNe require advanced simulation tools and massive computational resources, and to that end there are several production codes in existence; e.g., 
Aenus-Alcar~\citep{AenusAlcar}, 
Castro~\citep{Castro}, 
Chimera~\citep{bruenn:2020}, 
CoCoNuT-Vertex~\citep{CoconutVertex}, 
FLASH~\citep{fryxell:2000,dubey:2009,oConnor:2018}, 
Fornax~\citep{Fornax}, 
Prometheus-Vertex~\citep{PrometheusVertex}, and 
Zelmani~\citep{Zelmani,roberts:2016}, 
and the codes of \cite{sumiyoshi:2012,Nagakura2014}, and \cite{kuroda:2016}.  
To solve the equations of hydrodynamics --- with the aim of capturing shocks and resolving turbulent flows --- these codes use variations of either the finite-difference or the finite-volume high-resolution shock capturing method, in either an Eulerian or semi-Lagrangian framework.  
In particular, the finite-volume method divides the computational domain into finite cells (or volumes), formulates the hydrodynamics equations in integral form, and solves for physical quantities (e.g., mass density) in terms of cell averages.  
The cell averages are updated by accounting for (1) fluxes through the surface enclosing each cell and (2) volume sources (e.g., due to gravity). 
The integral formulation leads naturally to good conservation properties, and allows for discontinuous solutions (e.g., shocks).  
In computing the surface fluxes, local polynomials are \emph{reconstructed} using cell averages of the local cell and its neighbors.  
The local polynomials are then used to assign left and right states at each cell interface as inputs to a \emph{Riemann solver}, which provides the numerical flux.  
To avoid non-physical oscillations around shocks, limiters are applied to the reconstructed polynomial to enforce some degree of monotonicity, which can degrade the formal order of accuracy of the hydrodynamics scheme.  
(We refer to the above citations for further details on the hydrodynamics algorithms implemented in the specific codes listed.)  

As discussed above, turbulence is ubiquitous in the supernova environment and plays a role in the explosion mechanism.  
It is therefore desirable to maintain good spectral resolution to resolve as much of the turbulent spectrum as possible for a given spatial resolution, and this motivates the use of accurate Riemann solvers and high-order methods.  
On the other hand, due to their multiphysics nature, CCSN simulations with neutrino transport are computationally expensive, and must run efficiently on distributed memory architectures; e.g., using message passing interface (MPI).  
Furthermore, because of the high number of degrees of freedom involved in neutrino transport computations (a momentum space is attached to each spatial point), 
memory limitations require the number of spatial cells assigned to any given MPI process to not be large.
For a code to scale well, the number of ghost cells should be limited relative to the number of compute cells to manage the communication overhead, since each MPI process will have a halo region comprised of ghost cells populated with data from neighboring processes.  
While finite-difference and finite-volume methods can achieve high-order accuracy, the computational stencil width increases with increasing order of accuracy, thereby increasing the size of the halo region and the ratio of ghost cells to compute cells, thus impeding good scalability \citep[e.g.,][]{miller:2017}.

The discontinuous Galerkin (DG) method \citep[e.g.,][]{Cockburn:2001} is an alternative approach to solving the system of hydrodynamics equations (and many other systems).  
Similar to finite-volume methods, DG methods divide the computational domain into cells (or elements), and formulate the equations in integral form.  
However, contrary to finite-volume and finite-difference methods, in the DG method the solution is approximated by a local polynomial within each element, which implies that more local information is tracked in the solution process (i.e., not just the cell average).  
Because the full polynomial representation in each element is evolved, the reconstruction step needed in the finite-volume approach is not necessary.  
Meanwhile, Riemann solvers developed in the context of finite-volume methods can readily be used with DG methods to evaluate numerical fluxes on element interfaces.  
The DG method is a finite-element method, but does not demand continuity of the local polynomial approximation across element boundaries, and consequently, is well suited to capture shocks and other discontinuities.  
To prevent non-physical oscillations in the vicinity of a discontinuity, limiters are applied to the local polynomial to enforce monotonicity.  
More recently, so-called structure-preserving discretizations, which maintain fundamental physical properties of the system under consideration (e.g., positive mass density and pressure), have been developed within the DG framework \citep[e.g.,][]{zhangShu:2011PRSA}.  
Another advantage offered by the DG method is high-order spatial accuracy on a compact stencil.  
Only information from nearest neighbors is needed, independent of the order of accuracy.  
This makes the DG method well-suited for application on massively parallel architectures, since increasing the order of accuracy does not increase the communication overhead as much as other high-order methods \citep[e.g.,][]{miller:2017}.  
The desired combination of shock-capturing capabilities, high-order accuracy in smooth flows, and good scalability make DG methods an appealing choice. 
Additionally, DG methods are also amenable to $hp$-adaptivity \citep{remacle:2003}, wherein refinement of either the spatial mesh ($h$-refinement) or the local degree of the polynomial approximation ($p$-refinement) can be used to improve the accuracy of the method near shocks while maintaining high-order accuracy in regions of smooth flow.  
DG methods are also well-suited for problems involving curvilinear coordinates \citep{teukolsky:2016}.  

The DG method was introduced already in the 1970s by \citet{ReedHill1973} to solve the steady state neutron transport equation, and the initial framework for solving time-dependent problems with explicit Runge--Kutta time integration (commonly referred to as RKDG methods) was established in a series of papers by Cockburn \& Shu \citep{Cockburn:1989a,Cockburn:1989b,Cockburn:1990,Cockburn:1991,cockburn:1998}.  
Today, DG methods are widely used in science and engineering applications, and are rapidly gaining popularity in the computational astrophysics community \citep[see, e.g.,][and references therein]{radice:2011,schaal:2015a,teukolsky:2016,Spectre,fambri:2018}, but have so far not been applied to multiphysics CCSN simulations.

The \textbf{t}oolkit for \textbf{h}igh-\textbf{or}der \textbf{n}eutrino r\textbf{ad}iation hydr\textbf{o}dynamics\footnote{\url{https://github.com/endeve/thornado}} (\thornado) is being developed with the goal of realizing multiphysics simulations of CCSNe and related problems with high-order methods.  
To this end, the hydrodynamics and neutrino transport algorithms in \thornado\ are based on the DG method \cite[see, e.g.,][]{endeve:2019,chu:2019,laiu:2020}.  
It should be noted that, in addition to exhibiting favorable parallel scalability, DG methods are also an attractive choice for discretizing the neutrino transport equations because they recover the correct asymptotic behavior in the so-called diffusion limit \citep[e.g.,][]{larsen:1989,adams:2001}, which is characterized by frequent neutrino--matter interactions. 
Then, since the matter and neutrinos are strongly coupled in the CCSN environment, employing the DG method also for the hydrodynamics is most natural, as this enables treatment of the coupled physics in a unified mathematical framework.  
Currently, \thornado\ is being developed as a collection of modules, focusing on single-node performance for updating structured data blocks using CPUs and/or GPUs, with the future aim of leveraging an external framework --- e.g., AMReX\footnote{\url{https://amrex-codes.github.io}} \citep{zhang:2019} --- to support mesh adaptivity.  


This paper describes the DG algorithms for non-relativistic hydrodynamics in \thornado.  
We adapt a three-covariant formalism that is sufficiently general to accommodate Cartesian, spherical-polar, and cylindrical spatial coordinates.  
Although we presented preliminary results obtained with similar algorithms for non-relativistic and relativistic hydrodynamics in the context of an ideal EoS in \cite{endeve:2019}, this paper provides a more comprehensive description of the methods in \thornado, and, more important, develops the algorithms further in order to accommodate a nuclear EoS.  
Introducing a nuclear matter EoS leads to more realistic models, but also complicates the numerical procedure.  
For instance, when solving the conservation equations for mass, momentum, and energy, the implementation of a nuclear EoS requires an additional conservation law for electrons, \citep[see, e.g.,][for similar modifications]{colella:1985,zingale:2015}.  
Moreover, on-the-fly numerical evaluation of a realistic EoS is computationally expensive \citep[][]{swesty:1996}; thus, for computational expediency, EoSs are provided in tabulated form, and interpolations are used to access quantities away from table vertices, where a thermodynamically consistent interpolation scheme may be required \citep[see, e.g.,][for a discussion of such interpolation schemes]{swesty:1996,timmes:2000,fryxell:2000}.  
To limit the scope of this paper, we exclusively consider the SFHo EoS \citep{steiner:2013b}, which is provided in tabulated form by CompOSE\footnote{\url{https://compose.obspm.fr}}.  
In \thornado, the interface to the tabulated EoS is through the \weaklib\ library\footnote{\url{https://github.com/starkiller-astro/weaklib}}, which provides auxiliary functionality needed for computations (e.g., input/output and interpolation).  
As such, the EoS is currently treated as a black box.  

The Euler equations in curvilinear coordinates, extended to accommodate a nuclear EoS and self-gravity, are listed in Section~\ref{sec:model}.  
Then, in Section~\ref{sec:method}, we present the RKDG method in \thornado.  
Sections~\ref{sec:DG} and \ref{sec:time} provide the spatial and temporal discretizations, respectively, which are based on the standard framework from \cite{Cockburn:2001}.  
More specifically, we employ a nodal DG method \citep[e.g.,][]{hesthaven:2008} and adopt the spectral type nodal collocation approximation investigated by \cite{Bassi:2013}.  
Sections~\ref{sec:slope} and \ref{sec:boundEnforcing} discuss the slope limiter (to prevent non-physical oscillations) and the bound-enforcing limiter (to prevent non-physical states), respectively.  
The extension of these limiters to the case with a tabulated nuclear EoS is nontrivial.  
First, since slope limiting is most effective when applied to characteristic variables, we provide the characteristic decomposition of the flux Jacobian matrices for a nuclear EoS (Appendix~\ref{appendix:characteristic}).  
Second, since the domain of validity of the nuclear EoS is more complex than the ideal case, we develop an enhanced version of the bound-enforcing limiter of \cite{zhang:2010a}.  
Section~\ref{sec:poisson} describes the Poisson solver for use in spherically symmetric problems with self-gravity, which uses the finite-element method.  
Section~\ref{sec:tableInterpolation} provides details on the interpolation methods used to evaluate the tabulated EoS.  
We use basic trilinear interpolation, which is commonly employed in supernova simulation codes~\citep[e.g.,][]{bruenn:2020}.  
In Section~\ref{sec:results}, to demonstrate the efficacy of the algorithms, we present numerical results from basic test problems (advection and Riemann problems) in idealized settings in one and two spatial dimensions.  
We also include a test of the Poisson solver.  
Then, in Section~\ref{sec:collapse}, we apply the DG method to the problem of adiabatic collapse, shock formation, and shock propagation in spherical symmetry, using a 15~$M_\odot$ progenitor.  
Here we focus on aspects of the limiters, resolution dependence, and total energy conservation.  
Our major goals in this paper are to (1) present the key algorithmic components of the hydrodynamics in \thornado, (2) assess the implementation given the initial set of algorithmic choices, and (3) identify potential areas for improvement.  
This will clear the way for incorporating DG methods for neutrino transport and future neutrino radiation-hydrodynamics simulations with \thornado.  

\section{Physical Model}
\label{sec:model}

\subsection{Euler Equations}
\label{sec:euler}

In this paper we adopt the non-relativistic Euler equations of gas dynamics in a coordinate basis \citep[e.g.,][]{Rezzolla:2013}, supplemented with a nuclear equation of state (EoS), which are given by the mass conservation equation
\beq 
  \partial_{t} \rho + \frac{1}{\sqrt{\gamma}}\,\partial_{i} \big(\,\sqrt{\gamma}\,\rho\,v^{i}\,\big) = 0,
  \label{eq:massConservation}
\eeq
the momentum equation
\beq
  \partial_{t}(\rho\,v_{j}) + \frac{1}{\sqrt{\gamma}}\,\partial_{i}\big(\,\sqrt{\gamma}\,\Pi_{~j}^{i}\,\big) 
  = \f{1}{2}\,\Pi^{ik}\,\partial_{j}\gamma_{ik} - \rho\,\partial_{j}\Phi,
  \label{eq:momentumConservation}
\eeq
the energy equation
\beq
  \partial_{t} E + \frac{1}{\sqrt{\gamma}}\,\partial_{i}\big(\,\sqrt{\gamma}\,[\,E+p\,]\,v^{i}\,\big) 
  = -\rho\,v^{i}\,\partial_{i}\Phi,
  \label{eq:energyConservation}
\eeq
and the electron conservation equation
\beq
  \partial_{t} D_{\rm{e}} + \frac{1}{\sqrt{\gamma}}\,\partial_{i}\big(\,\sqrt{\gamma}\,D_{\rm{e}}\,v^{i}\,\big) = 0,
  \label{eq:electronConservation}
\eeq
where $\rho$ represents mass density, $v^{i}$ the components of the fluid three-velocity, $\Pi_{~j}^{i}=\rho\,v^{i}\,v_{j} + p\,\delta_{~j}^{i}$ the stress tensor, $p$ the fluid pressure, $D_{\rm{e}}=\rho \ye$, where $\ye$ is the electron fraction, $E=\epsilon \rho +\frac{1}{2}\rho v^2$ the total fluid energy density (internal plus kinetic), and $\epsilon$ is the specific internal energy.  
The Euler equations are closed with the EoS, where the pressure and specific internal energy are given functions of density, temperature $T$, and the electron fraction; e.g., $p = p(\rho, T, \ye)$.  
Thus, Equation~\eqref{eq:electronConservation} is necessary for the inclusion of a nuclear EoS.  
(Unless stated otherwise, we use the Einstein summation convention where repeated latin indices run from $1$ to $3$.)  
Included on the right-hand sides of Equations~\eqref{eq:momentumConservation} and \eqref{eq:energyConservation}, are gravitational sources from the Newtonian gravitational potential $\Phi$, which is obtained from the Poisson equation
\beq
  \f{1}{\sqrt{\gamma}}\pd{}{i}\big(\,\sqrt{\gamma}\,\gamma^{ij}\pd{\Phi}{j}\,\big) = 4\pi\,G\,\rho,
  \label{eq:poissonEquation}
\eeq
where $G$ is Newton's constant.  

The use of curvilinear coordinates is enabled through the spatial metric tensor $\gamma_{ik}$, which gives the squared proper spatial interval
\begin{equation}
  ds_{\vect{x}}^{2} = \gamma_{ik}\,dx^{i}\,dx^{k}.  
\end{equation}
The determinant of the spatial metric is denoted $\gamma$.  
The metric tensor is also used to raise and lower indices on vectors and tensors; e.g., $v_{i}=\gamma_{ik}\,v^{k}$.  
In this paper we only consider the commonly adopted Cartesian, cylindrical, and spherical-polar coordinate systems (see Table~\ref{tab:metricFunctions} for relevant quantities associated with each of these systems).  
Thus, the metric tensor is diagonal, and we assume that it is time independent.  
Note that we also list the scale factors $h_{1}$, $h_{2}$, and $h_{3}$ in Table~\ref{tab:metricFunctions}.  
By specifying the scale factors, components of the spatial metric are obtained from $\gamma_{11}=h_{1}h_{1}$, $\gamma_{22}=h_{2}h_{2}$, and $\gamma_{33}=h_{3}h_{3}$, and the square root of the metric determinant is $\sqrt{\gamma}=h_{1}h_{2}h_{3}$.  

For the discussion of the numerical method in Section~\ref{sec:method}, we rewrite Equations~\eqref{eq:massConservation}--\eqref{eq:electronConservation} in a more convenient way as a system of hyperbolic balance equations
\beq
  \partial_{t}\mathbf{U} + \frac{1}{\sqrt{\gamma}}\,\partial_{i}\big(\,\sqrt{\gamma}\,\mathbf{F}^{i}(\mathbf{U})\,\big) = \mathbf{S}(\mathbf{U},\Phi),
  \label{eq:eulerSystem}
\eeq
where
\beq
  \mathbf{U}
  =
  \left[
  \begin{array}{c}
    \rho \\ \rho v_{j} \\ E \\ D_{\rm{e}}
  \end{array}
  \right],
  \quad
  \mathbf{F}^{i}(\mathbf{U})
  =
  \left[
  \begin{array}{c}
    \rho\,v^{i} \\ \Pi^{i}_{~j} \\ (E+p)\,v^{i} \\ D_{\rm{e}}\,v^{i}
  \end{array}
  \right],
  \quad\text{and}\quad
  \mathbf{S}(\mathbf{U},\Phi)
  =
   \left[
  \begin{array}{c}
    0 \\ \f{1}{2}\,\Pi^{ik}\,\partial_{j}\gamma_{ik} - \rho\,\partial_{j}\Phi \\ -\rho\,v^{i}\,\partial_{i}\Phi \\ 0
  \end{array}
  \right]
  \label{eq:eulerSystemComponents}
\eeq
are the vector of evolved quantities, the flux vectors, and the source vector, respectively.  
We split the source vector further as $\mathbf{S}(\mathbf{U},\Phi)=\mathbf{S}^{\gamma}(\mathbf{U})+\mathbf{S}^{\Phi}(\mathbf{U},\Phi)$, where
\beq
  \mathbf{S}^{\gamma}(\mathbf{U})
  =
  \left[
  \begin{array}{c}
    0 \\ \f{1}{2}\,\Pi^{ik}\,\partial_{j}\gamma_{ik} \\ 0 \\ 0
  \end{array}
  \right]
  \quad\text{and}\quad
  \mathbf{S}^{\Phi}(\mathbf{U},\Phi)
  =
  -\rho\,
  \left[
  \begin{array}{c}
    0 \\ \partial_{j}\Phi \\ v^{i}\,\partial_{i}\Phi \\ 0
  \end{array}
  \right].  
  \label{eq:eulerSources}
\eeq

\begin{table}[h]
  \caption{Metric quantities for Cartesian, cylindrical, and spherical-polar coordinate systems.\label{tab:metricFunctions}}
  \small
  \vspace{-6pt}
  \begin{center}
  \begin{tabular}{cccccccccccccc}
    \midrule
    Coordinates & $x^{1}$ & $x^{2}$ & $x^{3}$ & $h_{1}$ & $h_{2}$ & $h_{3}$ & $\gamma_{11}$ & $\gamma_{22}$ & $\gamma_{33}$ & $\sqrt{\gamma}$
    & $\f{1}{\gamma_{22}}\pderiv{\gamma_{22}}{x^{1}}$ & $\f{1}{\gamma_{33}}\pderiv{\gamma_{33}}{x^{1}}$ & $\f{1}{\gamma_{33}}\pderiv{\gamma_{33}}{x^{2}}$ \\
    \midrule
    \midrule
    Cartesian & $x$ & $y$ & $z$ & 1 & 1 & 1 & 1 & 1 & 1 & 1 & 0 & 0 & 0 \\
    Cylindrical & $R$ & $z$ & $\phi$ & 1 & 1 & $R$ & 1 & 1 & $R^{2}$ & $R$ & 0 & $2/R$ & 0 \\
    Spherical & $r$ & $\theta$ & $\phi$ & 1 & $r$ & $r\sin\theta$ & 1 & $r^{2}$ & $r^{2}\sin^{2}\theta$ & $r^{2}\sin\theta$ & $2/r$ & $2/r$ & $2\cot\theta$ \\
    \midrule
    \midrule
  \end{tabular}
  \end{center}
\end{table}

\subsection{Equation of State}
\label{sec:eos}

The EoS provides thermodynamic quantities such as pressure, internal energy, and entropy (dependent variables) as functions of the independent variables; e.g., density, temperature, and electron fraction.  
(Other choices for the independent variables --- e.g., density, entropy, and electron fraction --- are of course also possible, but in the nuclear astrophysics modeling community it is perhaps most common to use $\rho$, $T$, and $\ye$.)  
These dependent variables, and in some cases their derivatives, are crucial for modeling hydrodynamics, nuclear reactions, and neutrino transport in core-collapse supernovae.  
Of particular importance for numerical methods for the hydrodynamics, is the relationship between the EoS and the well-posedness of the system given by Equation~\eqref{eq:eulerSystem}.  
Specifically, the system is said to be hyperbolic if the Jacobian matrices $\partial\mathbf{F}^{i}/\partial\mathbf{U}$ can be diagonalized with a set of real eigenvalues $\{\lambda_{1}^{i},\ldots,\lambda_{6}^{i}\}$ and has a set of linearly independent right eigenvectors $\{\mathbf{r}_{1}^{i},\ldots,\mathbf{r}_{6}^{i}\}$ such that \citep[cf.][]{leveque:1992,Rezzolla:2013}
\beq
  \big(\partial\mathbf{F}^{i}/\partial\mathbf{U}\big)\,\mathbf{r}_{j}^{i} = \lambda_{j}^{i}\,\mathbf{r}_{j}^{i},\quad\text{for}\quad j=1,\ldots,6.
  \label{eq:eigenDecomposition}
\eeq
(In Equation~\eqref{eq:eigenDecomposition}, repeated indices do not imply summation, but rather that it must hold for each of the three flux vectors.)  
For the system in Equation~\eqref{eq:eulerSystem}, the eigenvalues are given by $\{\,v^{i}-\CS\sqrt{\gamma^{ii}},v^{i},v^{i},v^{i},v^{i},v^{i}+\CS\sqrt{\gamma^{ii}}\,\}$, where $\CS$ is the sound speed; $\CS^{2}=\big(\partial p/\partial\rho\big)_{s,\ye}$, where $s$ is the entropy per baryon.  
A fundamental property of hyperbolic equations is that they are well-posed, which makes them suitable for numerical solution \citep[see, e.g.,][for a discussion]{Rezzolla:2013}.  
Thus, a necessary condition for our system to be suitable for numerical solution is $\CS^{2}>0$.  
When the independent variables are chosen to be $\rho$, $T$, and $\ye$, the square of the sound speed can be written explicitly in terms of thermodynamic derivatives as
\beq
  \CS^{2}=\Big(\pderiv{p}{\rho}\Big)_{s,\ye}
  =\Big(\pderiv{p}{\rho}\Big)_{T,\ye} - \Big(\pderiv{s}{T}\Big)_{\rho,\ye}^{-1}\Big(\pderiv{p}{T}\Big)_{\rho,\ye}\Big(\pderiv{s}{\rho}\Big)_{T,\ye}.  
  \label{eq:soundSpeedSquaredDTY}
\eeq
The sound speed, or a related quantity, is typically included with a tabulated EoS.  
In addition, advanced numerical methods make use of the eigenvectors in Equation~\eqref{eq:eigenDecomposition}, e.g., for the characteristic limiting described in Section~\ref{sec:slope}.  
These eigenvectors in turn depend on additional thermodynamic derivatives, whose estimation from the EoS table is discussed in \ref{sec:tableInterpolation}.
For use in computations, \thornado\ has been developed to use the EoS infrastructure provided by the \weaklib\ library.  
(Specifically, \weaklib\ supplies trilinear interpolation, and derivatives computed by analytic differentiation of the trilinear interpolation formula.)  

\section{Numerical Method}
\label{sec:method}

\subsection{Discontinuous Galerkin Method}
\label{sec:DG}

In \thornado\ we employ the Runge-Kutta discontinuous Galerkin (RKDG) method to solve the Euler equations given by Equation~\eqref{eq:eulerSystem}.  
(We refer to \citet{Cockburn:2001} for an excellent review on the RKDG method, and \cite{Shu:2016} for a summary of more recent developments.)  
To this end, the $d$-dimensional computational domain $D\subset\mathbb{R}^{d}$ is subdivided into the union $\mathcal{T}$ of non-overlapping elements $\vect{K}$ such that $D=\cup_{\vect{K}\in\mathcal{T}}\vect{K}$.  
We take each element to be a logically Cartesian box
\begin{equation}
  \vect{K} = \big\{\,\vect{x} : x^{i} \in K^{i} := (x_{\Lo}^{i},x_{\Hi}^{i}), i=1,\ldots,d \,\big\},
\end{equation}
where $x_{\Lo}^{i}$ and $x_{\Hi}^{i}$ are the low and high boundaries of the element in the $i$th dimension.  
We also define the surface elements $\tilde{\vect{K}}^{i}=\times_{j\ne i}^{d} K^{j}$ (so that $\vect{K}=\tilde{\vect{K}}^{i}\times K^{i}$), the set $\vect{x}=\{x^{i},\tilde{\vect{x}}^{i}\}$ to distinguish coordinates parallel and perpendicular to the $i$th dimension, and the element width $\Delta x^{i}=(x_{\Hi}^{i}-x_{\Lo}^{i})$ and center $x_{\Center}^{i}=\f{1}{2}(x_{\Lo}^{i}+x_{\Hi}^{i})$.  
We also define $|\vect{K}|=\prod_{i=1}^{d}\Delta x^{i}$ and $\tilde{\vect{K}}^{i}=\prod_{j=1,j\ne i}^{d}\Delta x^{j}$.  
We let the volume of an element be denoted
\beq
  V_{\vect{K}} = \int_{\vect{K}}dV_{h}, \quad\text{where}\quad dV_{h} = \sqrt{\gamma_{h}}\prod_{i=1}^{d}dx^{i},
  \label{eq:elementVolume}
\eeq
where $\gamma_{h}$ is the determinant of the approximate spatial metric $(\gamma_{h})_{ij}$.  
We will discuss the approximation to the spatial metric in more detail below.  

On each element, we define the approximation space consisting of functions $\psi_{h}$
\begin{equation}
  \mathbb{V}_{h}^{k} = \big\{\,\psi_{h} : \psi_{h}|_{\vect{K}}\in\mathbb{Q}^{k}(\vect{K}), \forall \vect{K}\in\mathcal{T}\,\big\},
\end{equation}
where $\mathbb{Q}^{k}$ is the tensor product space of one-dimensional polynomials of maximal degree $k$.  
In the DG method, the functions in $\mathbb{V}_{h}^{k}$ can be discontinuous across element interfaces.  
In \thornado\ we use Lagrange polynomials,
\begin{equation}
  \ell_{p}(\xi^{i}) = \prod_{\substack{q=1\\q\ne p}}^{N}\f{\xi^{i}-\xi_{q}^{i}}{\xi_{p}^{i}-\xi_{q}^{i}},
  \label{eq:lagrangePoly}
\end{equation}
where $N=k+1$ and the polynomials $\ell_{p}$ are defined on the unit reference interval $I^{i}=\{\,\xi^{i} : \xi^{i}\in(-\f{1}{2},\f{1}{2})\,\}$ ($i=1,\ldots,d$).  
The physical coordinate $x^{i}$ is related to the reference coordinate $\xi^{i}$ by the transformation $x^{i}(\xi^{i})=x_{\Center}^{i}+\Delta x^{i}\,\xi^{i}$.  
For the Lagrange polynomials, we define the set of interpolation points $S_{N}^{i}=\{\xi_{1}^{i},\ldots,\xi_{N}^{i}\}\subseteq I^{i}$.  
Note that for $\xi_{q}^{i}\in S_{N}^{i}$, we have $\ell_{p}(\xi_{q}^{i})=\delta_{pq}$, where $\delta_{pq}$ is the Kronecker delta.  
As an example, the multi-dimensional basis function $\phi_{\vect{i}}(\vect{x}(\vect{\xi}))\in\mathbb{V}_{h}^{k}$ takes the form
\begin{equation}
  \phi_{\vect{i}}(\vect{\xi}) = \phi_{\{i_{1},\ldots,i_{d}\}}(\xi^{1},\ldots,\xi^{d}) = \ell_{i_{1}}(\xi^{1}) \times\ldots\times \ell_{i_{d}}(\xi^{d}),
  \label{eq:basisFunctionMultiD}
\end{equation}
where we have introduced the multi-index $\vect{i}=\{i_{1},\ldots,i_{d}\}\in\mathbb{N}^{d}$ (a $d$-tuple) to achieve a more compact notation.  
To further illustrate, in each element $\vect{K}$ we approximate the solution to Equation~\eqref{eq:eulerSystem} by $\vect{U}_{h}$, which is given by an expansion of functions in $\mathbb{V}_{h}^{k}$ of the form
\begin{equation}
  \vect{U}_{h}(\vect{x},t)
  =\sum_{\vect{i}=\vect{1}}^{\vect{N}}\vect{U}_{\vect{i}}(t)\,\phi_{\vect{i}}(\vect{x}(\vect{\xi}))
  =\sum_{i_{1}=1}^{N}\ldots\sum_{i_{d}=1}^{N}\vect{U}_{\{i_{1},\ldots,i_{d}\}}(t)\,\ell_{i_{1}}(\xi^{1})\times\ldots\times\ell_{i_{d}}(\xi^{d}), 
  \label{eq:nodalExpansion}
\end{equation}
where $\vect{N}\in\mathbb{N}^{d}$ is the $d$-tuple $\{N,\ldots,N\}$. 
The DG method does not require that the approximate multidimensional solution is constructed from one-dimensional polynomials of the same degree $k$ in each dimension, but we make this choice.  
In the multidimensional setting, we denote the set of interpolation points in element $\vect{K}$ by $\vect{S}_{N}=\otimes_{i=1}^{d}S_{N}^{i}$.  
For $\vect{\xi}_{\vect{j}}\in\vect{S}_{N}$, we have $\phi_{\vect{i}}(\vect{\xi}_{\vect{j}})=\delta_{\vect{ij}}=\delta_{i_{1}j_{1}}\times\ldots\times\delta_{i_{d}j_{d}}$, which follows from the Kronecker delta property of the Lagrange polynomials emphasized above.  
Therefore, for $\vect{\xi}_{\vect{j}}\in\vect{S}_{N}$, a direct evaluation in Equation~\eqref{eq:nodalExpansion} shows that $\vect{U}_{h}(\vect{x}(\vect{\xi}_{\vect{j}}))=\vect{U}_{\vect{j}}(t)$; i.e., the expansion coefficients in Equation~\eqref{eq:nodalExpansion} --- the unknowns to be determined by the DG method --- are simply the evolved quantities evaluated in the interpolation points on each element.  

We are now ready to state the DG formulation, which forms the basis for the DG method implemented in \thornado.  
The semi-discrete DG problem is to find $\vect{U}_{h}\in\mathbb{V}_{h}^{k}$, which approximates $\vect{U}$ in Equation~\eqref{eq:eulerSystem}, such that
\begin{align}
  \langle\,\pd{\vect{U}_{h}}{t},\,\psi_{h}\,\rangle_{\vect{K}}
  =\mathcal{B}_{h}^{\sf{Flx}}\big(\vect{U}_{h},\psi_{h}\big)_{\vect{K}} + \langle\,\vect{S}(\vect{U}_{h},\Phi_{h}),\,\psi_{h}\,\rangle_{\vect{K}}
  \equiv\mathcal{B}_{h}\big(\vect{U}_{h},\Phi_{h},\psi_{h}\big)_{\vect{K}}
  \label{eq:dgSemidiscreteWeak}
\end{align}
holds for all test functions $\psi_{h}\in\mathbb{V}_{h}^{k}$ and all elements $\vect{K}\in\mathcal{T}$.  
In Equation~\eqref{eq:dgSemidiscreteWeak}, 
\beq
  \langle\,\pd{\vect{U}_{h}}{t},\,\psi_{h}\,\rangle_{\vect{K}}
  =\int_{\vect{K}}\pd{\vect{U}_{h}}{t}\,\psi_{h}\,dV_{h},
  \label{eq:dgSemidiscreteTimeDerivative}
\eeq
and we have defined the contributions from the fluxes as
\beq
  \mathcal{B}_{h}^{\sf{Flx}}\big(\vect{U}_{h},\psi_{h}\big)_{\vect{K}}
  =
  -\sum_{i=1}^{d}\int_{\tilde{\vect{K}}^{i}}
  \Big(\,\sqrt{\gamma_{h}}\,\widehat{\vect{F}}^{i}(\vect{U}_{h})\,\psi_{h}|_{x_{\Hi}^{i}}-\sqrt{\gamma_{h}}\,\widehat{\vect{F}}^{i}(\vect{U}_{h})\,\psi_{h}|_{x_{\Lo}^{i}}\,\Big)\,d\tilde{\vect{x}}^{i}
  +\sum_{i=1}^{d}\int_{\vect{K}}\vect{F}^{i}(\vect{U}_{h})\,\partial_{i}\psi_{h}\,dV_{h},
  \label{eq:dgSemidiscreteFluxes}
\eeq
and the contributions from the sources as
\beq
  \langle\,\vect{S}(\vect{U}_{h},\Phi_{h}),\,\psi_{h}\,\rangle_{\vect{K}}
  =\int_{\vect{K}}\vect{S}(\vect{U}_{h},\Phi_{h})\,\psi_{h}\,dV_{h}.  
  \label{eq:dgSemidiscreteSources}
\eeq
The approximation to the Newtonian gravitational potential, denoted $\Phi_{h}$ (not to be confused with the basis functions $\phi_{\vect{i}}$ in Equation~\eqref{eq:nodalExpansion}), is obtained by solving Equation~\eqref{eq:poissonEquation} using a finite element method.  
We discuss this in Section~\ref{sec:poisson}.  

In Equation~\eqref{eq:dgSemidiscreteFluxes}, the numerical flux $\widehat{\vect{F}}^{i}(\vect{U}_{h})$ is introduced to define a unique flux in the $i$th surface of $\vect{K}$.  
This numerical flux is computed from a numerical \emph{flux function} (obtained, e.g., from solving an approximate Riemann problem)
\beq
  \widehat{\vect{F}}^{i}(\vect{U}_{h};x^{i},\tilde{\vect{x}}^{i})
  = \vect{f}^{i}\big(\,\vect{U}_{h}(x^{i,-},\tilde{\vect{x}}^{i}),\vect{U}_{h}(x^{i,+},\tilde{\vect{x}}^{i})\,\big),
\eeq
where superscripts $-/+$ in the arguments of $\vect{U}_{h}(x^{i,-/+},\tilde{\vect{x}}^{i})$ indicate that the approximation is evaluated to the immediate left/right of the interface located at $x^{i}$.  
In \thornado\ we have implemented the HLL \citep{Harten:1983} and HLLC \citep{Toro:1994} flux functions, but in the numerical experiments in Sections~\ref{sec:results} and \ref{sec:collapse}, we use exclusively the HLL flux function given by
\beq
  \vect{f}^{i}\big(\vect{U}_{h}^{-},\vect{U}_{h}^{+}\big)
  = \f{\alpha^{i,+}\,\vect{F}^{i}(\vect{U}_{h}^{-})+\alpha^{i,-}\,\vect{F}^{i}(\vect{U}_{h}^{+})-\alpha^{i,-}\alpha^{i,+}\big(\vect{U}_{h}^{+}-\vect{U}_{h}^{-}\big)}{\alpha^{i,-}+\alpha^{i,+}},
\eeq
where $\vect{U}_{h}^{\pm}=\vect{U}_{h}(x^{i,\pm},\tilde{\vect{x}}^{i})$, and where $\alpha^{i,-}$ and $\alpha^{i,+}$ are wave speed estimates for the fastest (in absolute value; $\alpha^{i,\pm}\ge0$) left and right propagating waves, respectively.  
For these estimates we simply use \citep{Davis:1988}
\beq
  \alpha^{i,-} = \max_{j\in\{1,\ldots,6\}}\big(\,0,\,-\lambda_{j}^{i}(\vect{U}_{h}^{-}),\,-\lambda_{j}^{i}(\vect{U}_{h}^{+})\,\big)
  \quad\text{and}\quad
  \alpha^{i,+} = \max_{j\in\{1,\ldots,6\}}\big(\,0,\,+\lambda_{j}^{i}(\vect{U}_{h}^{-}),\,+\lambda_{j}^{i}(\vect{U}_{h}^{+})\,\big),
\eeq
where $\lambda_{j}^{i}$ are the eigenvalues of the flux Jacobian introduced in Equation~\eqref{eq:eigenDecomposition}.  

\begin{figure}[h]
  \centering
  \begin{minipage}[h]{0.4\textwidth}
    \includegraphics[width=\textwidth]{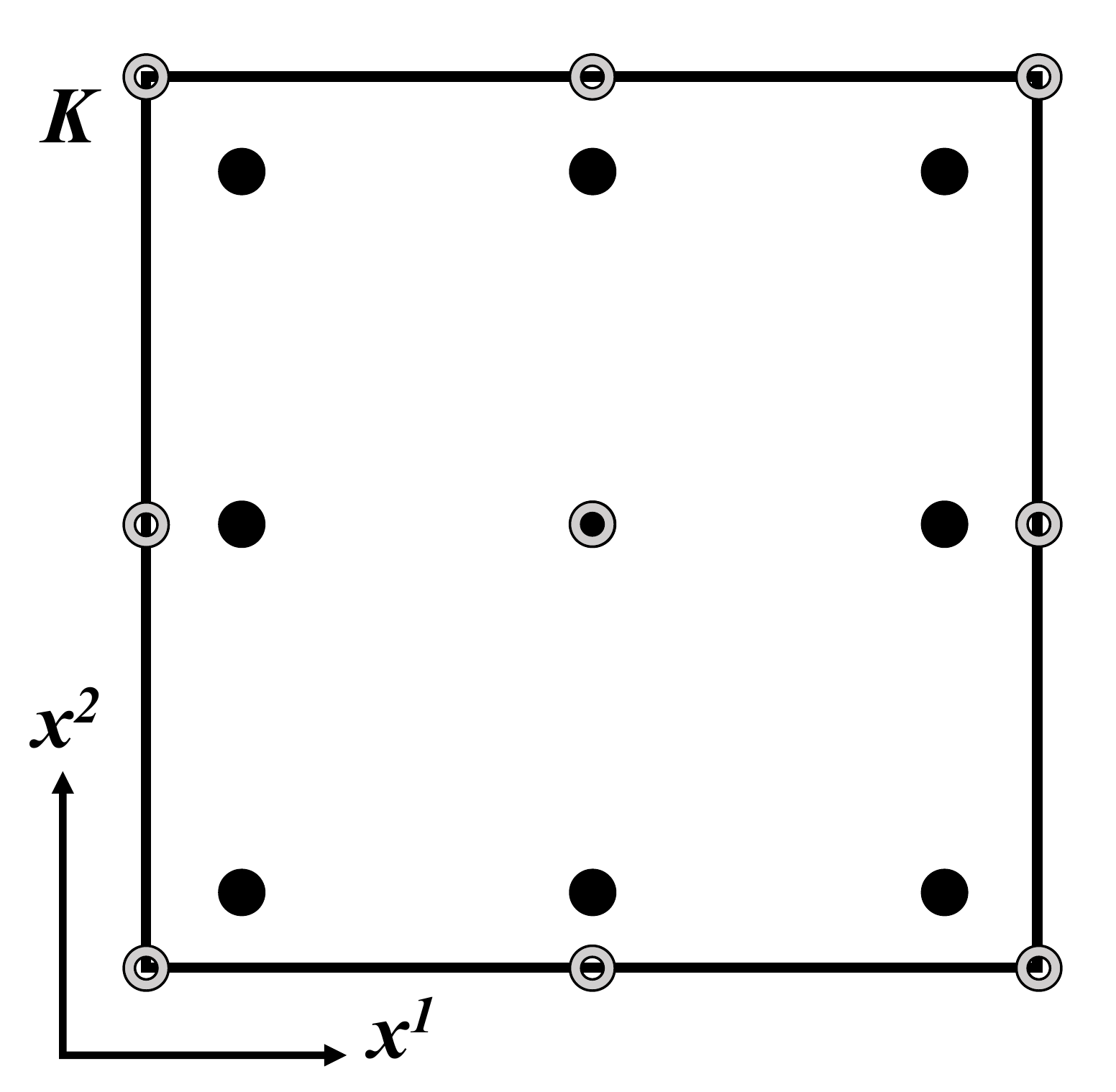}
  \end{minipage}
  \hspace{48pt}
  \begin{minipage}[h]{0.395\textwidth}
    \includegraphics[width=\textwidth]{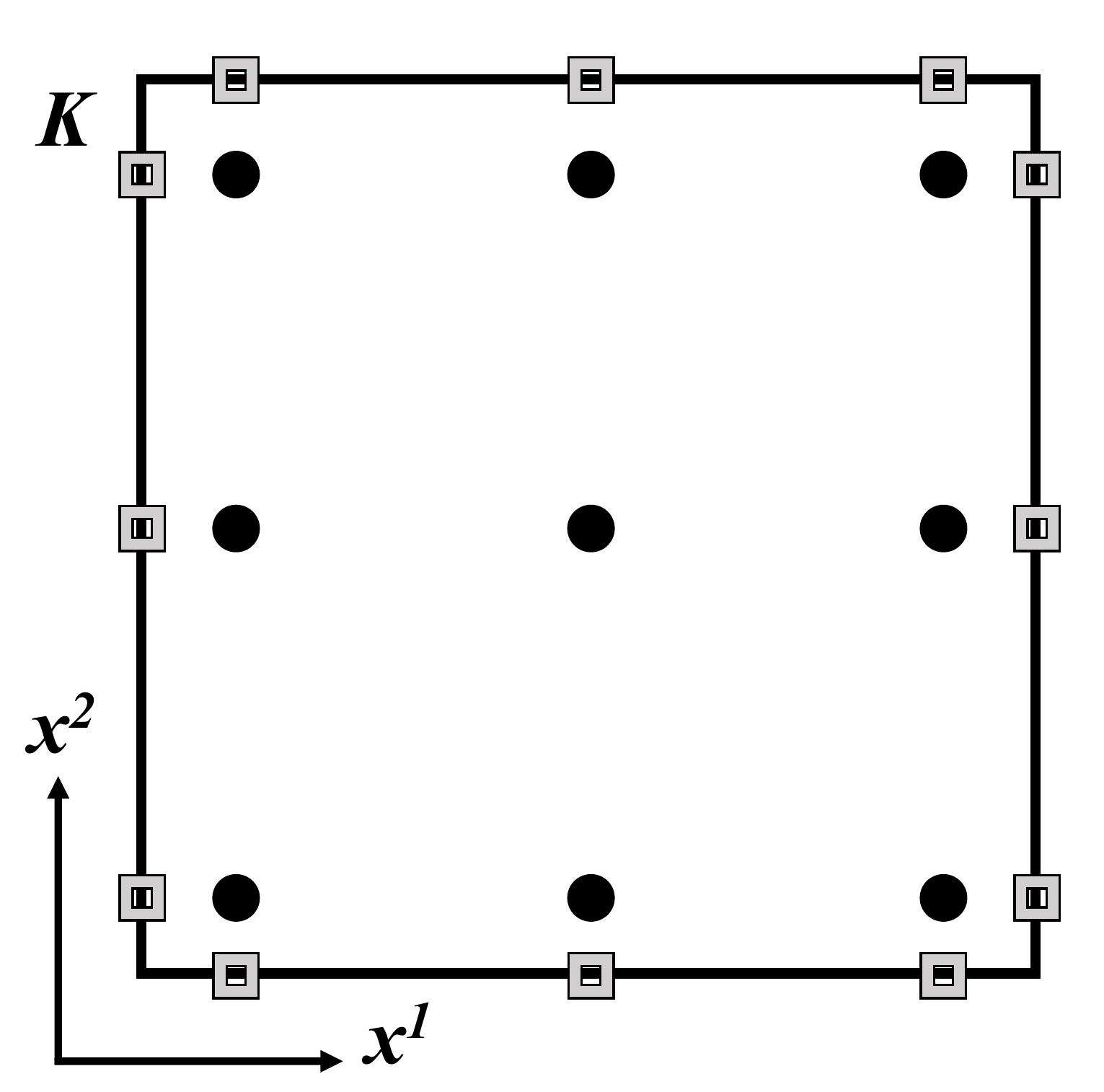}
  \end{minipage}
  \caption{Reference elements with interpolation and quadrature points used in the DG method implemented in \thornado\ for the two-dimensional case ($d=2$) with polynomials of degree $k=2$ ($N=3$).  In the left panel, interpolation points are shown for the hydrodynamics variables [$\vect{S}_{N}$ (based on LG quadrature points; black, filled circles)] and the geometry scale factors and the Newtonian gravitational potential [$\hat{\vect{S}}_{N}$ (based on LGL quadrature points; gray, open circles)].  In the right panel, quadrature points associated with volume integrals (black, filled circles) and surface integrals (gray, open squares) are shown.  Note that in the collocation nodal DG method, the interpolation points in the left panel, $\vect{S}_{N}$, coincide with the quadrature points in the right panel.  The quadrature points on the surface of the element are obtained as the projection of the quadrature points inside the element onto each surface.}
  \label{fig:ReferenceElements}
\end{figure}

Motivated by results presented by \cite{Bassi:2013}, we employ a spectral-type collocation nodal DG method in \thornado.  
To this end, we use Legendre--Gauss (LG) points to construct the interpolation points comprising $\vect{S}_{N}$.  
See the left panel of Figure~\ref{fig:ReferenceElements} for the distribution of the interpolation points $\vect{S}_{N}$ in the two-dimensional case with $k=2$ (black, filled circles).  
In the collocation nodal DG method, these interpolation points are also used as quadrature points to evaluate integrals in Equation~\eqref{eq:dgSemidiscreteWeak}.  
One of the benefits of this collocation method is computational efficiency since, even when using curvilinear coordinates, the mass matrix associated with the term in Equation~\eqref{eq:dgSemidiscreteTimeDerivative} is diagonal and easily invertible.  
On the other hand, demanding exact evaluation of integrals --- e.g., by using an extended quadrature set --- results in mass matrices that are non-diagonal and vary from element to element because of the spatially dependent metric determinant in $dV_{h}$ in Equation~\eqref{eq:dgSemidiscreteTimeDerivative}.  
The use of LG points, as opposed to Legendre--Gauss--Lobatto (LGL) points, provides better accuracy in evaluating the integrals.  
In the one-dimensional setting, the $N$-point LG quadrature evaluates polynomials of degree up to $2N-1$ exactly, while the corresponding LGL quadrature evaluates polynomials of degree up to $2N-3$ exactly.  
Let $Q_{N}^{i}$ denote the one-dimensional $N$-point LG quadrature on the interval $I^{i}$ with abscissas $\{\xi_{q}^{i}\}_{q=1}^{N}$ and weights $\{w_{q}^{i}\}_{q=1}^{N}$, normalized so that $\sum_{q=1}^{N}w_{q}^{i}=1$.  
Multidimensional integrals are evaluated by tensorization of one-dimensional quadratures.  
For volume integrals over the multidimensional reference element $\vect{I}=\times_{i=1}^{d}I^{i}$, we let $\vect{Q}_{N}=\otimes_{i=1}^{d}Q_{N}^{i}$ denote the tensorization of one-dimensional $N$-point LG quadrature rules with abscissas $\{\vect{\xi}_{\vect{q}}\}_{\vect{q}=\vect{1}}^{\vect{N}}$ and weights $\{w_{\vect{q}}\}_{\vect{q}=\vect{1}}^{\vect{N}}$, where $\vect{q}=\{q_{1},\ldots,q_{d}\}\in\mathbb{N}^{d}$, $\vect{\xi}_{\vect{q}}=\{\xi_{q_{1}}^{1},\ldots,\xi_{q_{d}}^{d}\}$, and $w_{\vect{q}}=w_{q_{1}}\times\ldots\times w_{q_{d}}$, so that the integral of a polynomial $P(\vect{x})\in\mathbb{V}_{h}^{k}$ in element $\vect{K}$ is evaluated as
\begin{align}
  &\int_{\vect{K}}P(\vect{x})\,d\vect{x}
  =|\vect{K}|\int_{\vect{I}}P(\vect{\xi})\,d\vect{\xi}
  =|\vect{K}|\,\vect{Q}_{N}\big[P(\vect{\xi})\big]
  =|\vect{K}|\sum_{\vect{q}=\vect{1}}^{\vect{N}}w_{\vect{q}}\,P(\vect{\xi}_{\vect{q}}) \nonumber \\
  &=\Delta x^{1}\times\ldots\times\Delta x^{d}\sum_{q_{1}=1}^{N}\ldots\sum_{q_{d}=1}^{N}w_{q_{1}}\times\ldots\times w_{q_{d}}\,P(\xi_{q_{1}}^{1},\ldots,\xi_{q_{d}}^{d}).  
  \label{eq:volumeQuadrature}
\end{align}
Similarly, for surface integrals over the reference surface element $\tilde{\vect{I}}^{i}=\times_{j=1,j\ne i}^{d}I^{j}$, we let $\tilde{\vect{Q}}_{N}^{i}=\otimes_{j=1,j\ne i}^{d}Q_{N}^{j}$ denote the tensorization of one-dimensional $N$-point LG quadrature rules with abscissas $\{\tilde{\vect{\xi}}_{\tilde{\vect{q}}_{i}}^{i}\}_{\tilde{\vect{q}}_{i}=\vect{1}}^{\vect{N}}$ and weights $\{w_{\tilde{\vect{q}}_{i}}\}_{\tilde{\vect{q}}_{i}=\vect{1}}^{\vect{N}}$, where $\tilde{\vect{q}}_{i}=\{q_{j}\}_{j=1,j\ne i}^{d}\in\mathbb{N}^{d-1}$, $\tilde{\vect{\xi}}_{\tilde{\vect{q}}_{i}}^{i}=\{\xi_{q_{j}}^{j}\}_{j=1,j\ne i}^{d}$, and $w_{\tilde{\vect{q}}_{i}}=\prod_{j=1,j\ne i}^{d}w_{q_{j}}$, so that for $P(x^{i},\tilde{\vect{x}}^{i})\in\mathbb{V}_{h}^{k}$, the integral over the surface element $\tilde{\vect{K}}^{i}$ is evaluated as
\begin{align}
  &\int_{\tilde{\vect{K}}^{i}}P(x^{i},\tilde{\vect{x}}^{i})\,d\tilde{\vect{x}}^{i}
  =|\tilde{\vect{K}}^{i}|\int_{\tilde{\vect{I}}^{i}}P(x^{i},\tilde{\vect{\xi}}^{i})\,d\tilde{\vect{\xi}}^{i}
  =|\tilde{\vect{K}}^{i}|\,\tilde{\vect{Q}}_{N}^{i}\big[P(x^{i},\tilde{\vect{\xi}}^{i})\big]
  =|\tilde{\vect{K}}^{i}|\sum_{\tilde{\vect{q}}_{i}=\vect{1}}^{\vect{N}}w_{\tilde{\vect{q}}_{i}}\,P(x^{i},\tilde{\vect{\xi}}_{\tilde{\vect{q}}_{i}}^{i}) \nonumber \\
  &\stackrel{(i=1)}{=}
  \Delta x^{2}\times\ldots\times\Delta x^{d}\sum_{q_{2}=1}^{N}\ldots\sum_{q_{d}=1}^{N}w_{q_{2}}\times\ldots\times w_{q_{d}}
  \,P(x^{1},\xi_{q_{2}}^{2},\ldots,\xi_{q_{d}}^{d}),
  \label{eq:surfaceQuadrature}
\end{align}
where the specific case with $i=1$ is given in the second line.  
The points used to evaluate volume integrals with the $\vect{Q}_{N}$ quadrature rule for the case with $d=k=2$ are shown as black, filled circles in the right panel in Figure~\ref{fig:ReferenceElements}.  
(Note that these points are identical to the interpolation points displayed as black, filled circles in the left panel in Figure~\ref{fig:ReferenceElements}.)  
The quadrature points used to evaluate surface integrals with $\tilde{\vect{Q}}^{1}$ and $\tilde{\vect{Q}}^{2}$ are shown as the gray, open squares on the boundary of the element.  

By inserting the expansion in Equation~\eqref{eq:nodalExpansion}, letting $\psi_{h}=\phi_{\vect{p}}$, where $\phi_{\vect{p}}$ is one of the basis functions in the expansion in Equation~\eqref{eq:nodalExpansion}, and using the quadrature rule in Equation~\eqref{eq:volumeQuadrature}, we can evaluate Equation~\eqref{eq:dgSemidiscreteTimeDerivative} as
\beq
  \langle\,\pd{\vect{U}_{h}}{t},\,\phi_{\vect{p}}\,\rangle_{\vect{K}}
  :=w_{\vect{p}}\,|\vect{K}|\,\sqrt{\gamma_{\vect{p}}}\,\pd{\vect{U}_{\vect{p}}}{t},
  \label{eq:dgTimeDerivativeNodal}
\eeq
where $w_{\vect{j}}\,|\vect{K}|\,\sqrt{\gamma_{\vect{j}}}$ are the elements of the diagonal mass matrix and $\gamma_{\vect{j}}=\gamma_{h}(\vect{x}_{\vect{j}})$.  
Similarly, using the quadrature in Equation~\eqref{eq:surfaceQuadrature}, the contributions from fluxes can be written as
\begin{align}
  \mathcal{B}_{h}^{\sf{Flx}}\big(\vect{U}_{h},\phi_{\vect{p}}\big)_{\vect{K}}
  :=&-\sum_{i=1}^{d}w_{\tilde{\vect{p}}_{i}}\,|\tilde{\vect{K}}^{i}|
  \Big(\,
    \sqrt{\gamma_{h}(x_{\Hi}^{i},\tilde{\vect{x}}_{\tilde{\vect{p}}_{i}}^{i})}\,\widehat{\vect{F}}^{i}(x_{\Hi}^{i},\tilde{\vect{x}}_{\tilde{\vect{p}}_{i}}^{i})\,\ell_{p_{i}}(x_{\Hi}^{i,-})
    -\sqrt{\gamma_{h}(x_{\Lo}^{i},\tilde{\vect{x}}_{\tilde{\vect{p}}_{i}}^{i})}\,\widehat{\vect{F}}^{i}(x_{\Lo}^{i},\tilde{\vect{x}}_{\tilde{\vect{p}}_{i}}^{i})\,\ell_{p_{i}}(x_{\Lo}^{i,+})
  \,\Big) \nonumber \\
  \hspace{24pt}
  &+\sum_{i=1}^{d}w_{\tilde{\vect{p}}_{i}}\,|\tilde{\vect{K}}^{i}|
  \sum_{q_{i}=1}^{N}w_{q_{i}}\sqrt{\gamma_{h}(x_{q_{i}}^{i},\tilde{\vect{x}}_{\tilde{\vect{p}}_{i}}^{i})}\,\vect{F}^{i}(x_{q_{i}}^{i},\tilde{\vect{x}}_{\tilde{\vect{p}}_{i}}^{i})\,\pderiv{\ell_{p_{i}}}{\xi^{i}}(\xi_{q_{i}}^{i}).  
  \label{eq:dgFluxesNodal}
\end{align}
Finally, the source term becomes
\beq
  \langle\,\vect{S}(\vect{U}_{h},\Phi_{h}),\,\phi_{\vect{p}}\,\rangle_{\vect{K}}
  :=w_{\vect{j}}\,|\vect{K}|\,\sqrt{\gamma_{\vect{p}}}\,\vect{S}_{\vect{p}},
  \label{eq:dgSourcesNodal}
\eeq
where $\vect{S}_{\vect{p}}$ is the source vector in Equation~\eqref{eq:eulerSystemComponents}, evaluated in $\vect{x}_{\vect{p}}$.  
Combining Equations~\eqref{eq:dgTimeDerivativeNodal}, \eqref{eq:dgFluxesNodal}, and \eqref{eq:dgSourcesNodal}, we can now write the spectral-type collocation DG approximation to the semi-discrete DG problem in Equation~\eqref{eq:dgSemidiscreteWeak} in terms of an evolution equation for the expansion coefficient $\vect{U}_{\vect{p}}$ in element $\vect{K}$ as
\begin{align}
  \pd{\vect{U}_{\vect{p}}}{t}
  &=-\sum_{i=1}^{d}\f{1}{w_{p_{i}}\Delta x^{i}\,\sqrt{\gamma_{\vect{p}}}}
  \Big(\,
    \sqrt{\gamma_{h}(x_{\Hi}^{i},\tilde{\vect{x}}_{\tilde{\vect{p}}_{i}}^{i})}\,\widehat{\vect{F}}^{i}(x_{\Hi}^{i},\tilde{\vect{x}}_{\tilde{\vect{p}}_{i}}^{i})\,\ell_{p_{i}}(x_{\Hi}^{i,-})
    -\sqrt{\gamma_{h}(x_{\Lo}^{i},\tilde{\vect{x}}_{\tilde{\vect{p}}_{i}}^{i})}\,\widehat{\vect{F}}^{i}(x_{\Lo}^{i},\tilde{\vect{x}}_{\tilde{\vect{p}}_{i}}^{i})\,\ell_{p_{i}}(x_{\Lo}^{i,+})
  \,\Big) \nonumber \\
  &\hspace{12pt}
  +\sum_{i=1}^{d}\f{1}{w_{p_{i}}\Delta x^{i}\,\sqrt{\gamma_{\vect{p}}}}
  \sum_{q_{i}=1}^{N}w_{q_{i}}\sqrt{\gamma_{h}(x_{q_{i}}^{i},\tilde{\vect{x}}_{\tilde{\vect{p}}_{i}}^{i})}\,\vect{F}^{i}(x_{q_{i}}^{i},\tilde{\vect{x}}_{\tilde{\vect{p}}_{i}}^{i})\,\pderiv{\ell_{p_{i}}}{\xi^{i}}(\xi_{q_{i}}^{i})
  +\vect{S}_{\vect{p}}.  
  \label{eq:dgSemidiscreteWeakNodal}
\end{align}
(For an example of Equation~\eqref{eq:dgSemidiscreteWeakNodal} in the simpler one-dimensional setting, see \citet{endeve:2019}; their Equation~(11).)  

The cell averages in element $\vect{K}$, defined as
\beq
  \vect{U}_{\vect{K}}
  =\f{1}{V_{\vect{K}}}\int_{\vect{K}}\vect{U}_{h}\,dV_{h}
  :=\f{\sum_{\vect{p}=\vect{1}}^{\vect{N}}w_{\vect{p}}\,\sqrt{\gamma_{\vect{p}}}\,\vect{U}_{\vect{p}}}{\sum_{\vect{p}=\vect{1}}^{\vect{N}}w_{\vect{p}}\,\sqrt{\gamma_{\vect{p}}}},
  \quad\text{where}\quad
  V_{\vect{K}}=|\vect{K}|\sum_{\vect{p}=\vect{1}}^{\vect{N}}w_{\vect{p}}\,\sqrt{\gamma_{\vect{p}}},
  \label{eq:cellAverage}
\eeq
play an important role in the analysis and implementation of the DG method given by Equation~\eqref{eq:dgSemidiscreteWeakNodal}.  
(Examples of the use of the cell averages are given in Sections~\ref{sec:slope} and \ref{sec:boundEnforcing}, where we discuss limiting techniques.)  
From the definition of the cell average in Equation~\eqref{eq:cellAverage} and from Equation~\eqref{eq:dgSemidiscreteWeakNodal}, the equation for the cell average can be written as
\beq
  \pd{\vect{U}_{\vect{K}}}{t}
  =-\f{|\vect{K}|}{V_{\vect{K}}}\sum_{i=1}^{d}\tilde{\vect{Q}}_{N}^{i}\big[\,\sqrt{\gamma_{h}}(x_{\Hi}^{i},\tilde{\vect{x}}^{i})\widehat{\vect{F}}^{i}(x_{\Hi}^{i},\tilde{\vect{x}}^{i})-\sqrt{\gamma_{h}}(x_{\Lo}^{i},\tilde{\vect{x}}^{i})\widehat{\vect{F}}^{i}(x_{\Lo}^{i},\tilde{\vect{x}}^{i})\,\big]/\Delta x^{i}
  +\vect{S}_{\vect{K}},
  \label{eq:dgSemidiscreteCellAverage}
\eeq
where we used the quadrature rule in Equation~\eqref{eq:surfaceQuadrature} to represent the surface integrals, while the source term can be written in terms of the quadrature rule in Equation~\eqref{eq:volumeQuadrature}
\beq
  \vect{S}_{\vect{K}} = \f{|\vect{K}|}{V_{\vect{K}}}\,\vect{Q}_{N}\big[\,\sqrt{\gamma_{h}}(\vect{x})\,\vect{S}(\vect{x})\,\big].  
  \label{eq:dgSemidiscreteCellAverageSource}
\eeq
To arrive at Equation~\eqref{eq:dgSemidiscreteCellAverage}, we used the property of the Lagrange polynomial in Equation~\eqref{eq:lagrangePoly} that $\sum_{p=1}^{N}\ell_{p}(\xi^{i})=1$ for any $\xi^{i}\in I^{i}$.  
Equation~\eqref{eq:dgSemidiscreteCellAverage} exhibits the expected conservation form, with quadrature rules replacing integrals over the surface of $\vect{K}$.  
In the absence of sources, the DG discretization in Equation~\eqref{eq:dgSemidiscreteWeakNodal} is conservative for mass, momentum, energy, and electron number.  
We also note that Equation~\eqref{eq:dgSemidiscreteCellAverage} is familiar from the literature on finite-volume (FV) methods, which only evolve the cell averages.  
The DG and FV methods are in fact equivalent in the first-order case, when $k=0$.  
However, for the extension to higher-order, FV methods reconstruct a local polynomial using cell averages in neighboring elements, while DG methods evolve all the degrees-of-freedom in the local polynomial representation, so that the reconstruction step is not needed.  
Thus, one benefit of avoiding the reconstruction step becomes clear in the high-order case: while the FV stencil width increases with increasing spatial order of accuracy, the DG method only requires data from the local element and its nearest neighbors, independent of the order of accuracy.  

We complete the specification of the basic DG method implemented in \thornado\ by discussing the source terms due to the use of curvilinear coordinates and gravitational fields.  
In particular, we write [cf. Equation~\eqref{eq:eulerSources}]
\beq
  \vect{S}_{\vect{p}} = \vect{S}_{\vect{p}}^{\gamma} + \vect{S}_{\vect{p}}^{\Phi}.  
\eeq

\subsubsection{Geometric Source Terms}

For the sources due to curvilinear coordinates, $\vect{S}_{\vect{p}}^{\gamma}$, the only nonzero components appear in the components of the momentum equation, which can be written in terms of the scale factors where, due to the diagonal metric, $\gamma_{ii}=h_{i}h_{i}$ and $\gamma^{ii}=1/\gamma_{ii}$
\begin{equation}
  \f{1}{2}\,\Pi^{ik}\pd{\gamma_{ik}}{j}
  =\f{1}{2}\,\Pi^{11}\pd{\gamma_{11}}{j} + \f{1}{2}\,\Pi^{22}\pd{\gamma_{22}}{j} + \f{1}{2}\,\Pi^{33}\pd{\gamma_{33}}{j}
  =\Pi^{1}_{\hspace{4pt}1}\f{1}{h_{1}}\pderiv{h_{1}}{x^{j}} + \Pi^{2}_{\hspace{4pt}2}\f{1}{h_{2}}\pderiv{h_{2}}{x^{j}} + \Pi^{3}_{\hspace{4pt}3}\f{1}{h_{3}}\pderiv{h_{3}}{x^{j}}.  
  \label{eq:geometrySources}
\end{equation}
For the coordinate systems we consider here, the scale factors are independent of $x^{3}$, and only the first and second components of Equation~\eqref{eq:geometrySources} are nonzero (i.e., $j=1,2$; cf.~Table~\ref{tab:metricFunctions}).  
Note that $h_{1}=1$ for all the coordinate systems; therefore, spatial derivatives of $h_{1}$ vanish.  
For Cartesian coordinates, the scale factors are unity, and all the components of $\vect{S}_{\vect{p}}^{\gamma}$ vanish.  
For cylindrical coordinates, only $h_{3}=R$ contributes, while for spherical-polar coordinates both $h_{2}=r$ and $h_{3}=r\sin\theta$ contribute.  
In \thornado, we approximate the scale factors by polynomials in each element.  
To this end, we define $\vect{h}=\big(h_{1},h_{2},h_{3}\big)^{T}$ and let the scale factors in $\vect{K}$ be given by the expansion
\begin{equation}
  \vect{h}_{h}(\vect{x}) = \sum_{\vect{i}=\vect{1}}^{\vect{N}}\vect{h}_{\vect{i}}\,\hat{\phi}_{\vect{i}}(\vect{x}) \in \mathbb{V}_{h}^{k},  
  \label{eq:nodalExpansionScaleFactors}
\end{equation}
where $\hat{\phi}_{\vect{i}}(\vect{x})$ are basis functions, similar to those defined in Equation~\eqref{eq:basisFunctionMultiD}.  
However, we demand that the scale factors are continuous across element interfaces.  
To achieve this we let $\hat{S}_{N}^{i}=\{\hat{\xi}_{1},\ldots,\hat{\xi}_{N}^{i}\}\subseteq I^{i}$ denote the set of LGL points in the unit reference interval, since the LGL points include the endpoints of $I^{i}$.  
For the scale factors (and, as discussed below, the Newtonian gravitational potential), we then let the interpolation points on $\vect{K}$ be given by $\hat{\vect{S}}_{N}=\otimes_{i=1}^{d}\hat{S}_{N}^{i}$.  
The distribution of the interpolation points $\hat{\vect{S}}_{N}$, used for the scale factors and the Newtonian gravitational potential, for the two-dimensional case with $k=2$ are shown in the left panel of Figure~\ref{fig:ReferenceElements} (gray, open circles).  
Hence, $\hat{\phi}_{\vect{i}}(\vect{x})$ is defined as in Equation~\eqref{eq:basisFunctionMultiD}, but with the Lagrange polynomials in Equation~\eqref{eq:lagrangePoly} constructed with the LGL points $\hat{\vect{S}}_{N}$, and the expansion coefficients $\vect{h}_{\vect{i}}$ are given by the exact value of the scale factors in the LGL points.  
Scale factors in the LG points $\vect{x}_{\vect{i}}\in\vect{S}_{N}$, which are needed, e.g., to compute the determinant of the spatial metric, are obtained from direct evaluation of Equation~\eqref{eq:nodalExpansionScaleFactors}, $\vect{h}_{h}(\vect{x}_{\vect{i}})$, so that $\gamma_{\vect{i}}=\gamma_{h}(\vect{x}_{\vect{i}}):=\gamma(\vect{h}_{h}(\vect{x}_{\vect{i}}))$.  
Derivatives of the scale factors, needed for the source terms in Equation~\eqref{eq:geometrySources}, are evaluated by analytic differentiation of Equation~\eqref{eq:nodalExpansionScaleFactors}.  
Since in the present case the metric is time independent, the needed scale factors and their derivatives can be precomputed at program startup and stored for later use.  
Note that scale factors are polynomials and at most linear functions of the spherical-polar or cylindrical radius, so the representation is exact in the $x^{1}$-dimension if $N\ge2$.  
However, for spherical-polar coordinates, $h_{3}$ is a trigonometric function in the $x^{2}$-dimension, and the representation in Equation~\eqref{eq:nodalExpansionScaleFactors} is only approximate.  

Next we consider a special case where the geometric source terms, $\f{1}{2}\Pi^{ik}\pd{\gamma_{ik}}{j}$, and the divergence of the stress tensor, $\f{1}{\sqrt{\gamma}}\pd{}{i}\big(\sqrt{\gamma}\,\Pi^{i}_{\hspace{2pt}j}\big)$, appearing in the components of the momentum equation, Equation~\eqref{eq:momentumConservation}, must balance each other.  
Specifically, for a fluid associated with an isotropic and spatially homogeneous stress tensor, i.e., $\Pi^{i}_{\hspace{2pt}k}=p_{0}\,\delta^{i}_{\hspace{2pt}k}$ ($p_{0}=\mbox{constant}$), the divergence of the stress tensor must balance the geometry source exactly to prevent inducing spurious flows.  

Considering Equation~\eqref{eq:dgSemidiscreteCellAverage}, with Equations~\eqref{eq:dgSemidiscreteCellAverageSource} and \eqref{eq:geometrySources}, in spherical-polar coordinates and in the absence of gravity, assuming an isotropic and spatially homogeneous stress tensor, the equation for the first component of the momentum density (cf. Equation~\eqref{eq:eulerSystemComponents}), in the sense of the cell-average, can be written as
\begin{align}
  \pd{(\rho v_{1})_{\vect{K}}}{t}
  &=-p_{0}\,\f{|\tilde{\vect{K}}^{1}|}{V_{\vect{K}}}\,
  \Big\{\,
    \tilde{\vect{Q}}_{N}^{1}\big[\,\sqrt{\gamma_{h}}(x_{\Hi}^{1},\tilde{\vect{x}}^{1})-\sqrt{\gamma_{h}}(x_{\Lo}^{1},\tilde{\vect{x}}^{1})\,\big]
    -\vect{Q}_{N}\big[\,2\sqrt{\gamma_{h}}(\vect{x})/x^{1}\,\big]
  \,\Big\} \nonumber \\
  &=-p_{0}\,\f{|\tilde{\vect{K}}^{1}|}{V_{\vect{K}}}\,
  \tilde{\vect{Q}}_{N}^{1}\big[\,(\sin\theta)_{h}\,\big(\,(r_{\Hi}^{2}-r_{\Lo}^{2})-2\,Q_{N}^{1}\big[r\big]\,\big)\,\big],
  \label{eq:momentumOneCellAverageSphericalPolar}
\end{align}
where $(\sin\theta)_{h}$ is the polynomial approximation to $\sin\theta$.  
Because the stress tensor is isotropic and spatially homogeneous, the numerical flux in the first component of the momentum equation is simply $\widehat{F}_{(\rho v_{1})}^{i}=p_{0}\delta^{i}_{\hspace{2pt}1}$.  
The right-hand side of Equation~\eqref{eq:momentumOneCellAverageSphericalPolar} vanishes because the LG quadrature, with $N\ge1$, is exact for the radial integral; i.e., $2\,Q_{N}^{1}\big[r\big]=(r_{\Hi}^{2}-r_{\Lo}^{2})$.  
Similarly, the second component of the momentum equation can be written as
\beq
  \pd{(\rho v_{2})_{\vect{K}}}{t}
  =-p_{0}\,\f{|\tilde{\vect{K}}^{2}|}{V_{\vect{K}}}\,
  \tilde{\vect{Q}}_{N}^{2}\big[\,r^{2}\,\big(\,(\sin\theta_{\Hi}-\sin\theta_{\Lo})-Q_{N}^{2}\big[\pd{}{\xi^{2}}(\sin\theta)_{h}\big]\,\big)\,\big].  
  \label{eq:momentumTwoCellAverageSphericalPolar}
\eeq
Since $(\sin\theta)_{h}$ is approximated by a polynomial of degree $k=N-1$, the $N$-point LG quadrature in the $\theta$-direction is evaluated exactly, so that $Q_{N}^{2}\big[\pd{}{\xi^{2}}(\sin\theta)_{h}\big]=(\sin\theta_{\Hi}-\sin\theta_{\Lo})$, which implies that the right-hand side of Equation~\eqref{eq:momentumTwoCellAverageSphericalPolar} vanishes.  
Note that these properties hold for polynomial approximations with $k\ge1$.  
The first-order accurate scheme ($k=0$) requires special treatment, and is not discussed here.  
(See, e.g., \citet{monchmeyer:1989} and \citet{blondin:1993}, for finite-volume schemes and associated challenges when using spherical-polar coordinates.)  

In cylindrical coordinates, the source term in Equation~\eqref{eq:geometrySources} contributes only to the first component of the momentum equation.  
In this case, the equation for the cell-average can be written as
\beq
  \pd{(\rho v_{1})_{\vect{K}}}{t}
  =-p_{0}\,\f{|\tilde{\vect{K}}^{1}|}{V_{\vect{K}}}\,
  \tilde{\vect{Q}}_{N}^{1}\big[\,(R_{\Hi}-R_{\Lo})-Q_{N}^{1}\big[\pd{R}{\xi^{1}}\big]\,\big].
  \label{eq:momentumOneCellAverageCylindrical}
\eeq
Again, since the quadrature in the $R$-direction is exact, $Q_{N}^{1}\big[\pd{R}{\xi^{1}}\big]=(R_{\Hi}-R_{\Lo})$, and the right-hand side of Equation~\eqref{eq:momentumOneCellAverageCylindrical} vanishes, as is desired under the conditions of an isotropic and spatially homogeneous stress tensor.  

\subsubsection{Gravitational Source Terms}

For the gravitational source terms appearing in the momentum and energy equations, our approach is similar to that used for the geometric sources discussed above.  
The gravitational potential in element $\vect{K}$ is approximated by the polynomial
\beq
  \Phi_{h}(\vect{x}) = \sum_{\vect{i}=\vect{1}}^{\vect{N}}\Phi_{\vect{i}}\,\hat{\phi}_{\vect{i}}(\vect{x}),
  \label{eq:nodalExpansionPotential}
\eeq
constrained to be continuous on the element interfaces, so that
\beq
  \Phi_{h}(x_{\Lo/\Hi}^{i,+},\tilde{\vect{x}}^{i}) = \Phi_{h}(x_{\Lo/\Hi}^{i,-},\tilde{\vect{x}}^{i}).  
\eeq
(Continuity of the potential on the element interfaces is guaranteed by the finite-element method in Section~\ref{sec:poisson}.)  
We then compute derivatives of the gravitational potential by analytic differentiation of the expansion in Equation~\eqref{eq:nodalExpansionPotential}, and write the momentum and energy sources in the interpolation point $\vect{x}_{\vect{p}}\in \vect{S}_{N}$ as
\beq
  \big(S_{\rho v_{j}}^{\Phi}\big)_{\vect{p}} = - \rho_{\vect{p}}\,(\pd{\Phi_{h}}{j})_{\vect{p}}
  \quad\text{and}\quad
  \big(S_{E}^{\Phi}\big)_{\vect{p}} = - \sum_{j=1}^{d}(\rho v^{j})_{\vect{p}}\,(\pd{\Phi_{h}}{j})_{\vect{p}},
  \label{eq:gravitationalSourcesDiscrete}
\eeq
where, $\rho_{\vect{p}}$, $(\rho v^{j})_{\vect{p}}$, and $(\pd{\Phi_{h}}{j})_{\vect{p}}$ are, respectively, the mass density, momentum density, and the derivative of Equation~\eqref{eq:nodalExpansionPotential}, evaluated in $\vect{x}_{\vect{p}}$.  
We note that the source terms in Equation~\eqref{eq:gravitationalSourcesDiscrete} are not well-balanced, i.e. designed specifically to capture steady states (e.g., hydrostatic equilibrium), which would require special treatment \citep[see, e.g.,][]{kappeli:2016,li:2018}.  

\subsection{Time Integration}
\label{sec:time}

After application of the DG spatial discretization, Equation~\eqref{eq:dgSemidiscreteWeak} can be viewed as a system of ordinary differential equations (ODEs), which can be written as
\begin{equation}
    \frac{d}{dt}\langle\,\vect{U}_{h},\,\psi_{h}\,\rangle_{\vect{K}} =\mathcal{B}_{h}\big(\vect{U}_{h},\Phi_{h},\psi_{h}\big)_{\vect{K}}.
\end{equation}
This system of ODEs is evolved with the explicit strong stability-preserving Runge-Kutta (SSP-RK) methods of \citet{shu:1989} \citep[see also][]{gottlieb:2001,Cockburn:2001}.  
Denoting the fluid fields and the gravitational potential at time $t^{n}$ by $\vect{U}_{h}^{n}$ and $\Phi_{h}^{n}$, respectively, the time stepping algorithm advancing the solution from $t^{n}$ to $t^{n+1}=t^{n}+\Delta t^{n}$ with $s$ stages is,
$\forall\psi_{h}\in\mathbb{V}_{h}^{k}$ and $\forall \vect{K}\in\mathcal{T}$,

\begin{algorithm}[h]
  $\langle\,\vect{U}_{h}^{\left(0\right)},\psi_{h}\rangle_{\vect{K}}:=\Lambda^{\textsc{be}}\left\{\Lambda^{\textsc{tvd}}\left\{\langle\,\vect{U}_{h}^{n},\psi_{h}\rangle_{\vect{K}}\right\}\right\}$ \\
  $\Phi_{h}^{\left(0\right)}\hspace{3.65em}:=\Phi_{h}^{n}$\\
  \For{$i=1,\ldots,s$}{
    $\langle\,\vect{U}_{h}^{\left(i\right)},\psi_{h}\rangle_{\vect{K}}:=\Lambda^{\textsc{be}}\left\{\,\Lambda^{\textsc{tvd}}\left\{\,\sum\limits_{j=0}^{i-1}\alpha_{ij}\left(\langle\,\vect{U}_{h}^{\left(j\right)},\psi_{h}\rangle_{\vect{K}}+\frac{\beta_{ij}}{\alpha_{ij}}\,\Delta t^{n}\,\mathcal{B}_{h}^{\left(j\right)}\right)\,\right\}\,\right\}$,\\
    \hspace{2em}where  $\mathcal{B}^{\left(j\right)}_{h}:=\mathcal{B}_{h}\left(\vect{U}^{\left(j\right)}_{h},\Phi^{\left(j\right)}_{h},\psi_{h}\right)$, with $\Phi_{h}^{\left(j\right)}:=\Phi\left(\vect{U}^{\left(j\right)}_{h}\right)$\\
    $\Phi_{h}^{\left(i\right)}:=\Phi\left(\vect{U}_{h}^{\left(i\right)}\right)$
}
  $\langle\,\vect{U}_{h}^{n+1},\psi_{h}\rangle_{\vect{K}}:=\langle\,\vect{U}_{h}^{\left(s\right)},\psi_{h}\rangle_{\vect{K}}$\\ $\Phi_{h}^{n+1}\hspace{3.65em}:=\Phi_{h}\left(\vect{U}_{h}^{n+1}\right)$
   \caption{Algorithm for SSP-RK time integration.} \label{alg.SSPRK}
\end{algorithm}
\noindent
Note that line~6 in Algorithm \ref{alg.SSPRK} invokes the Poisson solver for the gravitational potential.  
Details about the coefficients $\alpha_{ij}$ and $\beta_{ij}$ can be found in \citet{Cockburn:2001}.  
In order to for the evolution of the cell-average of the solution to be stable, the time step must satisfy the Courant--Friedrichs--Lewy (CFL) condition,
\begin{equation}
    \Delta t\leq \frac{C_{\textsc{cfl}}}{d\left(2k+1\right)}\times\min_{i\in\{1,\ldots,d\}}\left(\frac{\Delta x^{i}}{\left|\lambda^{i}\right|}\right),
    \label{eq:cflCondition}
\end{equation}
where $d$ is the number of spatial dimensions, $k$ is the maximal degree of the one-dimensional polynomials comprising $\mathbb{V}^{k}_{h}$, $C_{\textsc{cfl}}\lesssim1$ is the
CFL number, and $\lambda^{i}$ is the largest (in magnitude) eigenvalue of the flux Jacobian in Equation~\eqref{eq:eigenDecomposition}, corresponding to the fastest-moving wave in the $i$th spatial dimension. 

In principle, one would also need an additional restriction on the time step to guarantee that the solution remains in the set of physically admissible states (see Section~\ref{sec:boundEnforcing}).  
However, we do not enforce such a condition because in practice we find the CFL condition given by Equation~\eqref{eq:cflCondition} to be sufficient.

The operators $\Lambda^{\textsc{tvd}}$ and $\Lambda^{\textsc{be}}$ invoked in lines~1 and 4 in Algorithm \ref{alg.SSPRK} represent slope and bound-enforcing limiters, respectively, and play an important role in RKDG methods.  
In particular, the slope limiter is required in order for the SSP-RK method to guarantee stability when applied to non-linear problems \citep{Cockburn:2001}.  

\subsection{Slope Limiting}
\label{sec:slope}

To improve stability of the Runge-Kutta DG (RKDG) algorithm and prevent unphysical oscillations in the solutions around discontinuities, it is necessary to implement a limiting procedure for the polynomial $\mathbf{U}_{h}$.  
To this end, we use the basic minmod-type total variation diminishing (TVD) slope limiter \citep[see, e.g.,][]{cockburn:1998} in conjunction with the troubled-cell indicator (TCI) proposed by \citet{fu:2017}.  
The TCI prevents excessive limiting by only flagging elements where limiting is needed.  
When using the basic TVD limiter one assumes that any spurious oscillations are evident in the part of the solution that is represented by piecewise linear functions, and under- and over-shoots of the higher-order solution at inter-cell boundaries are detected by comparing local slopes with slopes constructed using cell averages of the target cell and its neighbors.  
Our implementation follows closely the description in \citet{schaal:2015a} for the case of an ideal EoS.  
Recall from Eq.~\eqref{eq:nodalExpansion} that in each cell the solution is expressed in the nodal form.  
It is convenient, however, for limiting purposes to express the solution in $\vect{K}$ using a modal representation
\beq
  \mathbf{U}_h (\vect{x},t) = \sum_{\vect{l}=\vect{1}}^{\vect{N}}\vect{C}_{\vect{l}}(t)\,P_{\vect{l}}(\vect{x}),
  \label{eq:modalExpansion}
\eeq
where the multidimensional basis functions $P_{\vect{l}}(\vect{x})\in\mathbb{V}_{h}^{k}$ are constructed from one-dimensional Legendre polynomials $\{P_{l}(\xi^{i})\}_{l=1}^{N}$ by tensorization.  
The Legendre polynomials are orthogonal on the unit interval $I^{i}$, and we use a normalization such that $P_{1}(\xi^{i})=1$ and $P_{2}(\xi^{i})=\xi^{i}$ (i.e., the polynomials are not orthonormal).  
Note that the case with $\vect{l}=\{l_{1},\ldots,l_{d}\}=\{1,\ldots,1\}=\vect{1}$ corresponds to $P_{\vect{1}}(\vect{x})=P_{1}(\xi^{1})\times\ldots\times P_{1}(\xi^{d})=1$; therefore, $\vect{C}_{\vect{1}}$ is equal to the cell average when Cartesian coordinates are used ($\gamma_{h}=1$); i.e.,
\beq
  \vect{C}_{\vect{1}} = \f{1}{|\vect{K}|} \int_{\vect{K}}\vect{U}_{h}\,d\vect{x}.  
  \label{eq:modalCoefficientOne}
\eeq
In our multi-index notation we define $|\vect{l}|=\sum_{i=1}^{d}l_{i}$, so that the basis functions $P_{\vect{l}}$ with $\vect{l}$ satisfying $|\vect{l}|=d+1$ are linear in one of the coordinates.  
For example, for the three-dimensional case ($d=3$) we have exactly three basis functions satisfying $|\vect{l}|=d+1=4$
\begin{align}
  P_{\{2,1,1\}}(\vect{x})&=P_{2}(\xi^{1})\times P_{1}(\xi^{2})\times P_{1}(\xi^{3})=P_{2}(\xi^{1})=\xi^{1}, \\
  P_{\{1,2,1\}}(\vect{x})&=P_{1}(\xi^{1})\times P_{2}(\xi^{2})\times P_{1}(\xi^{3})=P_{2}(\xi^{2})=\xi^{2},\quad\text{and} \\
  P_{\{1,1,2\}}(\vect{x})&=P_{2}(\xi^{1})\times P_{1}(\xi^{2})\times P_{2}(\xi^{3})=P_{2}(\xi^{3})=\xi^{3},
\end{align}
which are linear in the reference coordinates $\xi^{1}$, $\xi^{2}$, and $\xi^{3}$, respectively.  
From orthogonality of the Legendre polynomials, we can identify the expansion coefficients satisfying $|\vect{l}|=4$ in the modal representation in Equation~\eqref{eq:modalExpansion} as the average derivative of $\vect{U}_{h}$ with respect to the reference coordinates $\xi^{1}$, $\xi^{2}$, and $\xi^{3}$, respectively; i.e.,
\beq
  \vect{C}_{\{2,1,1\}} = \f{1}{|\bK|}\int_{\bK}(\pd{\vect{U}_{h}}{\xi^{1}})\,d\vect{x},\quad
  \vect{C}_{\{1,2,1\}} = \f{1}{|\bK|}\int_{\bK}(\pd{\vect{U}_{h}}{\xi^{2}})\,d\vect{x},\quad\text{and}\quad
  \vect{C}_{\{1,1,2\}} = \f{1}{|\bK|}\int_{\bK}(\pd{\vect{U}_{h}}{\xi^{3}})\,d\vect{x}.
  \label{eq:modalSlopes}
\eeq
These coefficients are here obtained by taking the derivative of Equation~\eqref{eq:modalExpansion} with respect to $\xi^{1}$, $\xi^{2}$, and $\xi^{3}$, respectively, and integrating over the element.  

The representations of the solution in Equations~\eqref{eq:nodalExpansion} and \eqref{eq:modalExpansion} are equivalent in the least squares sense
\beq
  \sum_{\vect{i}=\vect{1}}^{\vect{N}}\int_{\bK}
  \big(\, \vect{U}_{\vect{i}}(t)\, \phi_{\vect{i}}(\vect{x}) - \vect{C}_{\vect{i}}(t)\, P_{\vect{i}}(\vect{x}) \,\big)\, \psi_{h}(\vect{x})\, d\vect{x} = 0, \quad \forall\,\psi_{h} \in \mathbb{V}_{h}^{k},
  \label{eq:NodalModal}
\eeq
which provides a change of basis between Lagrange and Legendre polynomial representations, and relates the coefficients of nodal and modal representations by linear transformations.  
Setting $\psi_{h}=\phi_{\vect{j}}$ in Equation~\eqref{eq:NodalModal} gives the nodal coefficients in terms of the modal coefficients
\beq
  \vect{U}_{\vect{j}} = \sum_{\vect{i}=\vect{1}}^{\vect{N}}P_{\vect{i}}(\vect{\xi}_{\vect{j}})\,\vect{C}_{\vect{i}},
\eeq
while setting $\psi_{h}=P_{\vect{j}}$ in Equation~\eqref{eq:NodalModal} gives the modal coefficients in terms of the nodal coefficients
\beq
  \sum_{\vect{i}=\vect{1}}^{\vect{N}}\int_{\vect{I}}P_{\vect{j}}(\vect{\xi})\,P_{\vect{i}}(\vect{\xi})\,d\vect{\xi}\,\vect{C}_{\vect{i}}
  =\sum_{\vect{i}=\vect{1}}^{\vect{N}}\int_{\vect{I}}P_{\vect{j}}(\vect{\xi})\,\phi_{\vect{i}}(\vect{\xi})\,d\vect{\xi}\,\vect{U}_{\vect{i}},
\eeq
where the matrix on the left-hand side is diagonal and easily invertible.  
The matrix on the right-hand side is the same for all elements, and can be precomputed at program startup and stored with minimal storage requirements.
As illustrated in Equations~\eqref{eq:modalCoefficientOne} and \eqref{eq:modalSlopes}, the representation in terms of Legendre polynomials $P_{l}$ is more convenient for limiting because the polynomial degree increases with increasing $l$, and the identification of the expansion coefficients with average values and average derivatives is more straightforward.  
In the Lagrange basis, all the basis functions have the same polynomial degree.  

\begin{figure}[h]
  \centering
  \includegraphics[width=0.75\textwidth]{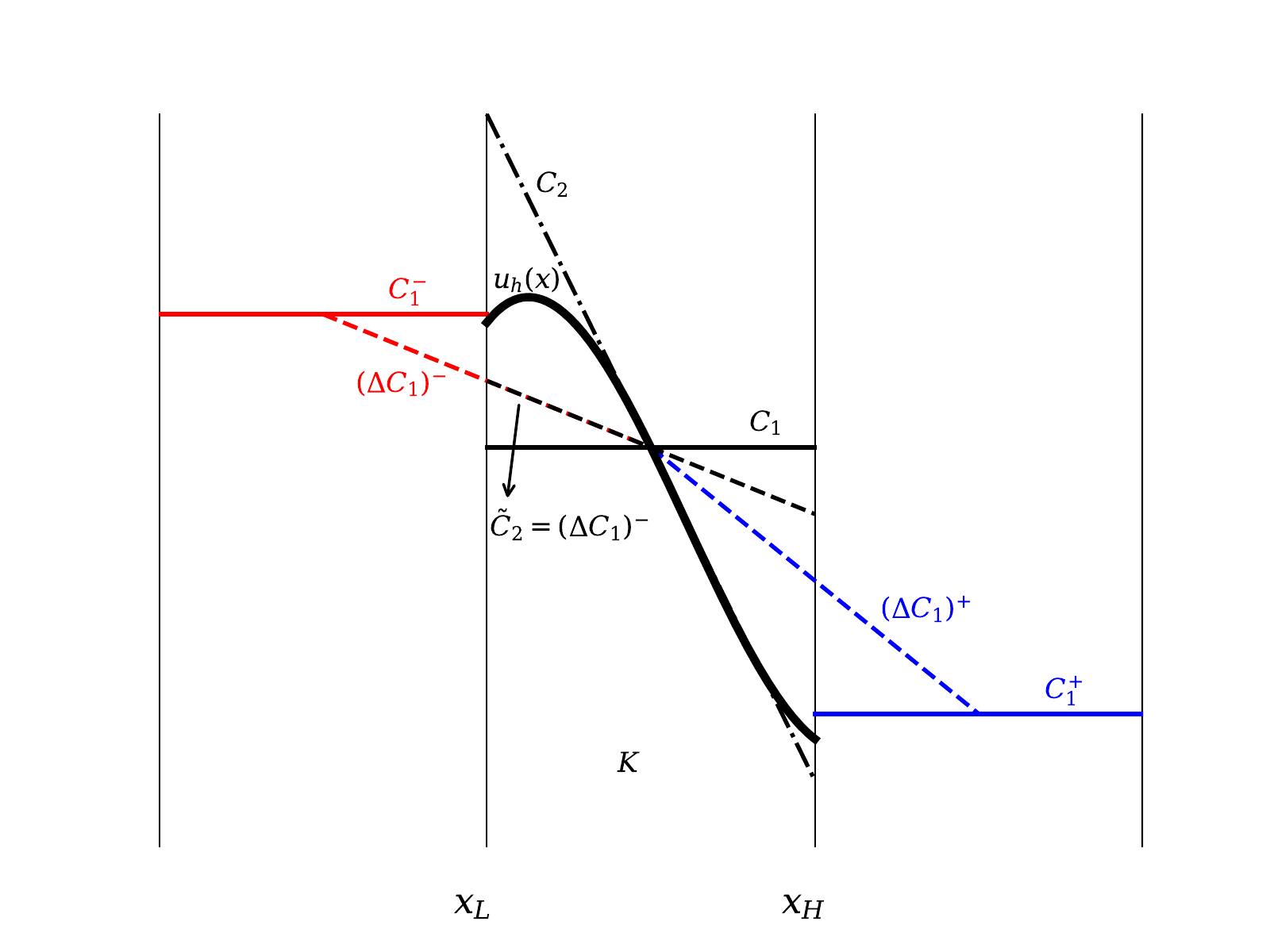}
  \caption{Illustration of how the minmod slope limiter works when applied to a one-dimensional, scalar field $U(x)$.  
  The original, high-order polynomial $U_{h}(x)=\sum_{l=1}^{N}C_{l}\,P_{l}(x)$ is represented by the thick solid black curve, while its constant and linear contributions, $C_{1}P_{1}(x)$ and $C_{2}P_{2}(x)$, are represented by solid and dash-dot black lines, respectively.  The slopes $(\Delta C_{1})^{+}=C_{1}^{+}-C_{1}$ and $(\Delta C_{1})^{-}=C_{1}-C_{1}^{-}$ --- the second and third argument in the minmod function in Equation~\eqref{eq:limiting}, respectively --- are represented by the blue and red dashed lines, respectively.  
  In this example, all three slopes have the same sign.  
  Then, since $(\Delta C_{1})^{-}<(\Delta C_{1})^{+}<C_{1}$, $\widetilde{C}_{2}:=(\Delta C_{1})^{-}$.}
  \label{fig:MinMod}
\end{figure}

We perform slope limiting by comparing the weights $\vect{C}_{\vect{l}}$ --- which for $|\vect{l}|=d+1$ and appropriate normalization of the Legendre polynomials are equal to the first derivatives of the solution in the cell --- with the limited weights $\widetilde{\vect{C}}_{\vect{l}}$, computed from
\beq
  \mathcal{M}\,\widetilde{\vect{C}}_{\vect{l}}
  = \text{minmod}
  \big(\,
    \mathcal{M}\,\vect{C}_{\vect{l}},\,
    \beta_{\TVD}\,\mathcal{M}\,(\vect{C}_{\vect{1}}^{+} - \vect{C}_{\vect{1}}),\,
    \beta_{\TVD}\,\mathcal{M}\,(\vect{C}_{\vect{1}} - \vect{C}_{\vect{1}}^{-})
  \,\big),\quad (\forall\vect{l}~\text{satisfying}~|\vect{l}|=d+1)
  \label{eq:limiting}
\eeq
where the multivariate minmod function is defined as
\beq
  \text{minmod}\big(\,a_{1},\, a_{2},\, a_{3}\,\big) 
  =
  \begin{cases}
    s \times \min\{\,\abs{a_{1}}, \abs{a_{2}}, \abs{a_{3}}\,\}, & \text{if}~ s = \text{sign}(a_{1})= \text{sign}(a_{2})= \text{sign}(a_{3}) \\
    0, & \text{otherwise.}
  \end{cases}
  \label{eq:minmod}
\eeq
The minmod function returns the minimum argument if they all have the same sign, and zero otherwise.  
In three spatial dimensions we estimate limited slopes independently for all the coefficients in Equation~\eqref{eq:modalSlopes}, and limiting is applied to a component of $\vect{U}_{h}$ whenever the corresponding linear coefficient in the modal expansion in Equation~\eqref{eq:modalExpansion} exceeds a given threshold value.  
Here we apply slope limiting when $|\widetilde{C}_{\vect{l}}-C_{\vect{l}}| > 10^{-6}\,C_{\vect{1}}$, for any $\vect{l}$ satisfying $|\vect{l}|=d+1$.  
($C_{\vect{l}}$ and $\widetilde{C}_{\vect{l}}$ are arbitrary components of the vectors $\vect{C}_{\vect{l}}$ and $\widetilde{\vect{C}}_{\vect{l}}$, respectively.)  
In Equation~\eqref{eq:limiting}, the parameter $\beta_{\TVD}$ takes values in the closed interval $\left[1,2\right]$, and determines how aggressively to apply limiting.  
The minimal $\beta_{\TVD}$ corresponds to a total variation diminishing scheme, which is more dissipative than a scheme with the maximal $\beta_{\TVD}$, which is potentially more oscillatory.  
Increasing $\beta_{\TVD}$ puts more weight on the neighboring cell averages, making the minmod function more likely to set $\widetilde{\vect{C}}_{\vect{l}} = \vect{C}_{\vect{l}}$, which results in no limiting being applied.  
The superscripts $-/+$ on the $\vect{C}_{\vect{1}}$ coefficients in the minmod function in Equation~\eqref{eq:limiting} indicate that the coefficient belongs to the expansion in the previous/next element in the coordinate direction of the slope to be limited.  
Figure~\ref{fig:MinMod} illustrates how the minmod limiter works in the one-dimensional case when applied to a scalar field $U(x)$.  
The transformation matrix $\mathcal{M}$ is included in Equation~\eqref{eq:limiting} to allow for limiting in characteristic fields (see discussion below).  
For \emph{component-wise limiting}, $\mathcal{M}$ is set to the identity matrix.  
Thus, when slope limiting is applied, the local solution is truncated as
\beq
  \vect{U}_{h}(\vect{x},t) := \widetilde{\vect{U}}_{h}(\vect{x},t) = \sum_{\substack{\vect{l}=\vect{1} \\ |\vect{l}|\le d+1}}^{\vect{N}}\widetilde{\vect{C}}_{\vect{l}}(t)\,P_{\vect{l}}(\vect{x}),
  \label{eq:NodalModalLimited}
\eeq
where $\widetilde{\vect{C}}_{\vect{l}}=0$ for all $\vect{l}$ with $|\vect{l}|>d+1$, and
\beq
  \widetilde{\vect{C}}_{\vect{1}} := \vect{U}_{\vect{K}} - \f{1}{V_{\vect{K}}}\sum_{\substack{\vect{l}=\vect{1} \\ |\vect{l}|=d+1}}^{\vect{N}}\int_{\vect{K}}P_{\vect{l}}(\vect{x})\,dV_{h}\,\widetilde{\vect{C}}_{\vect{l}}.
  \label{eq:conservativeCorrection}
\eeq
Thus, the minmod limiter reduces the local polynomial degree to at most $k=1$.  
If the arguments in the minmod function in Equation~\eqref{eq:limiting} have different signs, the minmod limiter further reduces the polynomial degree to $k=0$.  
Because of this, we use the TCI as discussed below.  
Although not considered for \thornado\ yet, we note that it is possible to generalize or improve the limiting strategy to maintain higher order of accuracy; see e.g., \cite{biswas:1994,krivodonova:2007,dumbser:2014}.  

The readjustment of $\widetilde{\vect{C}}_{\vect{1}}$ in Equation~\eqref{eq:conservativeCorrection}, which occurs after computing the limited slopes in Equation~\eqref{eq:limiting}, is necessary to preserve the cell average as defined in Equation~\eqref{eq:cellAverage}, and is due to the use of curvilinear coordinates \citep[see also related discussion by][their Section~C1]{radice:2011}.  
Preservation of the cell average in the limiting procedure is needed, e.g., to conserve mass.  
Without the `conservative correction' in Equation~\eqref{eq:conservativeCorrection}, the limiter preserves the cell average defined in Equation~\eqref{eq:modalCoefficientOne}, which is undesirable in curvilinear coordinates.  
Note that the second term on the right-hand side of Equation~\eqref{eq:conservativeCorrection} vanishes in Cartesian coordinates because of orthogonality of the Legendre polynomials.  
However, in curvilinear coordinates, this term does not vanish since the Legendre polynomials are not orthogonal with respect to the inner product weighted by $\sqrt{\gamma_{h}}$.  
In practice, we have found that the conservative correction is small, but necessary to maintain conservation to machine precision.  

We note that, in order to improve the evolution of the electron fraction, $\ye=D_{\rm{e}}/\rho$, we also apply the minmod limiter directly to the electron fraction, and enforce limiting of both $\rho_{h}$ and $D_{{\rm{e}},h}$ whenever oscillations in $\ye$ is detected by the minmod function.  

\begin{figure}[h]
  \centering
  \includegraphics[width=0.75\textwidth]{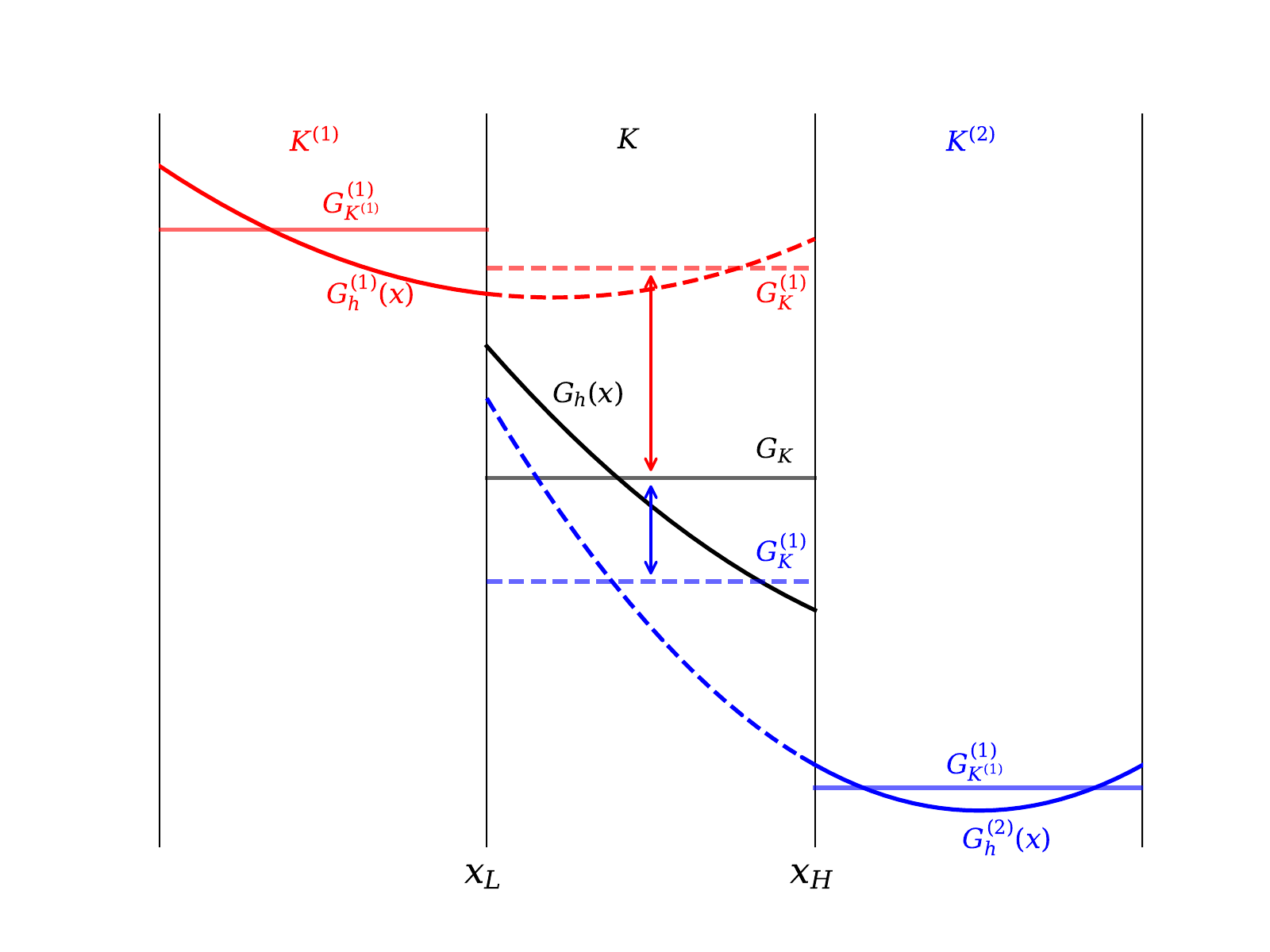}
  \caption{Illustration of how the troubled-cell indicator works in the one-dimensional case on a scalar field $G(x)$, to determine if limiting is needed in the target element $K$, where the polynomial representation is given by $G_{h}(x)$ (solid black curve), and the cell average is $G_{K}$ (solid gray line).  
  The polynomial representation in the left element, $K^{(1)}$, is given by $G_{h}^{(1)}(x)$ (solid red curve), with cell average $G_{K^{(1)}}^{(1)}$ (solid light red line).  
  Similarly, the polynomial representation in the right element, $K^{(2)}$, is given by $G_{h}^{(2)}(x)$ (solid blue curve), with cell average $G_{K^{(2)}}^{(2)}$ (solid light blue line).  
  The extrapolations of $G_{h}^{(1)}(x)$ and $G_{h}^{(2)}(x)$ into the target element are given by the dashed red and blue curves, respectively.  
  Finally, the cell averages of the extrapolations of $G_{h}^{(1)}(x)$ and $G_{h}^{(2)}(x)$, computed over the target cell, are denoted $G_{K}^{(1)}$ (dashed light red line) and $G_{K}^{(2)}$ (dashed light blue line), respectively.  
  The element is flagged for limiting if the difference in the cell averages, $|G_{K}-G_{K}^{(2)}|$ and/or $|G_{K}-G_{K}^{(2)}|$, becomes too large.}
  \label{fig:TCI}
\end{figure}

In order to determine where slope limiting is necessary, we use the TCI of \citet{fu:2017} to prevent excessive limiting.  
For example, it is well-known that the minmod limiter is overly diffusive around smooth extrema, where $\widetilde{\vect{C}}_{\vect{l}}=0$, which kills off all the high-order accuracy.  
We note in passing that other TCIs have been proposed \citep[see, e.g.,][]{qiu:2005}, but we have chosen the one by \citet{fu:2017} for its relative ease of implementation.  
This TCI is based on the function
\begin{equation}
  I_{\bK}(G_{h}) = \f{\sum_{j}|G_{\bK}-G_{\bK}^{(j)}|}{\max(\,\max_{j}|G_{\bK^{(j)}}^{(j)}|,\,|G_{\bK}|\,)},
  \label{eq:tci}
\end{equation}
where $G_{h}\in\vect{G}_{h}\subseteq\vect{U}_{h}$ is in the subset of fields used to detect troubled cells.  
In Equation~\eqref{eq:tci}, the sum in the numerator is taken over all the neighboring elements $\bK^{(j)}$ sharing a boundary with the target element $\bK$, while the max in the denominator is taken over neighboring elements $\bK^{(j)}$ and the target element $\bK$.  
The cell average of $G_{h}$ in $\bK$ is denoted $G_{\bK}$, and is here given by the right-hand side of Equation~\eqref{eq:modalCoefficientOne} --- i.e., without the weighting factor $\sqrt{\gamma_{h}}$ used in the proper definition of the cell average in Equation~\eqref{eq:cellAverage}.  
Computed in the same way, $G_{\bK}^{(j)}$ is the corresponding cell average computed by extrapolating the polynomial representation from the neighboring elements $\bK^{(j)}$ into the target $\bK$, and $G_{\bK^{(j)}}^{(j)}$ is the cell average native to the neighbor element $\bK^{(j)}$.  
An illustration of the troubled-cell indicator is given in Figure~\ref{fig:TCI} for the one-dimensional case applied to a single field $G(x)$.  

An element is flagged for limiting if, for any $G_{h}\in\vect{G}_{h}$, $I_{\bK}(G_{h})>C_{\TCI}(G)$, where $C_{\TCI}(G)$ is a user-defined threshold, which can be set differently for each $G$.  
In the numerical results presented in Section~\ref{sec:results}, we use the mass density, fluid energy, and electron fraction as the variables to detect troubled cells; i.e., $\vect{G} = (\rho,E,\ye)^{T}$.

When solving a system of hyperbolic conservation laws, experience has shown that the slope limiting described above is more efficient when performed on the so-called `characteristic variables', as opposed to the conserved variables $\mathbf{U}_h$ \citep[see, e.g.,][for a description]{cockburn:1998}.  
Because the Euler equations form a system of hyperbolic partial differential equations \citep[see, e.g.,][]{leveque:1992}, the flux Jacobian in Equation~\eqref{eq:eigenDecomposition} can be decomposed as
\beq
  \pderiv{\mathbf{F}^{i}(\mathbf{U})}{\mathbf{U}} =
  \mathcal{R}^{i}\, \Lambda^{i}\, (\mathcal{R}^{i})^{-1}\quad(i=1,\ldots,d),
\eeq
where the columns of the $6\times6$ matrix $\mathcal{R}^{i}$ contain the right eigenvectors of the flux Jacobian, the rows of $(\mathcal{R}^{i})^{-1}$ contain the left eigenvectors, and $\Lambda^{i}$ is a diagonal matrix containing the eigenvalues of the flux Jacobian.  
For hyperbolic systems, the eigenvalues are real and the eigenvectors form a complete set \citep[see e.g.,][]{leveque:1992}.  
At this point, we introduce the characteristic variable $\mathbf{w} = \mathcal{R}^{-1}\mathbf{U}$.  
Recall in Equation~\eqref{eq:limiting} that we introduced the transformation matrix $\mathcal{M}$.  
If we let $\mathcal{M}=(\mathcal{R}^{i})^{-1}$, limiting is performed on the characteristic variables.  
(For linear systems, the characteristic variables evolve independently, and limiting of one characteristic variable does not affect the others.)  
Once $\mathcal{M}\,\widetilde{\vect{C}}_{\vect{l}}$ is estimated in the characteristic variables as in Equation~\eqref{eq:limiting}, the limited slopes in the conserved variables are obtained by left multiplication with $\mathcal{M}^{-1}$ \citep[see e.g.,][]{cockburn:1998,schaal:2015a}, and the limiting process proceeds as in Equations~\eqref{eq:NodalModalLimited} and \eqref{eq:conservativeCorrection}.  
It should be noted that $\mathcal{R}^{i}$ and $(\mathcal{R}^{i})^{-1}$ are computed using cell averages of the conserved and metric variables.  

While this process of characteristic limiting has been done for an ideal EoS, and shown \citep[e.g.,][]{schaal:2015a} to give superior results when compared to component-wise limiting (especially for the high-order case; $k\ge1$), the extension to the tabulated nuclear EoS case is nontrivial.  
The reasons for this are (1) the increased complexity and dimensionality of the system due to the added electron conservation equation in Equation~\eqref{eq:electronConservation}, and (2) the additional care that must be taken when computing the thermodynamic derivatives associated with the flux Jacobian. 
In the case of an ideal, or other simplified EoS, the necessary thermodynamic derivatives (such as derivatives of pressure) are analytically defined. 
For a nuclear EoS, the derivatives do not have analytic expressions and the necessary eigenvectors must be constructed generally. 
We provide the characteristic decomposition of the flux Jacobian for the Euler system with a nuclear EoS in Appendix~\ref{appendix:characteristic}.  

\subsection{Bound-Enforcing Limiting}
\label{sec:boundEnforcing}

When solving the Euler equations of gas dynamics with an ideal EoS, the mass density $\rho$ and pressure $p$ (or, equivalently, internal energy density $e$) must remain positive.  
However, this property is not guaranteed by the basic DG method, which encourages the use of a more advanced procedure \citep[][]{zhang:2010a}.  
The internal energy density is given in terms of the conserved quantities as
\begin{equation} \label{eq:idealEoSPressureFunction}
  e(\vect{U})=E-\frac{m^{2}}{2\rho},
\end{equation}
where $m^{2}=m_{j}m^{j}$, $E$ is the fluid energy density, and $m_{j}=\rho v_{j}$ are the components of the momentum density.  
For the ideal EoS case, the set of physically admissible states is given by
\begin{equation} \label{eq:idealAdmissibleStates}
  \tilde{\mathcal{G}}=\big\{\, \vect{U}=(\rho,\vect{m},E)^{\rm{T}} \, | \, \rho > 0 ~ \text{and} ~ e(\vect{U})>0 \,\big\}.  
\end{equation}
If mass density is positive, the internal energy density is a concave function of $\vect{U}$, and $\tilde{\mathcal{G}}$ is a convex set \citep[][]{zhang:2010a}.  
For many EoSs (including the ideal EoS), where the pressure only depends on the mass density and internal energy density, $\vect{U}$ must remain in $\tilde{\mathcal{G}}$ as defined in Equation~\eqref{eq:idealAdmissibleStates}, otherwise the initial value problem is ill-posed.  
To maintain $\vect{U}\in\tilde{\mathcal{G}}$, the combination of a suitable time step restriction, a strong stability-preserving time integrator, and a bound-enforcing limiter is used \citep[e.g.,][]{zhang:2010a}.  
The time step restriction is derived as a sufficient condition to ensure that the updated \emph{cell average} satisfies $\vect{U}_{\bK}^{n+1}\in\tilde{\mathcal{G}}$, and requires $\vect{U}_{h}^{n}\in\tilde{\mathcal{G}}$ \emph{point-wise} within each element, while the limiter, which relies on $\vect{U}_{\bK}^{n+1}\in\tilde{\mathcal{G}}$ \emph{and} the convexity of $\tilde{\mathcal{G}}$, is used to again enforce point-wise $\vect{U}_{h}^{n+1}\in\tilde{\mathcal{G}}$ within each element.  
(We do not attempt to derive a sufficient time step restriction for the present setting in this paper, and simply use the condition in Equation~\eqref{eq:cflCondition}.)  
We note here that for two arbitrary elements $\vect{U}_{a},\vect{U}_{b}\in\tilde{\mathcal{G}}$, since the set $\tilde{\mathcal{G}}$ is convex, the convex combination $\vect{U}_{c}:=\vartheta\,\vect{U}_{a}+(1-\vartheta)\,\vect{U}_{b}$, where $\vartheta\in[0,1]$, is also in $\tilde{\mathcal{G}}$; i.e., $\vect{U}_{c}\in\tilde{\mathcal{G}}$.  
Moreover, $e(\vect{U})$ in Equation~\eqref{eq:idealEoSPressureFunction} is concave since Jensen's inequality --- $e(\vect{U}_{c})\ge\vartheta\,e(\vect{U}_{a})+(1-\vartheta)\,e(\vect{U}_{b})$ --- holds.  
The property of convex combinations is commonly used to design constraint-preserving numerical methods for systems where --- for physical reasons --- the dynamics is constrained to a convex set \cite[see, e.g.,][for examples beyond the non-relativistic Euler equations with an ideal EoS]{xing:2010,olbrant:2012,wu:2015,endeve:2015,chu:2019}.  


To maintain physically admissible states in the present setting with \thornado, we draw inspiration from the limiting strategy proposed for an ideal EoS by \cite{zhang:2010a}, which we have modified to work satisfactorily with a tabulated nuclear EoS.  
Specifically, thermodynamic quantities, including the specific internal energy $\epsilon=e/\rho$, are tabulated in terms of mass density, temperature, and electron fraction, which cover finite extents; i.e., $\rho\in[\rho_{\min},\rho_{\max}]$, $T\in[T_{\min},T_{\max}]$, and $\ye\in[Y_{e,\min},Y_{e,\max}]$.  
We use some of the table bounds to define the set of admissible states as
\begin{equation} \label{eq:nuclearAdmissibleStates}
  \mathcal{G}
  = \big\{\, \vect{U}=(\rho,\vect{m},E,D_{\rm{e}})^{\rm{T}}\, |\, (\rho,D_{\rm{e}})^{\rm{T}}\in\mathcal{G}_{\vect{u}} ~ \text{and} ~ \epsilon(\vect{U}) \geq \epsilon_{\rm{min}}\equiv\epsilon(\rho, T_{\mathrm{min}}, \ye) \,\big\},
\end{equation}
where we have defined the subset
\begin{equation}
  \mathcal{G}_{\vect{u}}
  =\big\{\,\vect{u}=(\rho,D_{\rm{e}})^{\rm{T}}\,|\,\rho_{\rm{min}} \le \rho \le \rho_{\rm{max}},~ D_{\rm{e}}>0, ~ \text{and} ~ Y_{\rm{e,min}} \le \ye(\vect{u}) \le Y_{\rm{e,max}}\,\big\},
\end{equation}
and seek to maintain $ \vect{U}_{h} \in \mathcal{G}$.  

First, we note that it is straightforward to show that the subset $\mathcal{G}_{\vect{u}}$ is convex.  
To do this, it is sufficient to show that a convex combination of two arbitrary elements of $\mathcal{G}_{\vect{u}}$ also belongs to $\mathcal{G}_{\vect{u}}$.  
To this end, let $\vect{u}_{a}\equiv(\rho_{a},D_{{\rm{e}},a})^{\rm{T}}\in\mathcal{G}_{\vect{u}},\vect{u}_{b}\equiv(\rho_{b},D_{{\rm{e}},b})\in\mathcal{G}_{\vect{u}}$, and define the convex combination $\vect{u}_{c}=\vartheta\,\vect{u}_{a}+(1-\vartheta)\,\vect{u}_{b}$, where $\vartheta\in[0,1]$.  
Then the first component of $\vect{u}_{c}$ is $\rho_{c}=\vartheta\,\rho_{a}+(1-\vartheta)\,\rho_{b}$.  
Since, by assumption, $\rho_{a},\rho_{b}\in[\rho_{\rm{min}},\rho_{\rm{max}}]$ and $\vartheta\in[0,1]$, it follows that $\rho_{c}\in[\rho_{\rm{min}},\rho_{\rm{max}}]$.  
Similarly, the second component of $\vect{u}_{c}$ is $D_{{\rm{e}},c}=\vartheta\,D_{{\rm{e}},a}+(1-\vartheta)\,D_{{\rm{e}},b}$.  
Then, since $D_{{\rm{e}},a},D_{{\rm{e}},b}>0$, it follows that $D_{{\rm{e}},c}>0$.  
Finally, we can write
\begin{equation}
  \ye(\vect{u}_{c}) = \f{D_{{\rm{e}},c}}{\rho_{c}} = \f{\vartheta\,D_{{\rm{e}},a}+(1-\vartheta)\,D_{{\rm{e}},b}}{\vartheta\,\rho_{a}+(1-\vartheta)\,\rho_{b}}
  =\alpha\,\f{D_{{\rm{e}},a}}{\rho_{a}}+(1-\alpha)\,\f{D_{{\rm{e}},b}}{\rho_{b}}
  =\alpha\,\ye(\vect{u}_{a})+(1-\alpha)\,\ye(\vect{u}_{b}),
\end{equation}
where
\begin{equation}
  \alpha = \f{\vartheta\,\rho_{a}}{\vartheta\,\rho_{a}+(1-\vartheta)\,\rho_{b}}.  
\end{equation}
Since $\rho_{a},\rho_{b}\ge\rho_{\rm{min}}>0$ and $\vartheta\in[0,1]$, we have $\alpha\ge0$.  
We also have $\alpha\le1$.  
Therefore, $\alpha\in[0,1]$, which implies $\ye(\vect{u}_{c})\in[Y_{\rm{e,min}},Y_{\rm{e,max}}]$ and $\vect{u}_{c}\in\mathcal{G}_{\vect{u}}$.  
Thus, the subset $\mathcal{G}_{\vect{u}}$ is convex.  

While, strictly speaking, the Euler equations in Section~\ref{sec:model} are valid for any mass density $\rho>0$, we note that there are physical reasons for maintaining the mass density within the finite table bounds, which are $\rho_{\min}\approx1.66\times10^{3}$~g~cm$^{-3}$ and $\rho_{\max}\approx3.16\times10^{15}$~g~cm$^{-3}$ for the tables used in this paper.  
Indeed, in CCSN simulations, it is possible for the cell averaged mass density to evolve outside these limits, which would require extending the table bounds.  
However, when the mass density approaches the upper bound, a relativistic description should be adopted, and when the mass density approaches the lower bound, the nuclear EoS adopted here is invalid because the matter is not in nuclear statistical equilibrium.  
These bounds must, however, also be enforced to avoid algorithm failure.  
For the purpose of the bound-enforcing limiter, the finite bounds on the mass density in Equation~\eqref{eq:nuclearAdmissibleStates} are included in case the bounds are violated for certain points within an element, e.g., in the vicinity of a shock, while the cell averaged mass density is still inside the table bounds.  
(The limiter developed here will not work if the cell averaged mass density exceeds the table bounds.)  
We have also equipped the set of admissible states $\mathcal{G}$ with the bounds $\ye\in[Y_{\rm{e,min}},Y_{\rm{e,max}}]\subseteq[0,1]$, which are also required to avoid algorithm failure.  
(In this work, $Y_{\rm{e,min}}=0.01$ and $Y_{\rm{e,max}}=0.7$).  
We note, however, that for the test problems in Section~\ref{sec:results} and the application in Section~\ref{sec:collapse}, we did not encounter a situation in which the mass density or the electron fraction exceeded their respective table bounds.  

\begin{figure}[h]
  \centering
  \includegraphics[width=0.75\textwidth]{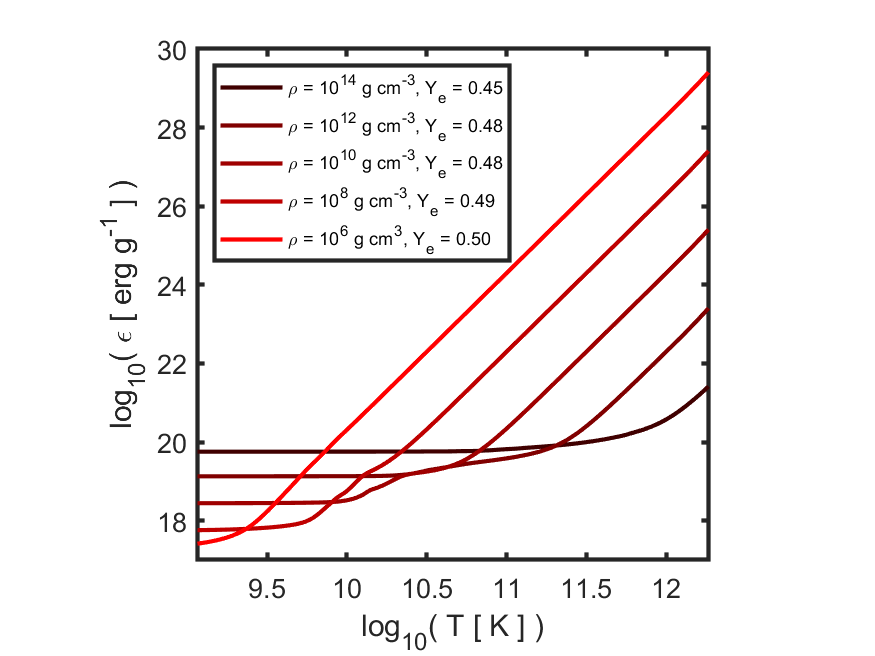}
  \caption{Relationship between specific internal energy and temperature from the SFHo EoS for select values of mass density and electron fraction. 
  (Note that the electron fraction is only sampled from the narrow range encountered in the adiabatic simulations discussed in Section~\ref{sec:collapse}.)  
  Due to degeneracy, the specific internal energy for all profiles demonstrates asymptotic behavior for low temperatures. 
  Thus, the lower boundary on $\epsilon$ would not change if the table was reconstructed with a lower temperature limit.}
  \label{fig:SpecificInternalEnergyProfiles}
\end{figure}

On the other hand, a complication that frequently arises in gravitational collapse simulations is that the specific internal energy falls below the minimum tabulated value (i.e., $\epsilon<\epsilon_{\mathrm{min}}$) --- especially around core bounce and shock formation, which we discuss in further detail in Section~\ref{sec:collapse}.  
When this happens, the EoS is not invertible for the temperature when given the state vector $(\rho,\epsilon,\ye)^{\rm{T}}$, and the algorithm fails since the temperature is needed to compute the pressure as well as other thermodynamic quantities.  
It is not feasible to merely generate tables with lower $T_{\rm{min}}$, since --- particularly for high mass densities --- the specific internal energy does not tend to zero as $T\to0$ due to the degeneracy (or zero temperature) contribution to the internal energy, as can be seen in Figure~\ref{fig:SpecificInternalEnergyProfiles}.
In CCSN simulations, where the iron core is degenerate at the onset of collapse, the initial specific internal energy is already close to the minimum value.  
Then, around core bounce and shock formation, where steep gradients in the evolved fields form, conditions with $\vect{U}_{h}\notin\mathcal{G}$ can easily arise within certain elements, and a limiting strategy is needed.  
Fortunately, we have observed that $\vect{U}_{\vect{K}}\in\mathcal{G}$ is always satisfied (although we do not seek to establish sufficient conditions to guarantee this here).  
This allows us to pursue the limiting strategy proposed by \citet{zhang:2010a}, which we detail below.

\begin{figure}[h]
  \centering 
  \includegraphics[width=0.75\textwidth]{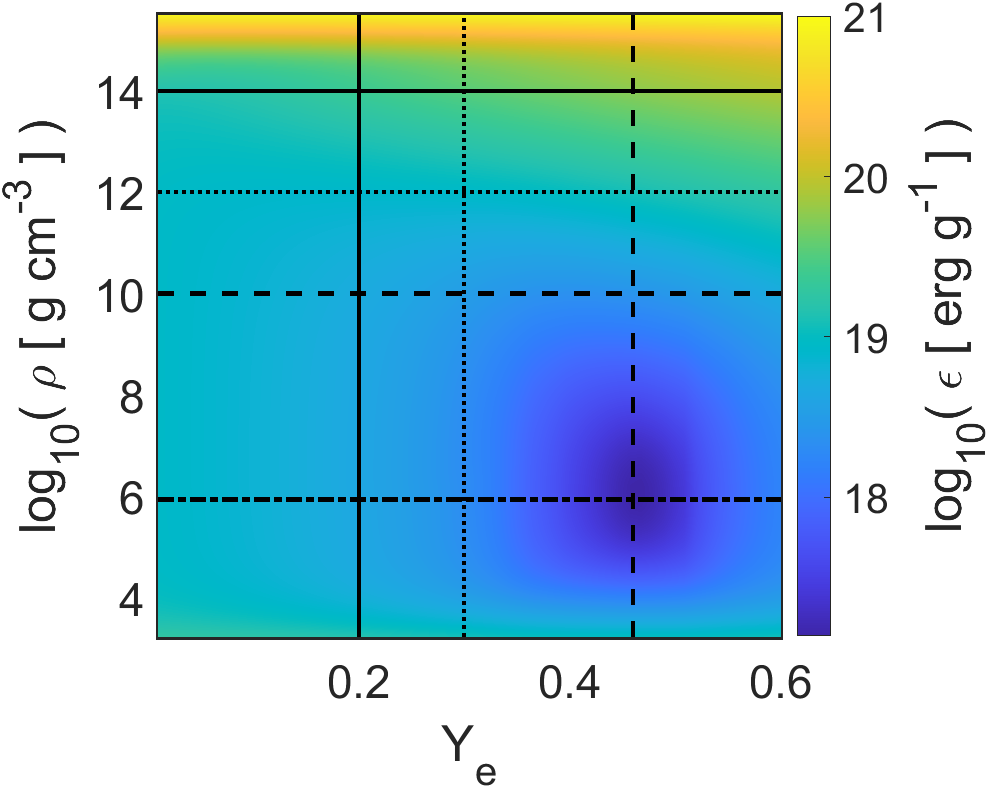}
  \begin{minipage}[h]{0.49\textwidth}
    \includegraphics[width=\textwidth]{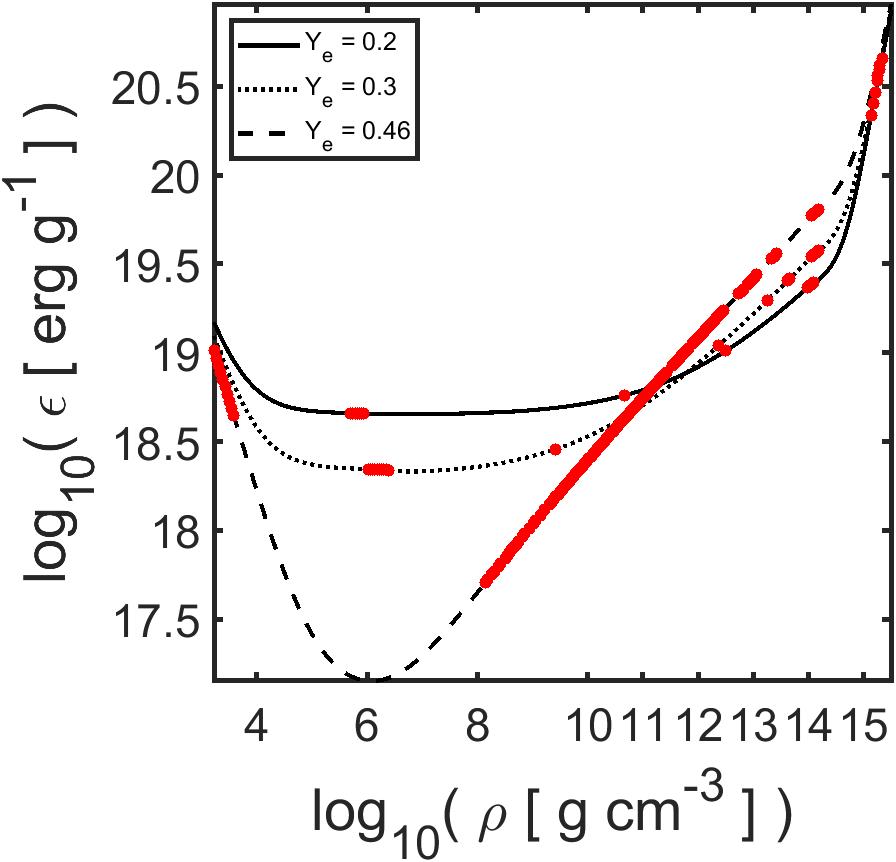}
  \end{minipage}
  \begin{minipage}[h]{0.49\textwidth}
   \includegraphics[width=\textwidth]{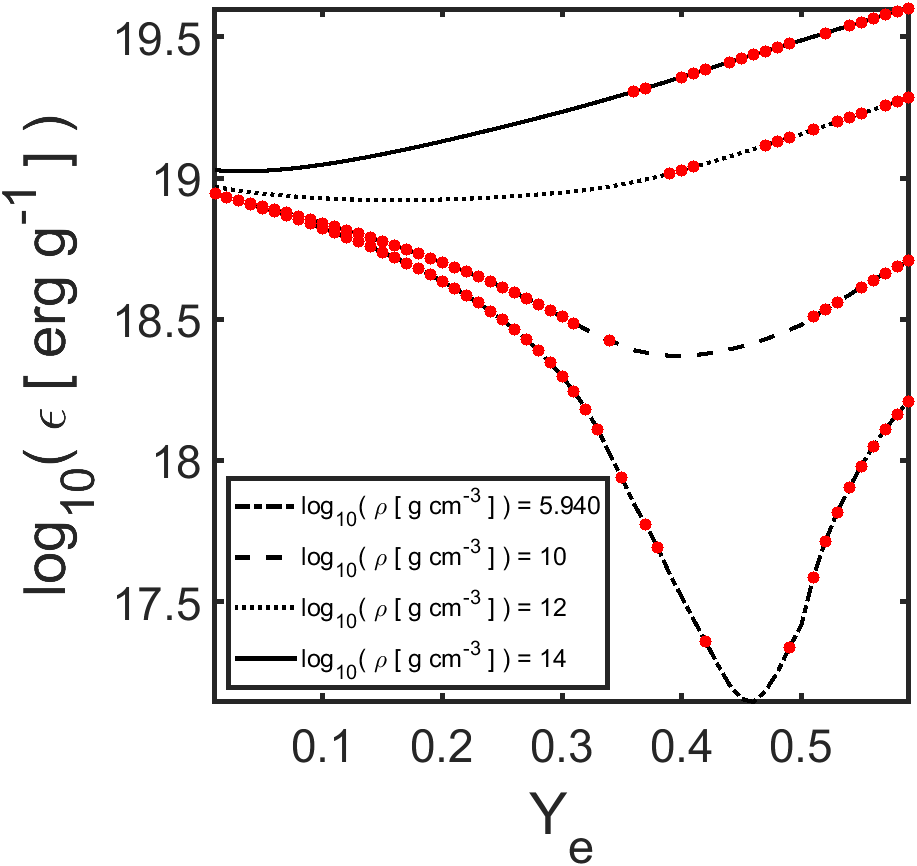}
  \end{minipage}
  \caption{The minimum specific internal energy, $\epsilon_{\rm{min}}=\epsilon(\rho,T_{\mathrm{min}},\ye)$ for the SFHo EoS is displayed in the top panel as a function of $\rho$ and $\ye$.  
  (Since the specific internal energy can be negative, we have added an offset of about $2.8 \times 10^{17}$~erg~g$^{-1}$ to the actual value returned by the EoS).  
  The vertical lines (constant $\ye$) correspond to lines plotted in the lower left panel, while horizontal lines (constant $\rho$) correspond to lines plotted in the lower right panel.  
  To further highlight the topology of the $\epsilon_{\rm{min}}$ surface, we plot $\epsilon_{\rm{min}}$ versus mass density for select values of $\ye$ in the lower left panel.  
  Similarly, we plot $\epsilon_{\rm{min}}$ versus electron fraction for select values of $\rho$ in the lower right panel.  
  These traces are selected to illustrate the non-convexity of $\mathcal{G}$. The red dots indicate non-convex regions, i.e. where the second derivative of the specific internal energy with respect to either  $\rho$ (left) or $\ye$ (right) is less than zero.
  From visual inspection, the constant $\ye$ profiles (lower left) appear to be convex; i.e., $(\partial^{2}\epsilon_{\min}/\partial\rho^{2})_{\ye}\ge0$.  
  However, the dashed line is not, and the dotted and solid lines are non-convex around $\rho=10^{14}$~g~cm$^{-3}$.  
  Meanwhile the constant $\rho$ profiles (lower right) are clearly not all convex since $(\partial^{2}\epsilon_{\min}/\ye^{2})_{\rho}$ can be negative (cf. dashed and dash-dot curves).}
  \label{fig:SFHoTable}
\end{figure}

There is, however, an additional complication that may cause the limiting strategy of \citet{zhang:2010a} to fail: the surface of specific internal energy at the minimum temperature $T_{\mathrm{min}}$ --- that is $\epsilon_{\min}(\rho,\ye)\equiv\epsilon(\rho,T_{\mathrm{min}},\ye)$ --- is not globally convex in the sense that the second derivatives $(\partial^{2}\epsilon_{\min}/\partial\rho^{2})_{\ye}$ and $(\partial^{2}\epsilon_{\min}/\partial\ye^{2})_{\rho}$ are not strictly positive everywhere, which implies that the set $\mathcal{G}$ in Equation~\eqref{eq:nuclearAdmissibleStates} is not strictly convex. 
We illustrate this in Figure \ref{fig:SFHoTable}, which shows $\epsilon(\rho,T_{\mathrm{min}},\ye)$ as a function of $\rho$ and $\ye$ for the SFHo EoS \citep[][]{steiner:2013a}.
Therefore, adopting the limiting procedure from the ideal EoS case to enforce $\vect{U}_{h}\in\mathcal{G}$ --- even if $\vect{U}_{\vect{K}}\in\mathcal{G}$ --- can compromise the robustness of the limiter.  
The reason is that the amount of limiting applied to the polynomial $\vect{U}_{h}$ is determined by finding the intersection point of the boundary of $\mathcal{G}$ and the straight line connecting the cell average $\vect{U}_{\vect{K}}$ and a non-physical point value $\vect{U}_{\vect{q}}\notin\mathcal{G}$.  
If $\mathcal{G}$ is not convex, there may be multiple intersection points, which can cause the limiter to fail.  
However, the issue of globally non-convex $\mathcal{G}$ is avoided if the limiter is only activated in regions for which $\mathcal{G}$ is locally convex.  
That is, for the elements that require limiting, the cell average $\vect{U}_{\vect{K}}$ and the DG solution $\vect{U}_{h}$, evaluated in the required quadrature points within each element $\vect{K}$, are in a locally convex region and sufficiently close to each other in $\mathcal{G}$.  
The latter is typically the case in regions of the flow characterized by small gradients, but may not be the case in the vicinity of a shock.  
Fortunately, as discussed further in Section~\ref{sec:collapse}, we do not encounter any situations in which the non-convexity of $\mathcal{G}$ causes the limiter to fail, but this needs to be further investigated in the context of multidimensional models with higher physical fidelity (i.e., models that include neutrino transport), which sample a larger part of the EoS than the simulations discussed in this paper. 

The bound-enforcing limiter is completely local to each element, and can thus be discussed in terms of a single element $\vect{K}$.  
As in \citep[][]{zhang:2010a}, we define a point set $\vect{S}^{+}$, which includes the volumetric nodal points in an element $\vect{K}$, as well as the points on the interface of $\vect{K}$. 
For the two-dimensional case with $k=2$, the point set is given by the union of all the points displayed in the right panel in Figure~\ref{fig:ReferenceElements}.  
Thus, $\vect{S}^{+}$ comprises the points where $\vect{U}_{h}$ is evaluated to construct the update for each $\vect{U}_{\vect{p}}$ in Equation~\eqref{eq:dgSemidiscreteWeakNodal}.  
Using Equation~\eqref{eq:nodalExpansion}, the solution is evaluated at all the points $\vect{x}_{\vect{q}} \in \vect{S}^{+}$, and limiting is applied if, for any point $\vect{x}_{\vect{q}} \in \vect{S}^{+}$, $\vect{U}_{\vect{q}} \equiv \vect{U}_{h}(\vect{x}_{\vect{q}}) \notin \mathcal{G}$.  
The step-by-step procedure for bound-enforcing limiting is described next, where it is assumed that the cell average satisfies $\vect{U}_{\vect{K}}\in\mathcal{G}$.  

\subsubsection{Step~1: Mass Density and Electron Density}
\label{sec:BELstep1}

The first step is to enforce $\rho_{\vect{q}}\in[\rho_{\rm{min}},\rho_{\rm{max}}]$ and $D_{{\rm{e}},\vect{q}}\ge\delta_{D_{\rm{e}}}>0$ for all $\vect{x}_{\vect{q}}\in\vect{S}^{+}$, where $\delta_{D_{\rm{e}}}$ is arbitrarily small.  
(The bound $D_{{\rm{e}},\vect{q}}>0$ is needed in Step~2 below.)  
Following \citet{zhangShu:2010Scalar}, we use the linear scaling limiter from \cite{liuOsher:1996}, and replace the polynomial $\vect{u}_{h}(\vect{x})=\big(\rho_{h}(\vect{x}),D_{{\rm{e}},h}(\vect{x})\big)^{\rm{T}}$ with the limited polynomial
\begin{equation}
  \vect{u}_{h}^{(1)}(\vect{x}) := (1-\vartheta_{1})\,\vect{u}_{\vect{K}} + \vartheta_{1}\,\vect{u}_{h}(\vect{x}) \quad (\vartheta_{1}\in[0,1]),
  \label{eq:limitedPolynomialStep1}
\end{equation}
where the limiter parameter $\vartheta_{1}\in[0,1]$ is found by a simple backtracing algorithm.  
Specifically, for any point $\vect{x}_{\vect{q}}\in\vect{S}^{+}$ with $\vect{u}_{\vect{q}}=\vect{u}_{h}(\vect{x}_{\vect{q}})\notin\mathcal{G}_{\vect{u}}$, we start with $\vartheta_{1,\vect{q}}=1$, which is recursively reduced (by 5\%) until
\begin{equation}
  \vect{u}_{\vect{q}}^{(1)} = (1-\vartheta_{1,\vect{q}})\,\vect{u}_{\vect{K}} + \vartheta_{1,\vect{q}}\,\vect{u}_{\vect{q}} \in \mathcal{G}_{\vect{u}}.  
\end{equation}
(In practice, to reduce the number of iterations, we set $\vartheta_{1,\vect{q}}=0$ whenever the backtracing algorithm has brought the value below $0.01$.)  
We then set $\vartheta_{1}:=\min_{\vect{q}}\vartheta_{1,\vect{q}}$, where the minimum is taken over all the points within the element where $\vect{u}_{h}$ was found to violate the bounds associated with Step~1.  
The limiter in Equation~\eqref{eq:limitedPolynomialStep1} simply scales $\vect{u}_{h}$ as evaluated in the points within the element towards the cell average, and the value for $\vartheta_{1}$ is determined in order to scale the solution in the points just enough to ensure that the bounds are satisfied for all $\vect{x}_{\vect{q}} \in \vect{S}^{+}$.  
In the worst case scenario, $\vartheta_{1}=0$, and the DG solution is set equal to the cell average everywhere within the element.  
Note that this step is conservative and does not change the cell averages; i.e., $\vect{u}_{\vect{K}}^{(1)}=\vect{u}_{\vect{K}}$.  
Also note that if the bounds on the mass density and electron density are not violated, then $\vartheta_{1}=1$ and $\vect{u}_{h}^{(1)}(\vect{x})=\vect{u}_{h}(\vect{x})$.  


\subsubsection{Step 2: Electron Fraction}
\label{sec:BELstep2}

In the second step, we enforce $Y_{{\rm{e}},\vect{q}}\equiv D_{{\rm{e}},h}(\vect{x}_{\vect{q}})/\rho_{h}(\vect{x}_{\vect{q}})\in[Y_{\rm{e,min}},Y_{\rm{e,max}}]$ for all $\vect{x}_{\vect{q}}\in\vect{S}^{+}$.  
To do this, we follow a procedure similar to the previous step, and replace $\vect{u}_{h}^{(1)}(\vect{x})$ with the limited polynomial
\begin{equation}
  \vect{u}_{h}^{(2)}(\vect{x}) := (1-\vartheta_{2})\,\vect{u}_{\vect{K}} + \vartheta_{2}\,\vect{u}_{h}^{(1)}(\vect{x}) \quad (\vartheta_{2}\in[0,1]),
  \label{eq:limitedPolynomialStep2}
\end{equation}
where
\begin{equation}
  \vartheta_{2} = \f{\alpha\,\rho_{\vect{K}}}{\alpha\,\rho_{\vect{K}}+(1-\alpha)\,\rho_{\alpha}^{(1)}}
  \quad\text{and}\quad
  \alpha=
  \min\Big\{\,
    1,\,\left|\f{Y_{\rm{e,min}}-Y_{{\rm{e}},\vect{K}}}{m_{\ye}-Y_{{\rm{e}},\vect{K}}}\right|,\,\left|\f{Y_{\rm{e,max}}-Y_{{\rm{e}},\vect{K}}}{M_{\ye}-Y_{{\rm{e}},\vect{K}}}\right|
  \,\Big\},
  \label{eq:thetaTwoAndAlpha}
\end{equation}
and where we have defined
\begin{equation}
  M_{\ye}=\max_{\vect{x}\in\vect{S}^{+}}\ye\big(\vect{u}_{h}^{(1)}(\vect{x})\big), \quad 
  m_{\ye}=\min_{\vect{x}\in\vect{S}^{+}}\ye\big(\vect{u}_{h}^{(1)}(\vect{x})\big), \quad\text{and}\quad
  Y_{{\rm{e}},\vect{K}}=D_{{\rm{e}},\vect{K}}/\rho_{\vect{K}}.  
\end{equation}
and with the cell average for mass density and electron density computed according to the definition in Equation~\eqref{eq:cellAverage}; i.e.
\begin{equation}
  \rho_{\vect{K}}=\f{\sum_{\vect{p}=\vect{1}}^{\vect{N}}w_{\vect{p}}\,\sqrt{\gamma_{\vect{p}}}\,\rho_{\vect{p}}}{\sum_{\vect{p}=\vect{1}}^{\vect{N}}w_{\vect{p}}\,\sqrt{\gamma_{\vect{p}}}}
  \quad\text{and}\quad
  D_{{\rm{e}},\vect{K}}=\f{\sum_{\vect{p}=\vect{1}}^{\vect{N}}w_{\vect{p}}\,\sqrt{\gamma_{\vect{p}}}\,D_{{\rm{e}},\vect{p}}}{\sum_{\vect{p}=\vect{1}}^{\vect{N}}w_{\vect{p}}\,\sqrt{\gamma_{\vect{p}}}},
\end{equation}
respectively.  
In the expression for $\vartheta_{2}$ in Equation~\eqref{eq:thetaTwoAndAlpha}, we simply set $\rho_{\alpha}^{(1)}=\max_{x\in\vect{S}^{+}}\rho_{h}^{(1)}(\vect{x})$, which is sufficient, but may not give the optimal value for $\vartheta_{2}$ (i.e., this choice may not give the largest $\vartheta_{2}$ while still maintaining $\vect{u}_{h}^{(2)}\in\mathcal{G}_{\vect{u}}$).  

Step~2 is also conservative and does not change the cell averages; i.e., $\vect{u}_{\vect{K}}^{(2)}=\vect{u}_{\vect{K}}^{(1)}=\vect{u}_{\vect{K}}$.  
Also, if the bounds on the electron fraction are not violated, $\vartheta_{2}=\alpha=1$ and $\vect{u}_{h}^{(2)}(\vect{x})=\vect{u}_{h}^{(1)}(\vect{x})$.  
After the completion of Steps~1 and 2, we have ensured $\vect{u}_{h}^{(2)}(\vect{x}_{\vect{q}})\in\mathcal{G}_{\vect{u}}$ for all $\vect{x}_{\vect{q}}\in\vect{S}^{+}$.  


\subsubsection{Step 3: Specific Internal Energy}
\label{sec:BELstep3} 

In the third, and final, step we enforce $\epsilon_{\vect{q}}\ge\epsilon_{{\rm{min}},\vect{q}}$ for all $\vect{x}_{\vect{q}}\in\vect{S}^{+}$.  
To this end, we define $\vect{U}_{h}^{(2)}=\big(\,\rho_{h}^{(2)},\vect{m}_{h},E_{h},D_{{\rm{e}},h}^{(2)}\,\big)^{\rm{T}}$, which is the full solution vector after steps~1 and 2.  
Using $\vect{U}_{h}^{(2)}$, the specific internal energy and electron fraction in each point $\vect{x}_{\vect{q}}\in\vect{S}^{+}$ are computed as
\begin{equation}
  \epsilon_{\vect{q}}^{(2)} = \epsilon\big(\vect{U}_{h}^{(2)}(\vect{x}_{\vect{q}})\big)
  = \Big(\,E_{\vect{q}}-\f{m_{\vect{q}}^{2}}{2\,\rho_{\vect{q}}^{(2)}}\,\Big)/\rho_{\vect{q}}^{(2)}
  \quad\text{and}\quad
  Y_{{\rm{e}},\vect{q}}^{(2)}
  =\ye\big(\vect{U}_{h}^{(2)}(\vect{x}_{\vect{q}})\big)
  =D^{(2)}_{{\rm{e}},\vect{q}}/\rho_{\vect{q}}^{(2)},
  \label{eq:specificInternalEnergyPPSolution}
\end{equation}
respectively.  
Then, if $\epsilon_{\vect{q}}^{(2)}<\epsilon_{{\rm{min}},\vect{q}}^{(2)}\equiv\epsilon\big(\rho_{\vect{q}}^{(2)},T_{\rm{min}},Y_{{\rm{e}},\vect{q}}^{(2)}\big)$ for any $\vect{x}_{\vect{q}}\in\vect{S}^{+}$, we replace $\vect{U}_{h}^{(2)}$ with the limited polynomial
\begin{equation}
  \vect{U}_{h}^{(3)}(\vect{x}) := (1-\vartheta_{3})\,\vect{U}_{\vect{K}} + \vartheta_{3}\,\vect{U}_{h}^{(2)}(\vect{x}) \quad (\vartheta_{3}\in[0,1]).
  \label{eq:limitedPolynomialStep3}
\end{equation}
Here, the polynomial representation of the full solution is written as a convex combination of the cell average and the polynomial representation after Step~2.  
Since we assume $\vect{U}_{\vect{K}}\in\mathcal{G}$, setting $\vartheta_{3}=0$ will ensure $\vect{U}_{h}^{(3)}(\vect{x})\in\mathcal{G}$.  
However, setting $\vartheta_{3}=0$, so that $\vect{U}_{h}^{(3)}=\vect{U}_{\vect{K}}$, kills off all the high-order accuracy of the polynomial representation, which is undesirable.  
Instead, one would want to find the largest value for $\vartheta_{3}$ to retain as much high-order accuracy as possible \emph{and} enforce $\vect{U}_{h}^{(3)}(\vect{x})\in\mathcal{G}$ for all $\vect{x}_{\vect{q}}\in\vect{S}^{+}$.  
As discussed above, this is complicated by the fact that $\mathcal{G}$ is not strictly convex.  
It is further complicated by the fact that the surface $\epsilon_{\rm{min}}\big(\rho,\ye\big)$ is only available at discrete points from the EoS table.  
Because of this, we will assume that $\mathcal{G}$ is locally convex and first obtain $\vartheta_{3,\vect{q}}$ by solving
\begin{equation}
  \epsilon\big(\,\vect{s}(\vartheta_{3,\vect{q}})\,\big)
  = (1-\vartheta_{3,\vect{q}})\,\epsilon_{{\rm{min}},\vect{K}} + \vartheta_{3,\vect{q}}\,\epsilon_{{\rm{min}},\vect{q}}^{(2)},
  \label{eq:equationForTheta3}
\end{equation}
for each $\vect{x}_{\vect{q}}$ where $\epsilon_{\vect{q}}^{(2)}<\epsilon_{{\rm{min}},\vect{q}}^{(2)}$.  
On the left-hand side of Equation~\eqref{eq:equationForTheta3} we have defined
\begin{equation}
  \vect{s}(\vartheta_{3,\vect{q}}) = (1-\vartheta_{3,\vect{q}})\,\vect{U}_{\vect{K}} + \vartheta_{3,\vect{q}}\,\vect{U}_{\vect{q}}^{(2)},
  \label{eq:straightLineTheta3}
\end{equation}
while on the right-hand side of Equation~\eqref{eq:equationForTheta3} we have defined $\epsilon_{{\rm{min}},\vect{K}} = \epsilon(\rho_{\vect{K}},T_{\rm{min}},Y_{{\rm{e}},\vect{K}})$.  
Then we set
\begin{equation}
  \vartheta_{3} := \min_{\vect{q}}\vartheta_{3,\vect{q}},
\end{equation}
where the minimum is taken over all the points in $\vect{S}^{+}$ where the specific internal energy fell below the minimum value.  

We note that the limiter in Equation~\eqref{eq:limitedPolynomialStep3} is conservative in all the fields in the sense that the cell average is preserved; i.e.,
\begin{equation}
  \f{1}{V_{\vect{K}}}\int_{\vect{K}}\vect{U}_{h}^{(3)}\,dV_{h}
  =(1-\vartheta_{3})\,\vect{U}_{\vect{K}} + \vartheta_{3}\,\f{1}{V_{\vect{K}}}\int_{\vect{K}}\vect{U}_{h}^{(2)}\,dV_{h}
  =(1-\vartheta_{3})\,\vect{U}_{\vect{K}} + \vartheta_{3}\,\vect{U}_{\vect{K}}
  =\vect{U}_{\vect{K}}.  
\end{equation}

\begin{figure}[h]
  \centering
  \includegraphics[width=0.75\textwidth]{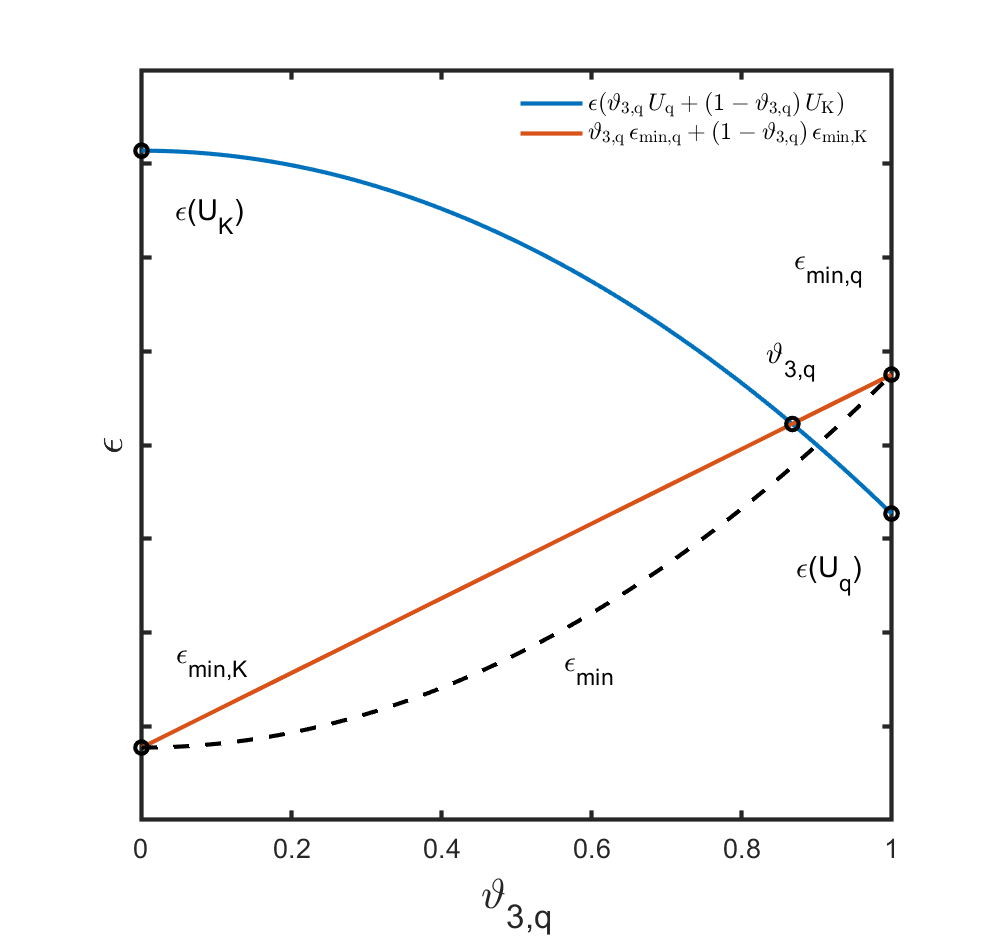}
  \caption{Illustration of the bisection problem used to find $\vartheta_{3,\vect{q}}$ in Equation~\eqref{eq:equationForTheta3} to determine the extent of limiting needed to ensure that the specific internal energy does not fall below the table boundary $\epsilon_{\rm{min}}$ (dashed black curve). In the example depicted here, $\epsilon(\vect{U}_{\vect{q}})$, the right endpoint of the blue curve, is below the table boundary, and limiting is needed.  We find $\vartheta_{3,\vect{q}}$ as the intersection point between the blue curve connecting $\epsilon(\vect{U}_{\vect{K}})$ and $\epsilon(\vect{U}_{\vect{q}})$, and the red curve connecting $\epsilon_{{\rm{min}},\vect{K}}$ and $\epsilon_{{\rm{min}},\vect{q}}$.  In this case, $\vartheta_{3,\vect{q}}\approx0.87$.}
  \label{fig:Bisection}
\end{figure}

The motivation for solving Equation~\eqref{eq:equationForTheta3} is as follows \citep[cf.][for the ideal EoS case]{zhang:2010a}:  
$\vect{s}(\vartheta_{3,\vect{q}})$ is the parametrized straight line connecting the cell average $\vect{U}_{\vect{K}}$ and the point value $\vect{U}_{\vect{q}}^{(2)}$.  
Since $\vect{U}_{\vect{K}}\in\mathcal{G}$, we know that
\begin{equation}
  \epsilon_{\vect{K}}\equiv\Big(\,E_{\vect{K}}-\frac{m_{\vect{K}}^{2}}{2\rho_{\vect{K}}}\,\Big)/\rho_{\vect{K}}\ge\epsilon_{{\rm{min}},\vect{K}}.
  \label{eq:specificInternalEnergyCellAverage}
\end{equation}
On the other hand, if $\vect{U}_{\vect{q}}^{(2)}\notin\mathcal{G}$, there is at least one intersection point of the line $\vect{s}(\vartheta_{3,\vect{q}})$ and the boundary of $\mathcal{G}$; i.e. the surface $\epsilon_{\rm{min}}(\rho,\ye)$.  
(If $\mathcal{G}$ is convex, which we assume in this step, there is exactly one intersection point.)  
Since we do not know the exact shape of the surface, we approximate it by the line segment connecting the boundary points $\epsilon_{{\rm{min}},\vect{K}}$ and $\epsilon_{{\rm{min}},\vect{q}}^{(2)}$, and by the convexity assumption, this line lies above the surface $\epsilon_{\rm{min}}(\rho,\ye)$.  
Thus, in Equation~\eqref{eq:equationForTheta3}, the solution $\vartheta_{3,\vect{q}}$ provides the intersection point between the line connecting the points $\epsilon_{\vect{K}}$ and $\epsilon_{\vect{q}}^{(2)}$ and the line connecting the points $\epsilon_{{\rm{min}},\vect{K}},\epsilon_{{\rm{min}},\vect{q}}^{(2)}$.  
See Figure~\ref{fig:Bisection} for an illustration.  

Equation~\eqref{eq:equationForTheta3} is solved for $\vartheta_{3,\vect{q}}$ with a simple bisection algorithm, using the end points $\vartheta_{3,\vect{q}}=0$ and $\vartheta_{3,\vect{q}}=1$ as starting points.  
We note that, in practice, the solution to Equation~\eqref{eq:equationForTheta3} does not have to be accurate to many significant digits, and the bisection algorithm can be terminated after a few iterations.  
We also note that since $\epsilon_{\rm{min}}(\rho,\ye)$ is not strictly convex, as is shown in Figure~\ref{fig:SFHoTable}, Equation~\eqref{eq:equationForTheta3} can have multiple roots, and the bisection algorithm may result in a limited solution that is still outside $\mathcal{G}$.  
We have, however, not encountered a situation where this happens.  
On the contrary, in the numerical examples presented in Section~\ref{sec:results}, we find that the limiting procedure discussed in this section significantly improves the robustness of the DG algorithm.  
As can be seen by looking ahead to Figure~\ref{fig:Gravitational Collapse, Bound Enforcing Limiter Activation Sites} in Section~\ref{sec:collapse}, the bound-enforcing limiter is continuously activated, with $\vartheta_{3}\in[0.4,1]$, in a short time interval around core bounce in an adiabatic collapse simulation.  

Finally, we have assumed that the cell average satisfies $\vect{U}_{\vect{K}}\in\mathcal{G}$ when the limiter is applied.  
If this assumption does not hold, the bound-enforcing limiter will fail.  
By considering the equation for the cell average in Equation~\eqref{eq:dgSemidiscreteCellAverage}, in combination with forward Euler time stepping, it may be possible to derive a sufficient restriction on the time step such that $\vect{U}_{\vect{K}}^{n+1}\in\mathcal{G}$, provided $\vect{U}_{\vect{K}}^{n}\in\mathcal{G}$ and $\vect{U}_{\vect{q}}^{n}\in\mathcal{G}$ (possibly with additional points included in the set $\vect{S}^{+}$).  
We do, however, not pursue this endeavor here.  
Instead, we use the time step restriction given in Equation~\eqref{eq:cflCondition}, which may not be sufficient.  
In the absence of an explicit expression for a sufficient time step restriction (assuming one exists), one may design a time step control algorithm where the step size is recursively reduced, and the time step retaken, until a physically admissible cell average is obtained.  
On the other hand, we have yet to encounter an application in which a solution with cell average $\vect{U}_{\vect{K}}\notin\mathcal{G}$ is passed to the bound-enforcing limiter.

\subsection{Poisson Solver}
\label{sec:poisson}

In \thornado, the approximate Newtonian gravitational potential, $\Phi_{h}$, is obtained using the Poseidon code (Roberts et al., in preparation).  
Poseidon solves Equation~\eqref{eq:poissonEquation} on a spherical-polar grid with a combination of an angular spectral expansion using spherical harmonics and a radial finite element solution method. Here, we discuss the case of spherical symmetry, and therefore limit the angular expansion to the monopole harmonic function.  Therefore we will focus only on the finite element method \citep{Larson:2013} used in the radial expansion. 

Poseidon represents the approximate solution, $\Phi_{h}$, to Equation~\eqref{eq:poissonEquation} as a continuous expansion of functions of the form
\beq
  \Phi_{h}(r, t) 
  = \sum_{i=1}^{N_{D}} \Phi_{i}(t) v_{i}(r), 
  \label{eq:PotentialExpansion}
\eeq
where $N_{D}$ is the total number of interpolation nodes on the domain $D$, and $\Phi_{i}(t)$ are spatially constant expansion coefficients. As the method used to solve the Poisson equation is a purely spatial in nature, we will omit the time parameter, $t$, for the rest of this section. 
The basis functions $v_{i}(r)$ belong to the approximation space, $V_{h}$, defined by 
\beq
  V_{h} = \big\{ \psi_{h} : \psi_{h}|_{K^{(j)}} \in P^{k}(K^{(j)}), ~ j=1,\ldots,N_{e}\,\big\}, 
\eeq
where $P^{k}$ is a space of one-dimensional piecewise polynomials of degree $k$, and $K^{(j)}$ are the radial elements of the same decomposition of the computational domain as expressed in Section~\ref{sec:DG}.  
Given this choice of approximation space and domain decomposition, $N_{D}$ is given by $N_{D} = N_{e} k + 1$, where $N_{e}$ is the number of radial elements on the domain. 

\begin{figure}[h]
  \centering
  \includegraphics[width=0.75\textwidth]{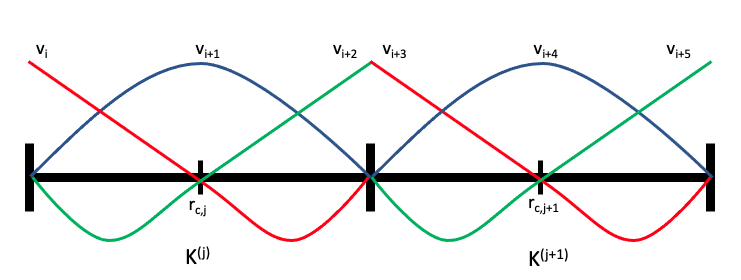}
  \caption{Illustration of the basis functions, $v_{i}$, used in the finite element solution method of Poseidon for the case $k=2$. Each function is associated with a specific element $K^{(j)}$ and a node $\xi_{j,m}$ within that element, such that $v_{i}\big( r(\xi_{j,m}) \big) = 1$.  Outside of the associated element, $v_{i} = 0$, and is therefore not depicted.}
  \label{fig:Poisson_LagPoly}
\end{figure}

The Newtonian gravitational potential is a continuous function in space; therefore, we require the approximate solution, $\Phi_{h}$, to be $C^{0}$ continuous across element interfaces.
This is achieved through the choice of interpolation points and approximation space polynomials.  
Within a specific element $K^{(j)}$, the interpolation points, $\hat{S}_{j} = \{\xi_{j,1},\ldots,\xi_{j,m}, \ldots, \xi_{j,k+1}\}\subset I = [-1,1]$, are chosen to be the Legendre--Gauss--Lobatto (LGL) points.  
The physical coordinate $r$ is related to the reference coordinate $\xi$ by the transformation 
\beq
	r \big(\xi \big) = r_{{\rm{c}},j} + \frac{\Delta r_{j}}{2} \xi,
\eeq where $r_{{\rm{c}},j}$ is the physical coordinate for the center of element $K^{(j)}$ and is such that 
\beq
	r\big(\xi_{j,k+1}\big) = r\big(\xi_{j+1,1}\big).
\eeq The inverse relationship, 
\beq
	\xi(r) = \frac{2}{\Delta r_{j}}\big(r - r_{{\rm{c}},j}\big),
\eeq
 allows us to express the chosen approximation space polynomials $v_{i}(r) \in V_{h}$ as 
\begin{align}
	v_{i}(r) = 
	\begin{cases}
		\ell_{j,m}\big( \xi(r) \big)  & \text{for } r \in K^{(j)} \\
		0 					 & \text{else}
	\end{cases},
\end{align}
where $\ell_{j,m}$ are the Lagrange polynomials in Equation~\eqref{eq:lagrangePoly} constructed with the LGL points, $\hat{S}_{j}$.  
Each approximation function $v_{i}(r)$ is associated with a node $\xi_{j,m}$ such that $v_{i}\big( r(\xi_{j,m}) \big) = 1$ by the Kronecker delta property of the Lagrange polynomials. 
This choice of interpolation points and approximation functions enforces the $C^{0}$ continuity of the solution.  
See Figure~\ref{fig:Poisson_LagPoly} for an illustration of elements and associated basis functions in the finite element method for the case with $k=2$.  

The finite element method seeks to find $\Phi_{h} \in V_{h}$, which approximates $\Phi$ in Equation~\eqref{eq:poissonEquation} such that
\beq
  \langle\ \nabla^2 \Phi_{h}, \psi_{h} \rangle_{D}  = \langle\ 4 \pi G \rho, \psi_{h} \rangle_{D}
  \label{eq:FEMWeak_A}
\eeq
holds for all test functions $\psi_{h} \in V_{h}$. In Equation~\eqref{eq:FEMWeak_A},
\beq
	\langle\ \nabla^2 \Phi_{h}, \psi_{h} \rangle_{D} = 4\pi\int_{R_{\Lo}}^{R_{\Hi}} \nabla^2 \Phi_{h}\, \psi_{h}\,r^{2}dr,
	\label{eq: FEMWeak_Left}
\eeq
and
\beq
\langle\ 4 \pi G \rho, \psi_{h} \rangle_{D} = 16 \pi^2 G \int_{R_{\Lo}}^{R_{\Hi}} \rho\, \psi_{h}\,r^{2}dr,
\eeq
where $R_{\Lo}$ and $R_{\Hi}$ are the low and high radial boundary locations of the domain, respectively.  
Using integration by parts on Equation~\eqref{eq: FEMWeak_Left}, Equation~\eqref{eq:FEMWeak_A} becomes the weak form of Equation~\eqref{eq:poissonEquation},
\beq
  -\langle\ \pd{\Phi_{h}}{r}, \pd{\psi_{h}}{r} \rangle_{D} + \big(\pd{\Phi_{h}}{r}\big) \psi_{h}|^{R_{\Hi}}_{R_{\Lo}} = \langle\ 4 \pi G \rho, \psi_{h} \rangle_{D}. 
  \label{eq:FEMWeak_B}
\eeq

\begin{figure}[h]
  \centering
  \includegraphics[width=0.3\textwidth]{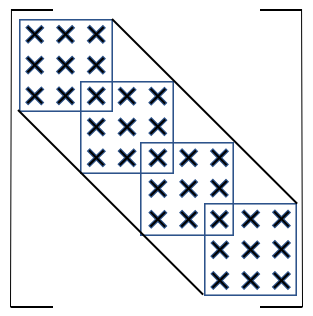}
  \caption{Nonzero structure of the matrix $S$ used in Poseidon for the case of $k = 2$, and $N_{e} = 4$. The diagonal lines denote the band structure of the matrix.  The squares denote the overlapping block structure within the band.  The single overlapping element shared by the consecutive blocks comes from the shared interpolation node at element interfaces. }
  \label{fig:Poisson_Matrix}
\end{figure}

For the gravitational collapse problem discussed in Section~\ref{sec:collapse}, we impose the Neumann boundary condition,
\beq
  \pd{\Phi_{h}}{r}(R_{\Lo}) = 0,		\label{eq:PoissonNeumannBC}
\eeq
on the inner boundary ($R_{\Lo}=0$) to preserve the symmetry of the solution, and the Dirichlet boundary condition,
\beq
  \Phi_{h}(R_{\Hi}) = -\frac{GM_{\rm{enc}}}{R_{\Hi}},	\label{eq:PoissonDirichletBC}
\eeq
on the outer boundary, where $M_{\rm{enc}}$ is the total enclosed mass given by
\beq
	M_{\rm{enc}} = 4\pi\int_{R_{\Lo}}^{R_{\Hi}} \rho_{h}(r)\,r^{2}dr.
\eeq
The Neumann condition in Equation~\eqref{eq:PoissonNeumannBC} reduces Equation~\eqref{eq:FEMWeak_B} to 
\beq
-\langle\ \pd{\Phi_{h}}{r}, \pd{\psi_{h}}{r} \rangle_{D} + \big(\pd{\Phi_{h}}{r}(R_{\Hi})\big) \psi_{h}(R_{\Hi}) = \langle\ 4 \pi G \rho, \psi_{h} \rangle_{D}. \label{eq:FEMWeak_C}
\eeq
Next, the expansion in Equation~\eqref{eq:PotentialExpansion} and $\psi_{h} = v_{j}$ are substituted into Equation~\eqref{eq:FEMWeak_C} to give
\beq
  \sum_{i=1}^{N_{D}}  \Phi_{i} \Big( -\langle\ \pd{v_{i}}{r}, \pd{v_{j}}{r} \rangle_{D} + \big( \pd{v_{i}}{r}(R_{\Hi})\big) v_{j}(R_{\Hi}) \Big) = \langle\ 4 \pi G \rho, v_{j} \rangle_{D},
  \hspace{1em} j=1,\cdots,N_{D}		 \label{eq:FEMWeak_D}.
\eeq
To enforce the Dirichlet condition, the expansion coefficient $\Phi_{N}$ is set to the boundary value given by Equation~\eqref{eq:PoissonDirichletBC}, and the dimensionality of the problem is reduced to $N_{D}-1$, eliminating the $ \big( \pd{v_{i}}{r}(R_{\Hi})\big) v_{j}(R_{\Hi})$ term as $v_{j}(R_{\Hi}) = 0$,  $\forall$ $j \ne N_{D}$.   
Equation~\eqref{eq:FEMWeak_D} then becomes
\beq
  \sum_{i=1}^{N_{D}-1}  -\Phi_{i} \langle\ \pd{v_{i}}{r}, \pd{v_{j}}{r} \rangle_{D} = \langle\ 4 \pi G \rho, v_{j} \rangle_{D}, \quad  j=1,\cdots,N_{D}-1. \label{eq:FEMWeak_E}
\eeq
Defining the stiffness matrix as
\beq
S = \left\{ s_{ij} \right\}_{i,j=1}^{N_{D}-1}, \hspace{10pt} s_{ij} = -\langle\ \pd{v_{i}}{r}, \pd{v_{j}}{r} \rangle_{D}, 
\eeq
the load vector as
\beq
  L = \left\{ \langle\ 4 \pi G \rho, v_{j} \rangle_{D} \right\}_{j=1}^{N_{D}-1},
\eeq
and the unknown coefficient vector as
\beq
  C = \left\{ \Phi_{i} \right\}_{i=1}^{N_{D}-1},
\eeq
the system in Equation~\eqref{eq:FEMWeak_E} can then be written in matrix form as
\begin{align}
SC = L.
\end{align}
The matrix $S$ is a sparse symmetric band matrix, with bandwidth equal to $k$.  When $k=1$, the matrix $S$ is tridiagonal. When $k>1$, an overlapping block structure occurs within the diagonal band of $S$, see \figref{fig:Poisson_Matrix}.  
The sparsity of the matrix is given by 
\beq
 {\rm{Sparsity}} = \frac{N_{e}k^2 - N_{e} + 1}{N_{e}^2k^2 + 2N_{e}k + 1}.
 \eeq  
To reduce memory overhead, $S$ is stored in compressed column storage (CCS) format.  The system is then solved using a CCS compatible Cholesky factorization. 
Once these coefficients are known the approximate solution can be reconstructed anywhere within the domain using Equation~\eqref{eq:PotentialExpansion}.    

\subsection{Table Interpolation}
\label{sec:tableInterpolation}

As in \cite{bruenn:1985}, \cite{mezzacappa:1999}, and \cite{bruenn:2020}, we obtain a thermodynamic quantity $F$ and its derivatives from the tabulated EoS through trilinear interpolation in the space spanned by $\left( \log_{10}(\rho), \, \log_{10}(T), \, \ye \right)$.  
The software to compute these are provided by the \weaklib\ library, and, for completeness, we restate formulas here.  
To simplify the notation, let $X = \log_{10}(\rho)$, $Y = \log_{10}(T)$, and $Z = \ye$. 
Then, $\bar{F} = \bar{F}(X, Y, Z)$, where $\bar{F}$ is related to the thermodynamic quantity by $F = 10^{\bar{F}} - F_{\text{offset}}$.  
That is, trilinear interpolations are performed on logged quantities, and the offset is used to ensure that $\bar{F}$ is well-defined when $F$ is negative.  
Obtaining $F$ first requires the eight points $(X_{a}, Y_{b}, Z_{c}) : a, b, c \in \{0, 1\}$ from the table that make up the corners of a "cube" of the points closest to $\left(X, Y, Z\right)$. 
These points then satisfy
\begin{equation}
    X_{1} - X_{0} = \frac{1}{N_{\rho}}, \quad
    Y_{1} - Y_{0} = \frac{1}{N_{T}}, \quad\text{and}\quad
    Z_{1} - Z_{0} = \frac{1}{N_{\ye}},
\end{equation}
where $N_{\rho}$ and $N_{T}$ are the number of the points per decade in $\rho$ and $T$, respectively, and $N_{\ye}$ is the number of points per unit interval in $\ye$.  
$\bar{F}$ is then given by the trilinear interpolation formula, e.g., found in Eq.~(32) in \cite{mezzacappa:1999}, which, in multi-index notation, can be written compactly as
\begin{equation}
   \bar{F}(\vect{X}) = \sum_{\vect{i} = \vect{0}}^{\vect{1}}w_{\vect{i}}(\vect{X})\bar{F}_{\vect{i}},
    \label{eq:TrilinearInterpolationQuantity}
\end{equation}
where $\vect{X} = (X, Y, Z)$. 
In this context, the weights $w_{\vect{i}}(\vect{X})$ are given by
\begin{equation}
	w_{\vect{i}}(\vect{X}) = B_{i_{1}}(X)B_{i_{2}}(Y)B_{i_{3}}(Z),
	\label{eq:TrilinearInterpolationWeights}
\end{equation}
where $B_{i_{1}}(X)$ ($i_{1}\in\{0,1\}$) are linear Lagrange polynomials
\begin{equation}
    B_{0}(X) = \frac{X_{1} - X}{X_{1} - X_{0}} \quad\text{and}\quad
    B_{1}(X) = \frac{X - X_{0}}{X_{1} - X_{0}},
\end{equation}
and $B_{i_{2}}(Y)$ and $B_{i_{3}}(Z)$ are similarly defined by replacing $X$ with $Y$ or $Z$, respectively.  

As in \cite{mezzacappa:1999}, derivatives with respect to $\rho$, $T$, and $\ye$ are calculated directly from this expression; i.e.
\begin{equation}
    \left(\pderiv{F}{\rho}\right)_{T, \ye} = \frac{(F + F_{\text{offset}})}{\rho}\left(\pderiv{\bar{F}}{X}\right)_{Y, Z} = 
	\frac{(F + F_{\text{offset}})}{\rho}\sum_{\vect{i} = \vect{0}}^{\vect{1}}\pderiv{w_{\vect{i}}}{X}\bar{F}_{\vect{i}},
    \label{eq:TrilinearInterpolationRhoDerivative}
\end{equation}

\begin{equation}
    \left(\pderiv{F}{T}\right)_{\rho, \ye} = \frac{(F + F_{\text{offset}})}{T}\left(\pderiv{\bar{F}}{Y}\right)_{X, Z} = 
    \frac{(F + F_{\text{offset}})}{T}\sum_{\vect{i} = \vect{0}}^{\vect{1}}\pderiv{w_{\vect{i}}}{Y}\bar{F}_{\vect{i}},
    \label{eq:TrilinearInterpolationTDerivative}
\end{equation}

\begin{equation}
    \left(\pderiv{F}{\ye}\right)_{\rho, T} = (F + F_{\text{offset}})\left(\pderiv{\bar{F}}{Z}\right)_{X, Y} = 
    (F + F_{\text{offset}})\sum_{\vect{i} = \vect{0}}^{\vect{1}}\pderiv{w_{\vect{i}}}{Z}\bar{F}_{\vect{i}}.
    \label{eq:TrilinearInterpolationYDerivative}
\end{equation}

We note that this interpolation scheme does not, by construction, satisfy the Maxwell relations of thermodynamics. While this may impact the ability to resolve adiabatic flows (see \cite{swesty:1996} and \cite{timmes:2000} for further discussion), we do not observe any clear evidence of this being a problem in our computations.  
In addition, while we believe that the low-order accuracy of the trilinear interpolation scheme may play a role in both the convergence rates observed with the high-order RKDG scheme in Section~\ref{sec:advection} and the issues with characteristic limiting around the phase transition observed in Section~\ref{sec:collapse}, additional investigations are required.

\section{Numerical Results}
\label{sec:results}

In this section, we present results obtained with the DG method as implemented in $\thornado$ for various test problems relevant to CCSNe and other astrophysical phenomena.  
With the exception of few reference calculations obtained using an ideal EoS in Section~\ref{eq:advectionConvergence}, all the results were obtained using a tabulated version of the SFHo EoS of \citet{steiner:2013a}, which covers the ranges $\rho\in[1.66\times10^{3},3.16\times10^{15}]$~g~cm$^{-3}$, with $N_{\rho}=25$, $T\in[1.16\times10^{9},1.84\times10^{12}]$~K, with $N_{T}=50$, and $\ye\in[0.01,0.7]$, with $N_{\ye}=100$.  
(See, however, \cite{endeve:2019} for a documentation of results obtained with \thornado\ using an ideal EoS.)  
In the first two subsections, we begin by presenting results from one-dimensional advection tests using Cartesian coordinates, and one- and two-dimensional Riemann problems using Cartesian, spherical-polar, and cylindrical coordinates (Sections~\ref{sec:advection} and \ref{sec:riemann}, respectively).  
These tests serve as an initial gauge of the implementation of the DG algorithm in \thornado\ with a nuclear EoS.  
Using Riemann problems with initial conditions adapted from their ideal EoS counterparts, we aim to investigate the performance of our implementation in curvilinear coordinates, as well as the slope limiter presented in Section~\ref{sec:slope} and the bound-enforcing limiter presented in Section~\ref{sec:boundEnforcing}.  
The Poisson solver is tested in Section~\ref{sec:gravpotential}.  
Then, in Section~\ref{sec:collapse}, our focus turns to the main application, adiabatic gravitational collapse in spherical symmetry, where we investigate the performance of \thornado's DG algorithm by investigating various aspects of the solver with an eye towards future spherically symmetric --- and eventually multidimensional --- supernova simulations with neutrino transport.  
In all the tests, the CFL number in Equation~\eqref{eq:cflCondition} is set to $C_{\textsc{cfl}}=0.5$.  



\subsection{Advection Tests}
\label{sec:advection}

\subsubsection{Rate of Convergence}
\label{eq:advectionConvergence}

The accuracy of the DG method can be manipulated by changing the number of nodes per cell $N=k+1$ and/or the total number of cells $K$.  
The number of nodes per cell (or element) governs the expected order of accuracy of the method.  
($N$th order spatial accuracy is expected with $N$ nodes.)  
This section covers the rate at which changing the number of degrees of freedom $n_{\mathrm{DOF}}=(k+1) \times K$ impacts the accuracy; i.e. the convergence rate.  
Inspired by \citet{Suresh:1997}, this test is performed over the 1D computational domain $D= [-100,\,100]$~km, with smooth initial conditions, and periodic boundary conditions.  
The initial state for the tabulated EoS case is set with the primitive state vector $\mathbf{P}$ as
\begin{equation*}
    \mathbf{P} = \big(\rho,u,p,\ye\big)^{\rm{T}} = \big(\,\rho_{0}\,\big(\,1 + 0.1 \, \sin^{4}( \pi x / L )\,\big), \, v_{0}, \, p_{0}, \, 0.3 \,\big)^{\rm{T}},
\end{equation*}
where $\rho_{0}=10^{12} \, \mathrm{g} \, \mathrm{cm}^{-3}$ is the background density, $v_{0}=0.1\, c$ is the velocity, $p_{0}=0.01\, \rho_{0} \, c^{2}$ the background pressure, and $L=200$~km is the domain length.  
In this test, the mass density, a quartic sine wave, is advected for one period without any limiting, while the velocity, pressure, and electron fraction remain constant.  
The error in mass density between the initial and final states is then calculated in the $L_{1}$ error norm,
\begin{equation}
    L_{1} \equiv \sum_{j=1}^{n_{\mathrm{DOF}}}\,\left|\rho_{j,\mathrm{final}}-\rho_{j,\mathrm{initial}}\right|.
\end{equation}
In Figure~\ref{fig:Advection, Convergence} we plot this quantity, scaled by both $n_{\mathrm{DOF}}$ and a background density $\rho_{0}$, versus $n_{\mathrm{DOF}}$ (crosses).  
(For reference, we also plot results obtained with an ideal EoS case with $\Gamma = 1.4$; open circles.)  
The solutions are obtained using $N=2$ (black symbols) and $N=3$ (red symbols) nodes with second and third-order time integration schemes, respectively.  
For each $N$, we use seven different values of $K:\,8, \, 16, \, 32, \, 64, \, 128, \, 256, \, \mathrm{and} \, 512$. For this smooth problem we always observe that for a fixed $n_{\mathrm{DOF}}$ the scheme with $N=3$ is significantly more accurate than the scheme with $N=2$.  
For the nuclear EoS case, the $L_{1}$ error for the second-order scheme ($N=2$) crosses zero and generates a cusp at $n_{\mathrm{DOF}}=512$.  
Otherwise, the results obtained with the second-order method agree well with the expected convergence rate for both the tabulated and ideal EoS cases.  
For $N=3$, the ideal EoS case exhibits third-order accuracy throughout.  
However, for $n_{\mathrm{DOF}}>96$, the results for $N=3$ with the tabulated EoS appear to undergo a transition from third-order to second-order accuracy.  
We suspect that the trilinear interpolation method discussed in \ref{sec:tableInterpolation} may be the cause of the loss of accuracy for large $n_{\mathrm{DOF}}$, but this requires further investigation.  

\begin{figure}[h]
  \centering
  \includegraphics[width = 0.75\textwidth]{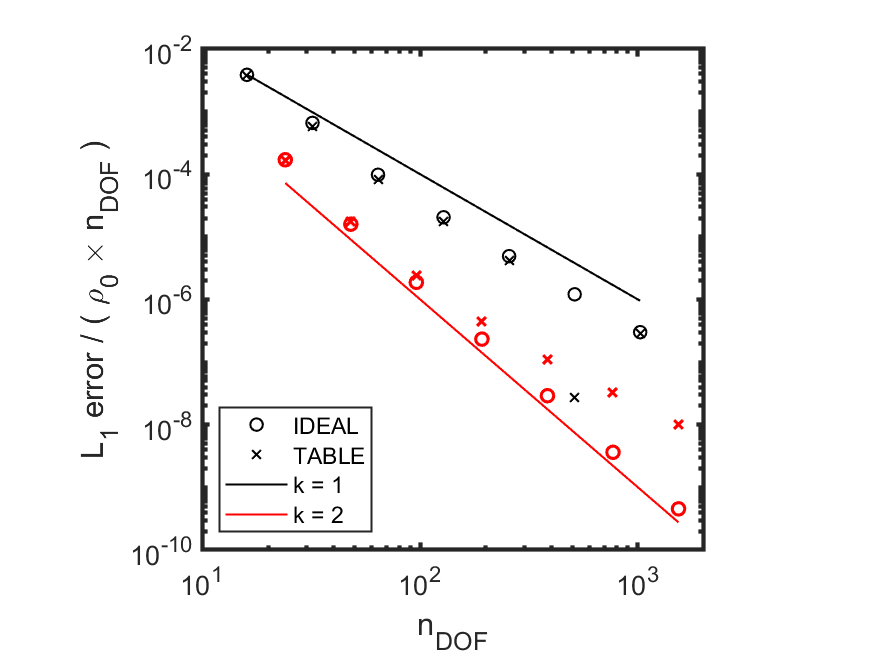}
  \caption{$L_{1}$ error between the initial and final states of an advected quartic sine wave, adopted from \citet{Suresh:1997}. The results are scaled by the number of degrees of freedom to obtain the average error per node. For the tabulated EoS results, the background density $\rho_{0}=10^{12} \, \mathrm{g} \, \mathrm{cm}^{-3}$ is also used to scale the error, but $\rho_{0}=1$ for the ideal case, which is run in dimensionless mode. The solid lines are proportional $n_{\mathrm{DOF}}^{k+1}$, and serve as references for the convergence rates of solutions represented by polynomials of degree $k=1$ and $k=2$.}
  \label{fig:Advection, Convergence}
\end{figure}

\subsubsection{Discontinuous Multi-Wave}

This test from \citet{Suresh:1997} involves the advection of a discontinuous initial state for mass density, which includes a Gaussian wave, a square wave, a triangular wave, and a semi-ellipse (see light red lines in Figure~\ref{fig:Discontinuous Multi-wave, Density}). 
This test is performed over a periodic 1D domain $D = [-100, \, 100]$~km, with the initial state given as
\begin{equation*}
    \mathbf{P} = \big(\rho,u,p,\ye\big)^{\rm{T}} = \big(\,\rho(x,0), \, v_{0}, \, p_{0}, \, 0.3 \,\big)^{\rm{T}},
\end{equation*}
where $v_{0}$ and $P_{0}$ are given the same values as in the previous test, and $\rho\left(x,0\right)$ is a piece-wise function defined as
\begin{table}[H]
    \centering
    \begin{tabular}{l l}
         $\rho(x,0)=\rho_{0}\left(1 + 0.1 \, \mathrm{exp}\left( -\log \left(2\right) \left( x/L + 0.7 \right)^{2}/\left(0.0009\right)\right) \right)$ & if $-80 \, \mathrm{km} \, \leq x \leq -60 \, \mathrm{km}$  \\
         $\rho(x,0)=\rho_{0}\left(1 + 0.1\right)$ & if $-40 \, \mathrm{km} \, \leq x \leq -20 \, \mathrm{km}$ \\
         $\rho(x,0)=\rho_{0}\left(1+0.1\left(1-\left|10\left(x/L - 0.1\right)\right|\right)\right)$ & if $\,\,\,\,\,\,\,\,0  \, \mathrm{km} \, \leq x \leq \,\,\,\,20 \, \mathrm{km}$ \\
         $\rho(x,0)=\rho_{0}\left(1+0.1\left(1-100\left(x/L-0.5\right)^{2}\right)^{1/2}\right)$ & if $\,\,\,\,\,40 \, \mathrm{km} \, \leq x \leq \,\,\,\,60 \, \mathrm{km}$ \\
         $\rho(x,0)=\rho_{0} $ & otherwise,
    \end{tabular}
\end{table}

\noindent
where $L=100$~km.  

We compare the performance of second- and third-order schemes in this test.  
Thus, a second- and third-order SSP-RK time integration scheme was used for $k=1$ and $k=2$, respectively. 
This test used the characteristic limiting procedure described in Section~\ref{sec:slope} with a TCI threshold $C_{\TCI}=1.0 \times 10^{-3}$ and a total variation diminishing parameter  $\beta_{\mathrm{TVD}}=1.5$. Figure~\ref{fig:Discontinuous Multi-wave, Density} shows the initial density profile (light red lines) along with four different cases of the mass density being evolved one (medium red lines) and ten (dark red lines) times across the periodic domain.  
Results obtained with the second-order method are displayed in the top panels, while results obtained with the third-order method are displayed in the bottom panels.  
Note that the results displayed in the top left and top right panels where obtained using the same total number of degrees of freedom as the results displayed in the bottom left and bottom right panels, respectively.  
Analytically, the evolved solution should match up exactly with the initial condition after each full domain crossing.  
However, the numerical solution is distorted by dissipation and dispersion.  
For fixed $n_{\mathrm{DOF}}$, the third-order method appears to provide more accurate results.  
As the solution is evolved in the $k=1$ case, accuracy is lost primarily around sharp edges, namely for the Gaussian and triangular waveforms.  
For the $k=1$ case with 384 elements, the solution is not well-resolved around the base of each waveform, but some accuracy is gained around the maxima.  
Loss of accuracy around sharp edges is also observed with the third-order method using $128$ elements (bottom left panel).  
However, as is seen in the bottom right panel, the features of the solution are better captured with the third-order method using $256$ elements.  
For the third-order method, we note that most of the distortion of the initial profile occurs in the first domain crossing, as the profiles after one and ten crossings are almost on top of each other.  
This is not so much the case for the second-order scheme, where the results after one and ten crossings are more easily distinguished.  
However, there is a trade-off between numerical accuracy and computational expense.  

\begin{figure}[h]
    \centering
    \includegraphics[width=.50\textwidth]{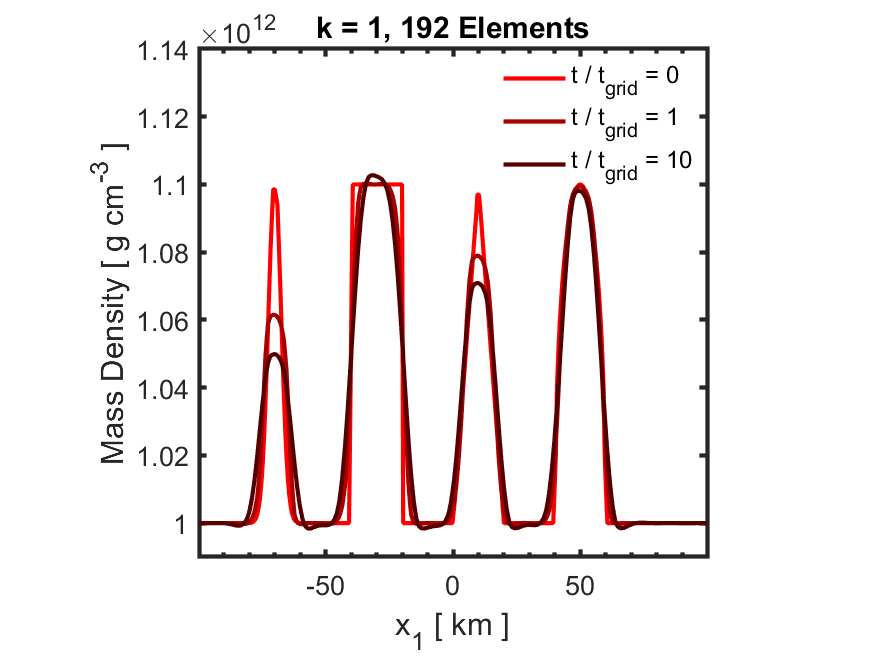}\hfill
    \includegraphics[width=.50\textwidth]{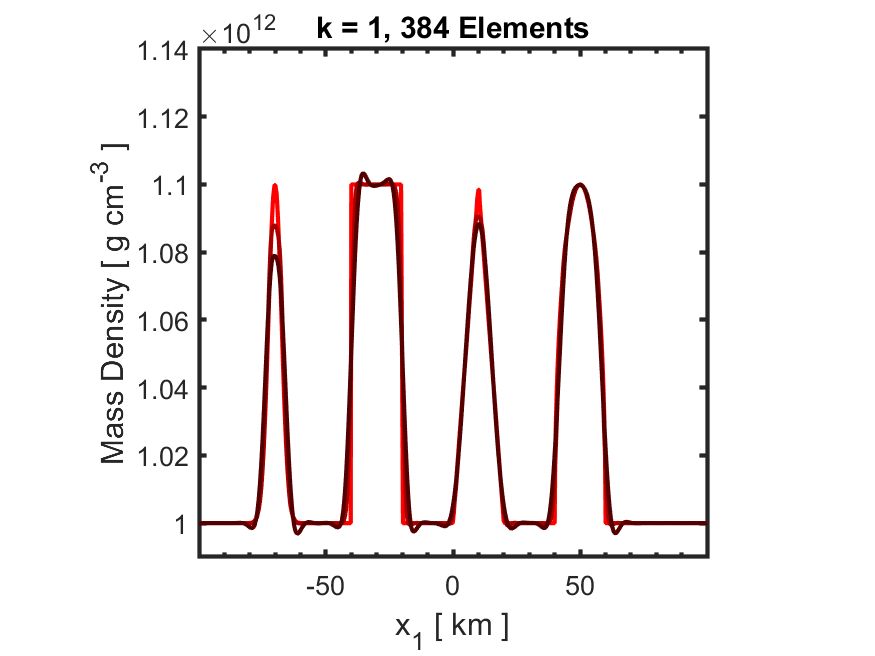}\hfill
    \includegraphics[width=.50\textwidth]{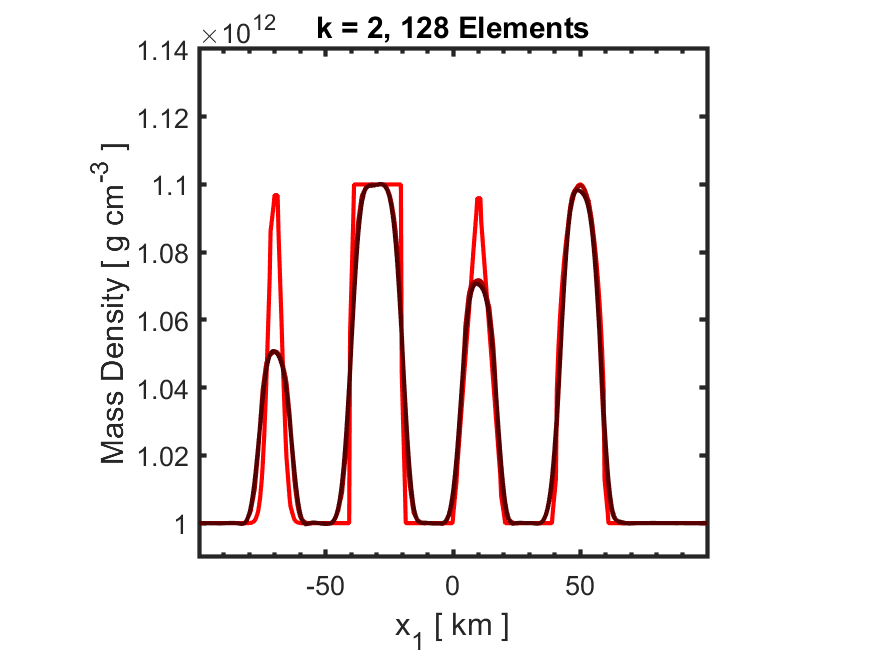}\hfill
    \includegraphics[width=.50\textwidth]{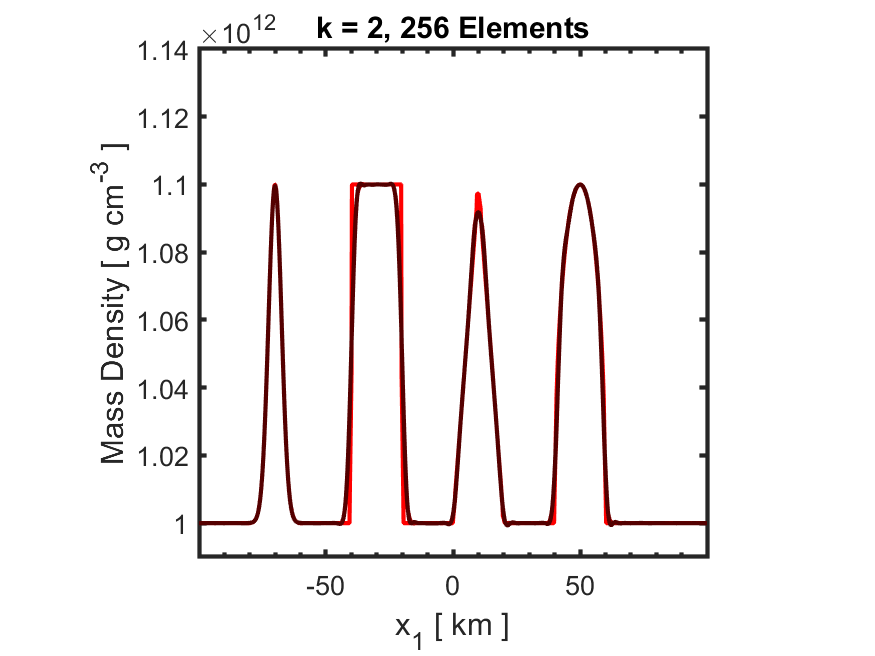}
    \caption{Mass density profiles for the discontinuous multi-wave advection test adopted from \citet{Suresh:1997}.  In each panel we plot the initial condition ($t/t_{\mathrm{grid}}=0$; light red), the solution after one period ($t/t_{\mathrm{grid}}=1$; medium red), and after ten periods ($t/t_{\mathrm{grid}}=10$; dark red).  
    ($t_{\mathrm{grid}}$ is the physical time required for one grid crossing.)  
    In the top panels we plot results obtained with the second-order method ($k=1$ and second-order SSP-RK time stepping) using $192$ (left panel) and $384$ (right panel) elements.  
    In the bottom panels we plot results obtained with the third-order method ($k=2$ and third-order SSP-RK time stepping) using $128$ (left panel) and $256$ (right panel) elements.  
    Increasing the number of nodes and/or elements results in better resolution around sharp peaks.}
    \label{fig:Discontinuous Multi-wave, Density}
\end{figure}

\subsection{Riemann Problems}
\label{sec:riemann}

\subsubsection{Sod Shock Tube: Cartesian Coordinates}
\label{sec:riemannSod}

This test is based on the classic Riemann problem from Sod \citep{Sod:1978}.  
It involves an initially stationary fluid with a discontinuity separating two states -- left and right -- with high pressure and density on the left and low pressure and density on the right. 
This initial state evolves into a shock propagating into the low density region, followed by a contact discontinuity, and a rarefaction wave propagating back into the high density state.  
Shock tube problems such as this stress a method's ability to capture shocks and contact discontinuities without smearing or introducing unphysical oscillations.  
Given the importance of shocks in CCSNe, this serves as a critical first test for any method designed to model these explosions.

Here, the problem is modified to use physical units in a regime realizable in simulations of CCSNe.  
The computational domain is $D = \left[ -5, 5\right]$~km with the discontinuity initially at $x=0$~km, separating the left and right states
\begin{align}
  \mathbf{P}_{\rm{L}} = (\,\rho,\,v,\,p,\,\ye\,)_{\rm{L}}^{\rm{T}}&= \big(\,10^{12}~\text{g~cm}^{-3}, 0\,, 10^{32}~\text{erg~cm}^{-3}, 0.4\,\big)^{\rm{T}}\,\,\, \label{eq:sodLeft} \\
  \mathbf{P}_{\rm{R}} = (\,\rho,\,v,\,p,\,\ye\,)_{\rm{R}}^{\rm{T}}&= \big(\,1.25\times10^{11}~\text{g~cm}^{-3}, 0\, , 10^{31}~\text{erg~cm}^{-3}, 0.3\,\big)^{\rm{T}}. \label{eq:sodRight}
\end{align}
(Note that the initial $\ye$ profile is also discontinuous.)

\begin{figure}[h]
  \centering
  \includegraphics[width = 0.75\textwidth]{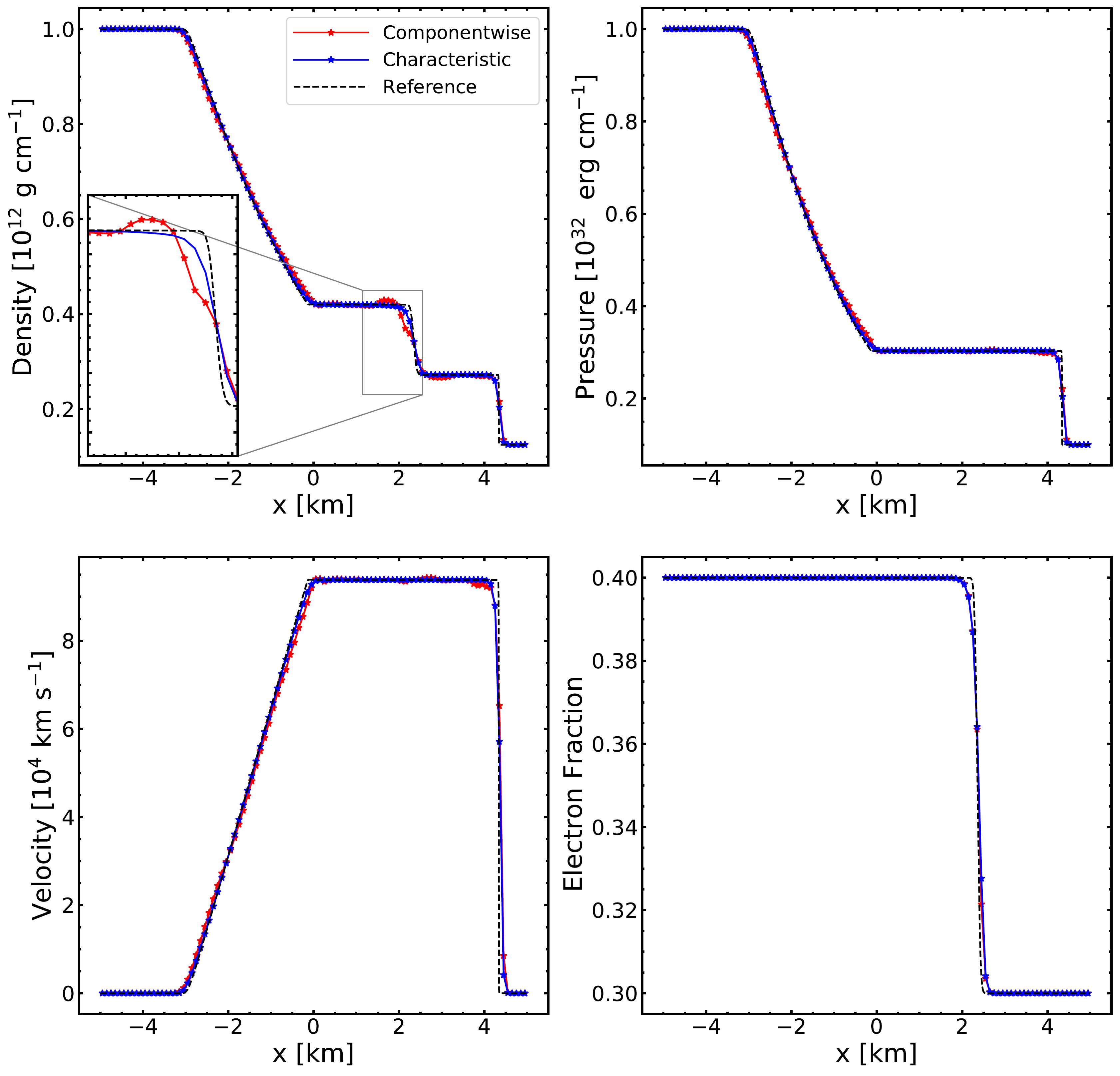}
  \caption{Numerical solution of the Sod shock tube at $t = 0.021$~ms using 100 elements and third-order accurate methods with characteristic (blue) and component-wise limiting (red) for density (upper left), pressure (upper right), velocity (lower left), and electron fraction (lower right), compared to a reference solution (black) using 10000 elements, obtained with first-order spatial discretization and third-order time integration.}
  \label{fig:sod}
\end{figure}

The problem is evolved until $t = 0.021$~ms, using 100 uniform elements with $\beta_{\TVD}=1.75$, and no troubled-cell indicator ($C_{\TCI}=0$), so that limiting is applied everywhere. 
We use third-order spatial discretization ($k=2$) and third-order temporal integration (SSP-RK3).  
A main focus with this test is to compare results obtained with component-wise and characteristic limiting (discussed in Section~\ref{sec:slope}).  
Figure~\ref{fig:sod} shows results for mass density (upper left), pressure (upper right), velocity (lower left), and electron fraction (lower right), using both characteristic (blue) and component-wise limiting (red), compared to a reference solution (black) computed using the first-order accurate spatial method ($k=0$), third-order time integration, and 10000 elements.  
We note that both limiting schemes capture the general nature of the solution, including the rarefaction wave, which extends from about $-3$ to $0$~km, the contact discontinuity, which is located at about $2$~km, and the shock, located at about $4$~km.  
The scheme based on characteristic limiting, however, is better at suppressing oscillations, and is less dissipative across the contact discontinuity.  
These observations are consistent with those made by \citet{schaal:2015a} in the ideal EoS case.  

\subsubsection{Sod Shock Tube: Spherical-Polar and Cylindrical Coordinates}

As a test of \thornado's ability to work with non-Cartesian coordinate systems, we also solve a spherically symmetric version of the Sod shock tube problem in 1D spherical-polar and 2D cylindrical coordinates.  
For spherical-polar coordinates, the domain is $D = \left[ 0, 10 \right]$~km, with the initial discontinuity placed at $r = 5$~km, while, for cylindrical coordinates, our domain is $D = \left[ 0, 10 \right]$~km $\times \left[ -10, 10 \right]$~km, and the discontinuity is placed at $r = \sqrt{R^{2} + z^{2}} = 5$~km.  
For the initial left and right states, we use those given in the 1D Cartesian Sod test in Equations~\eqref{eq:sodLeft}-\eqref{eq:sodRight}, with the exception that the electron fraction is given a constant value of $\ye = 0.4$ across the entire domain.  
We evolve both tests until $t = 0.025$~ms using 100 elements in the spherical case and 100 $\times$ 200 elements in the cylindrical case.  
Both tests use the third-order methods ($k=2$ and SSP-RK3), characteristic limiting with $\beta_{\TVD}=1.75$, and no troubled cell indicator ($C_{\TCI}=0$).  
We note that for the 2D test with cylindrical coordinates, we used \thornado's interface to AMReX to take advantage of AMReX's MPI infrastructure.  

\begin{figure}[h]
  \centering
  \includegraphics[width=0.95\textwidth]{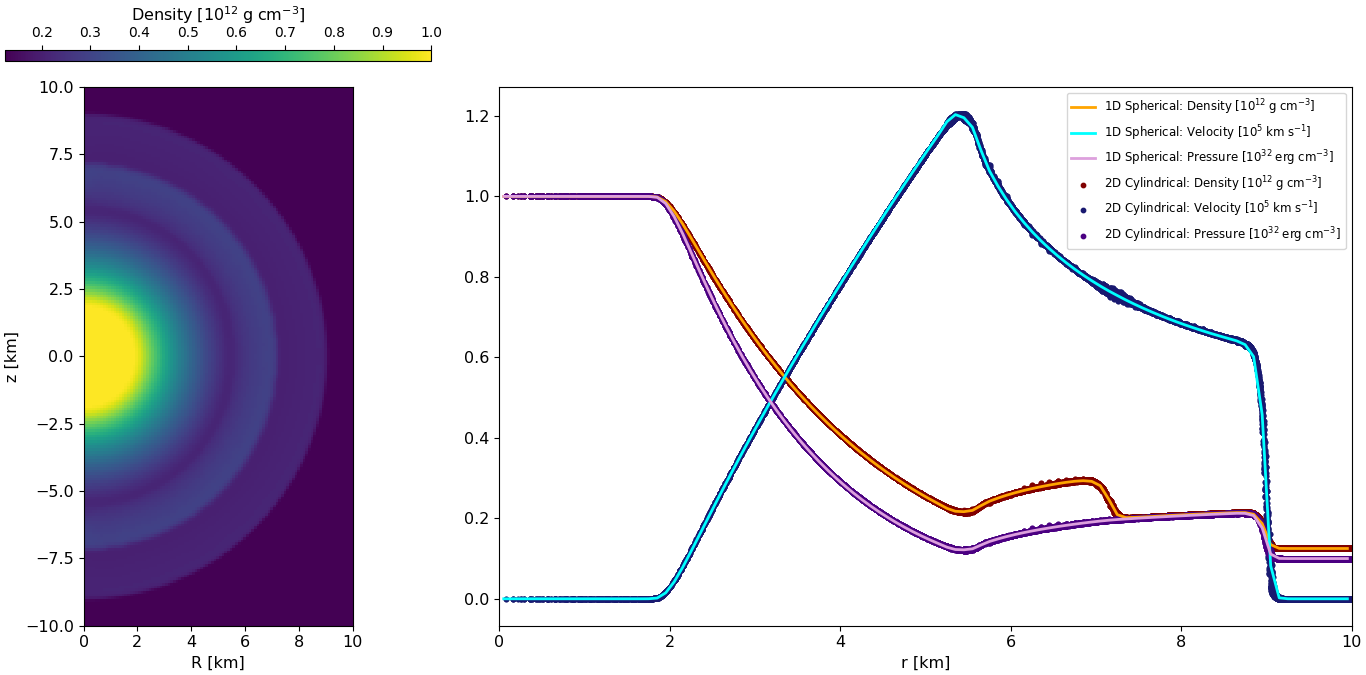}
  \caption{Two-dimensional density distribution (left panel), along with radial density, velocity, and pressure profiles (right panel) for the spherically symmetric Sod shock tube problem evolved to $t = 0.025$~ms using both 1D spherical-polar and 2D cylindrical coordinates; solid lines and scatter plots, respectively.}
  \label{fig:curvilinearsod}
\end{figure}

Results are shown in Figure~\ref{fig:curvilinearsod}.  
In the left panel of Figure~\ref{fig:curvilinearsod}, we show the 2D density distribution for the cylindrical test.  
In the right panel of Figure~\ref{fig:curvilinearsod}, we show the density, velocity, and pressure profiles of the spherical-polar test (solid lines), along with scatter plots of the corresponding quantities from the cylindrical test versus spherical-polar radius $r$ for comparison.  
We note that the characteristics of the solution profiles are similar to those obtained by others using an ideal EoS \citep[e.g.,][]{omang:2006}.  
There is also good agreement between the results obtained with spherical-polar and cylindrical coordinates.  
As in the Cartesian test, we note the clear resolution of the shock and contact discontinuity with no discernible oscillations.  
Furthermore, we note some spread in the scatter plots from the cylindrical solution, most notably in the velocity profile across the contact discontinuity.  
However, despite the truly multidimensional setup in the cylindrical case, there is decent preservation of the spherical symmetry inherit in the test.  

\subsubsection{Shock Tube Provoking the Bound-Enforcing Limiter}

This test, performed in 1D with Cartesian coordinates, is similar to the Sod shock tube discussed in Section~\ref{sec:riemannSod}, but with initial conditions tuned to provoke the bound-enforcing limiter developed in Section~\ref{sec:boundEnforcing}.  
The goal is to demonstrate that the limiter keeps the solution within the set of admissible states (specifically that $\epsilon\ge\epsilon_{\min}$) while also conserving the total mass, energy, and electron number in time, given, respectively, by
\begin{equation}
  \int_{D}\big\{\,\rho_{h}(x,t),\,E_{h}(x,t),\,D_{{\rm{e}},h}(x,t)\,\big\}\,dx.  
\end{equation}
The computational domain is $D = \left[-5, \, 5\right]$~km, and a discontinuity is placed at $x=0$~km, which separates the left and right states of the Riemann problem
\begin{align*}
    \mathbf{P}_{\rm{L}} = (\rho,v,p,\ye)_{\rm{L}}^{\rm{T}} &= 
    \left(1.00 \times 10^{13} \, \mathrm{g} \, \mathrm{cm}^{-3}, \, 0 \, \mathrm{m} \, \mathrm{s}^{-1}, \, 1.070 \times 10^{31} \, \mathrm{erg} \, \mathrm{cm}^{-3}, \, 0.04 \right)^{T} \\
    \mathbf{P}_{\rm{R}} = (\rho,v,p,\ye)_{\rm{R}}^{\rm{T}} &= 
    \left( 1.25 \times 10^{12} \, \mathrm{g} \,  \mathrm{cm}^{-3}, \, 0 \, \mathrm{m} \, \mathrm{s}^{-1}, \, 1.023 \times 10^{30} \, \mathrm{erg} \, \mathrm{cm}^{-3}, \, 0.10 \right)^{\rm{T}}.
\end{align*}
The numerical solution is evolved to $t=0.2$~ms, using 256 elements with polynomial degree $k=2$ and SSP-RK3 time integration.  
To fully test the bound-enforcing limiter, we run this test without the slope limiter discussed in Section~\ref{sec:slope}.
Moreover, it is possible to design an initial state for the Sod shock tube problem that does not place the solution close to or below the minimum table boundary. 
An example of this is seen in section \ref{sec:riemannSod}, where the bound enforcing limiter is not required to keep the solution within the set of admissible states.
However, we note that this particular test (using the initial condition described immediately above) fails without the bound enforcing limiter, regardless of whether or not the slope limiter is implemented.\footnote{Even when slope limiting is used, the bound enforcing limiter is required for this test, but we decide to deactivate the slope limiter to provoke the bound enforcing limiter even more.}
Thus, the bound enforcing limiter allows for a wider selection of initial states that would otherwise cause the algorithm to fail. 

\begin{figure}[h]
  \centering
  \includegraphics[width = 0.31\textwidth]{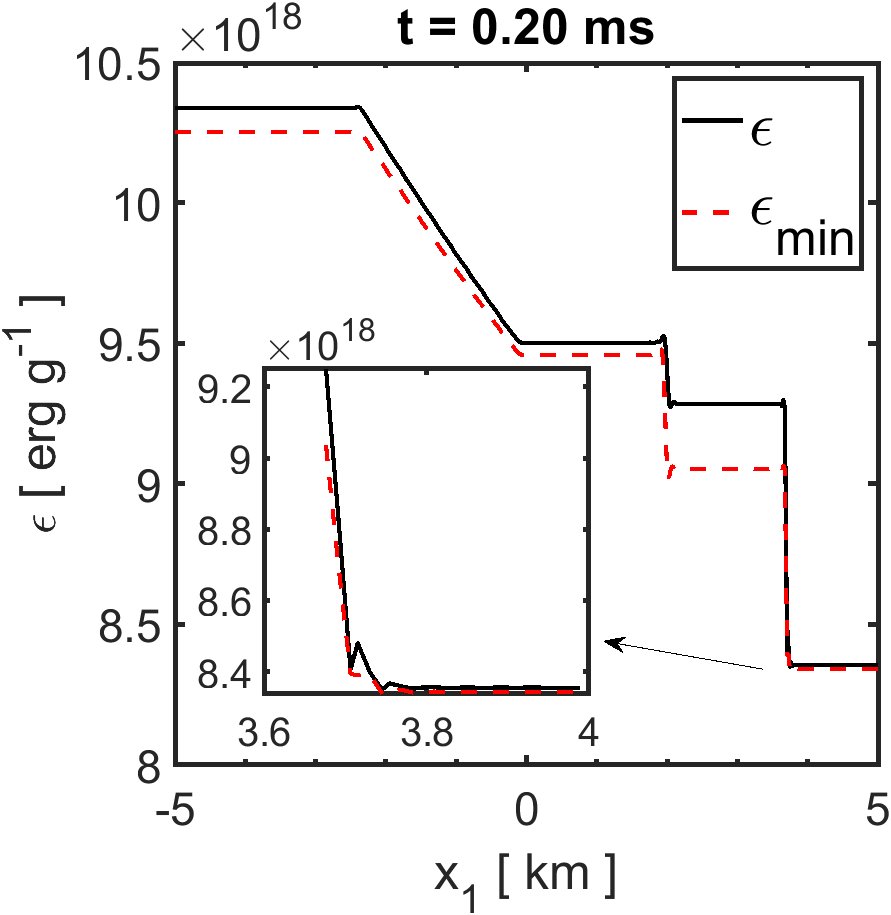}
  \hfill
  \centering
  \includegraphics[width = 0.31\textwidth]{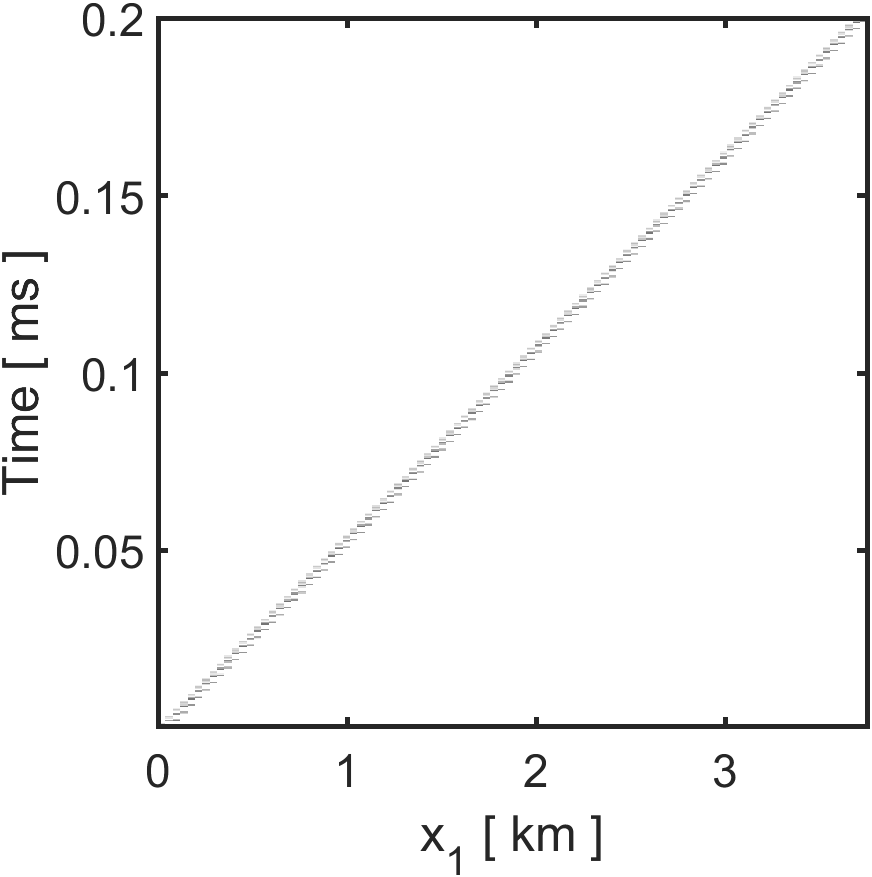}
  \hfill
  \centering
  \includegraphics[width = 0.31\textwidth]{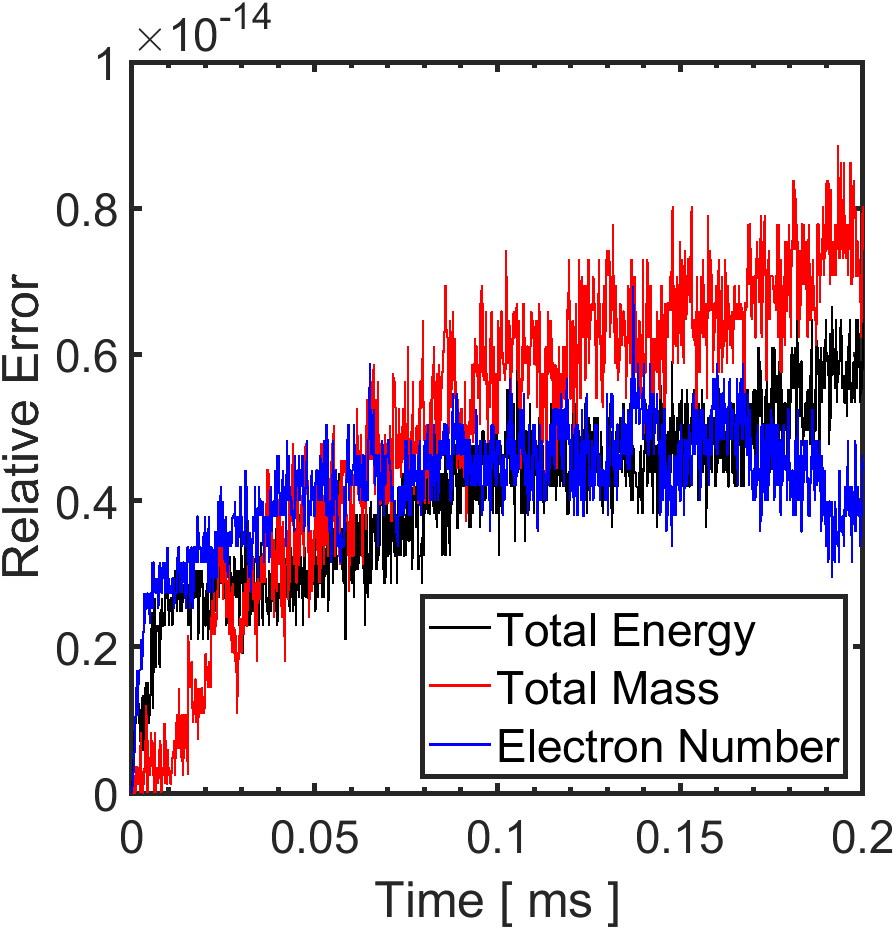}
  \caption{Numerical results for the shock tube provoking the bound-enforcing limiter. In the left panel we plot the specific internal energy (solid black line) and the minimum specific internal energy (dashed red line) versus position $x$.  The middle panel shows the activation sites of the bound-enforcing limiter as indicated by a space-time plot of $\vartheta_{3}(x,t)$.  The solution is closest to the boundary just ahead of the shock, as indicated by the inset in the left panel.  In the right panel we plot the relative change in the conserved quantities; i.e., total fluid energy (black), mass (red), and electron number (blue).}
  \label{fig:Pochik Shock Tube}
\end{figure}

Numerical results from this test are displayed in Figure~\ref{fig:Pochik Shock Tube}.  
In the left panel, we plot the specific internal energy versus position at the end of the simulation (solid black curve).  
We also plot the minimum internal energy $\epsilon_{\min}(\rho,\ye)$ (dashed red curve).  
Around the shock, $\epsilon$ is very close to the minimum value, as can be seen in the inset in left panel of Figure~\ref{fig:Pochik Shock Tube}.  
In fact, the specific internal energy remains very close to the minimum value throughout this test.  
The middle panel displays a space-time plot of the limiter parameter $\vartheta_{3}(x,t)$ in Equation~\eqref{eq:limitedPolynomialStep3}, and shows the activation sites for the bound-enforcing limiter, where the average value for $\vartheta_{3}$ when limiting is required is $0.9$ and it ranges from $0.60 < \vartheta_{3} < 0.99$. 
The bound-enforcing limiter is activated due to small oscillations slightly ahead of the shock, and produces a trace of the shock trajectory as seen in the middle panel in Figure~\ref{fig:Pochik Shock Tube}.
The slope of the prominent trace in $\vartheta_{3}(x,t)$ indicates a shock velocity of $v_{\mathrm{shock}} \approx 1800$~km~s$^{-1}$.  
Finally, the right panel in Figure~\ref{fig:Pochik Shock Tube} shows the relative change in the conserved quantities versus time.  
The change in these quantities are due to machine roundoff, indicating that the bound-enforcing limiter is sufficiently conservative for this test.  

\subsubsection{Shu-Osher Shock Tube}
\label{sec:riemannShuOsher}

This test adopted from \citet{shu:1989} involves a Mach=3 shock interacting with a lower density region with a sinusoidal perturbation.  
As the shock propagates and interacts with the density perturbations, the perturbations move upstream, forming high frequency oscillations just behind the shock.  
This problem tests the ability of a shock-capturing method to limit unphysical oscillations without destroying physical, small-scale features of the post-shock flow. 
We note that small-scale features resulting from hydrodynamical instabilities, such as turbulence and convection, are crucial to CCSN explosion dynamics \citep[e.g.,][]{murphy:2011, murphy:2013, couch:2015a, radice:2016, mabanta:2018, couch:2020} and many other astrophysical applications.  

Here, the problem is modified to use physical units in a regime relevant to CCSNe.  
The computational domain is $D = \left[ -5, 5\right]$~km, with a discontinuity initially located at $x=1$~km separating the left and right states
\begin{align*}
  \mathbf{P}_{\rm{L}} =(\,\rho,\,v,\,p,\,\ye\,)_{\rm{L}}^{\rm{T}} &= \big(\,3.60632\times10^{12}~\text{g~cm}^{-3},\, 7.425\times10^{4}~\text{km}~\text{s}^{-1},\,1.333\times10^{32}~\text{erg~cm}^{-3},\, 0.5\,\big)^{\rm{T}}\,\,\, \\
  \mathbf{P}_{\rm{R}} =(\,\rho,\,v,\,p,\,\ye\,)_{\rm{R}}^{\rm{T}} &= \big(\,\big[1 + 0.2\times\sin(5~\text{km}^{-1}~x)\big]\times10^{12}~\text{g~cm}^{-3},\, 0,\, 1.0\times10^{31}~\text{erg~cm}^{-3},\, 0.5\,\big)^{\rm{T}}.
\end{align*}
The fluid is evolved until $t = 0.0625$~ms, using 256 uniform elements and $\beta_{\TVD}=2.0$.  
We use third-order spatial discretization ($k=2$) and third-order temporal integration (SSP-RK3).  
In this test we compare results obtained with characteristic and component-wise limiting, and, for each limiting method, we show results for various values of the TCI threshold.  

\begin{figure}[h]
  \centering
  \includegraphics[width = 0.85\textwidth]{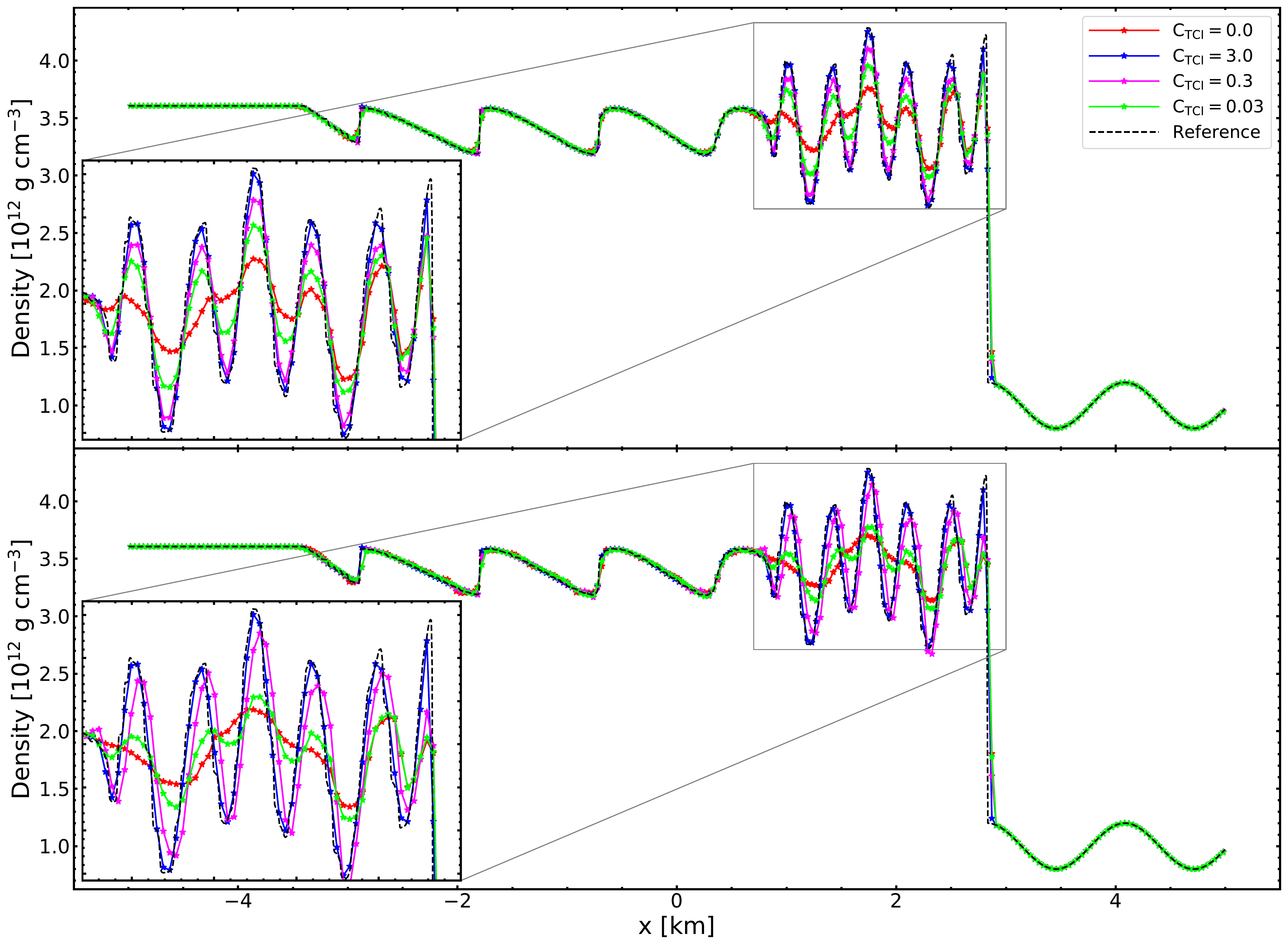}
  \caption{Numerical solution of the Shu-Osher shock tube with nuclear EoS at $t = 0.062$~ms, using 256 elements and third-order accurate methods with characteristic (top) and component-wise (bottom) limiting.  In each panel, we plot the mass density versus position, obtained with various values of the troubled cell indicator threshold $C_{\TCI}$: 0.0 (full limiting, red), 0.03 (green), 0.3 (magenta), and 3.0 (blue), compared to a reference solution (black) obtained using 2048 elements, third-order spatial discretization, and third-order time integration.}
  \label{fig:shu-osher}
\end{figure}

In Figure~\ref{fig:shu-osher}, we show the density obtained using characteristic (top) and component-wise (bottom) limiting for various values of the troubled-cell indicator threshold $C_{\TCI}$: 0.0 (full limiting, red), 0.03 (green), 0.3 (magenta), and 3.0 (blue); i.e., the same values that were used in \citet{endeve:2019} for the ideal EoS case.  
Larger values of $C_{\TCI}$ imply less slope limiting.  
These results are compared to a reference solution obtained using 2048 elements (black), with third-order spatial and temporal discretization, and $C_{\TCI}=0.0$.  
In both limiting schemes, full limiting washes out the density variations behind the shock, while increasing the TCI threshold allows for these features to be better captured. 
The results obtained with $C_{\TCI}=3.0$ are very close to the reference solution.  
However, for reasons discussed in Section~\ref{sec:collapseTCI}, we do not recommend using such a high value for $C_{\TCI}$ in general, since some amount of limiting --- even in smooth regions --- seems to be required.  
For all values of the threshold (except perhaps the case with $C_{\TCI}=3.0$, which applies little limiting away from the shock), the characteristic limiting scheme better captures the shape and amplitude of the oscillations behind the shock (see insets in each panel, focusing on the higher frequency oscillations just behind the shock).

\subsubsection{Two-Dimensional Riemann Problem}

Here we consider a two-dimensional Riemann problem, adapted from \citet{lax:1998}, which involves a fluid with a different initial state in each quadrant given by
\begin{alignat*}{2}
  \mathbf{P_{\mathrm{NW}}}
  &=\big(\,\rho,\,v^{1},\,v^{2},\,p,\,\ye\,\big)_{\mathrm{NW}}^{\rm{T}}
  &&=\big(\, 10^{12}~\text{g~cm}^{-3},\, 7.275\times10^{4}~\text{km~s}^{-1},\, 0,\, 10^{32}~\text{erg~cm}^{-3},\, 0.3 \,\big)^{\rm{T}}, \\
  \mathbf{P_{\mathrm{NE}}}
  &=\big(\,\rho,\,v^{1},\,v^{2},\,p,\,\ye\,\big)_{\mathrm{NE}}^{\rm{T}}
  &&=\big(\, 5.313\times10^{11}~\text{g~cm}^{-3},\, 0,\, 0,\, 4.0\times10^{31}~\text{erg~cm}^{-3},\, 0.3 \,\big)^{\rm{T}}, \\
  \mathbf{P_{\mathrm{SE}}}
  &=\big(\,\rho,\,v^{1},\,v^{2},\,p,\,\ye\,\big)_{\mathrm{SE}}^{\rm{T}}
  &&=\big(\, 10^{12}~\text{g~cm}^{-3},\, 0,\, 7.275\times10^{4}~\text{km~s}^{-1},\, 10^{32}~\text{erg~cm}^{-3},\, 0.3 \,\big)^{\rm{T}}, \\
  \mathbf{P_{\mathrm{SW}}}
  &=\big(\,\rho,\,v^{1},\,v^{2},\,p,\,\ye\,\big)_{\mathrm{SW}}^{\rm{T}}
  &&=\big(\, 8.0\times10^{11}~\text{g~cm}^{-3},\, 0,\, 0,\, 10^{32}~\text{erg~cm}^{-3},\, 0.3 \,\big)^{\rm{T}},
\end{alignat*}
on a domain $D = \left[0,1.0 \right]$~km $\times \left[0,1.0 \right]$~km.
This test, which corresponds to ``Configuration~12'' in \citet{lax:1998}, involves two shocks moving into the northeastern quadrant and contact discontinuities (or slip lines) at the northern and eastern boundaries of the southwestern quadrant.  
It is adapted from the original works to use physical units in a regime relevant to CCSNe with a nuclear EoS.  
The initial configuration presented here is one of many possible configurations of 2D Riemann problems presented in \citet{lax:1998}.  
The fluid is evolved until $t = 0.0025$~ms using 400$^2$ uniform elements, $\beta_{\TVD}=1.75$, and $C_{\TCI}=0$ (i.e., limiting is applied everywhere).  
We use third-order spatial discretization ($k=2$) and third-order temporal integration (SSP-RK3).  
To run this test, we used \thornado's interface to AMReX in order to take advantage of AMReX's MPI parallelization. 

\begin{figure}[h]
  \label{fig:2d_riemann}
  \centering
  \includegraphics[width=1.0\textwidth]{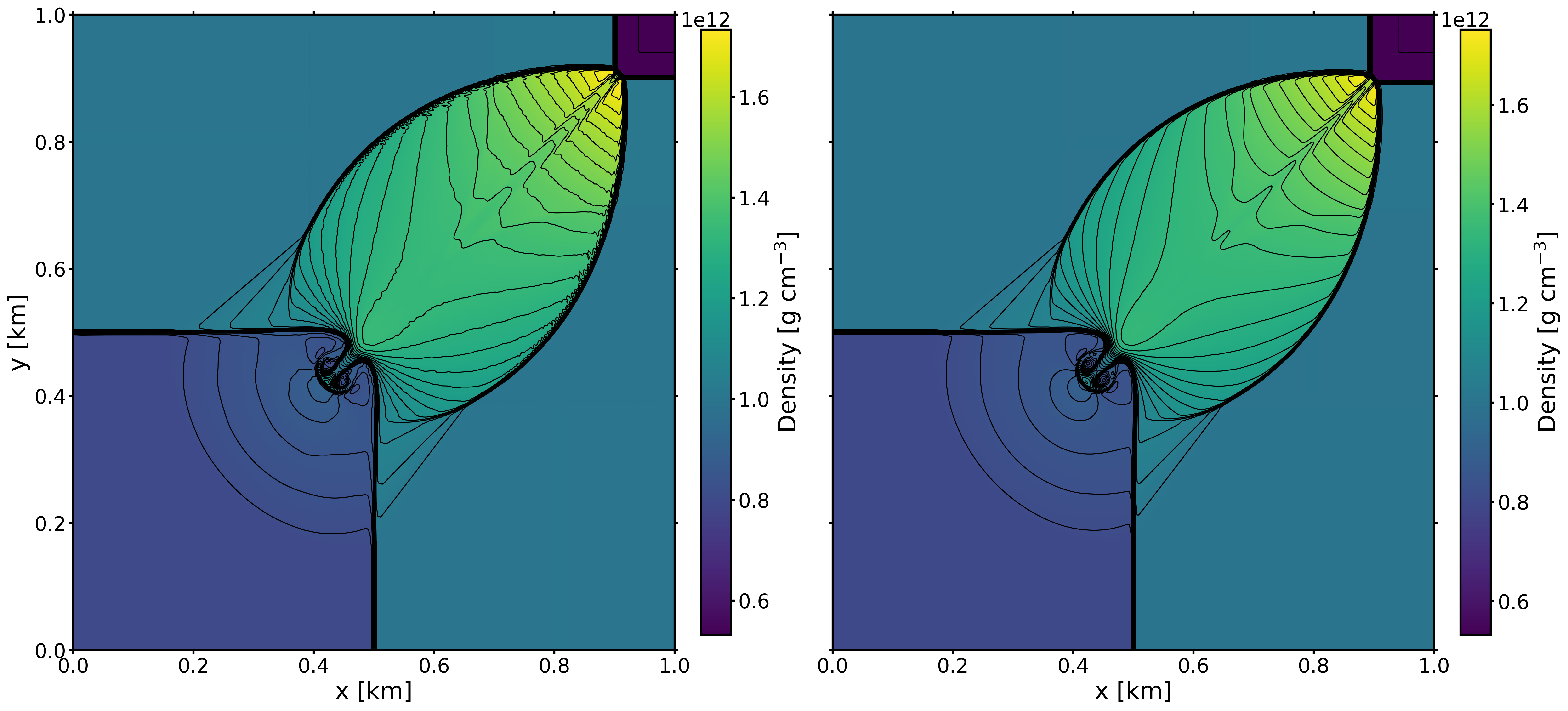}\hfill
  \includegraphics[width=1.0\textwidth]{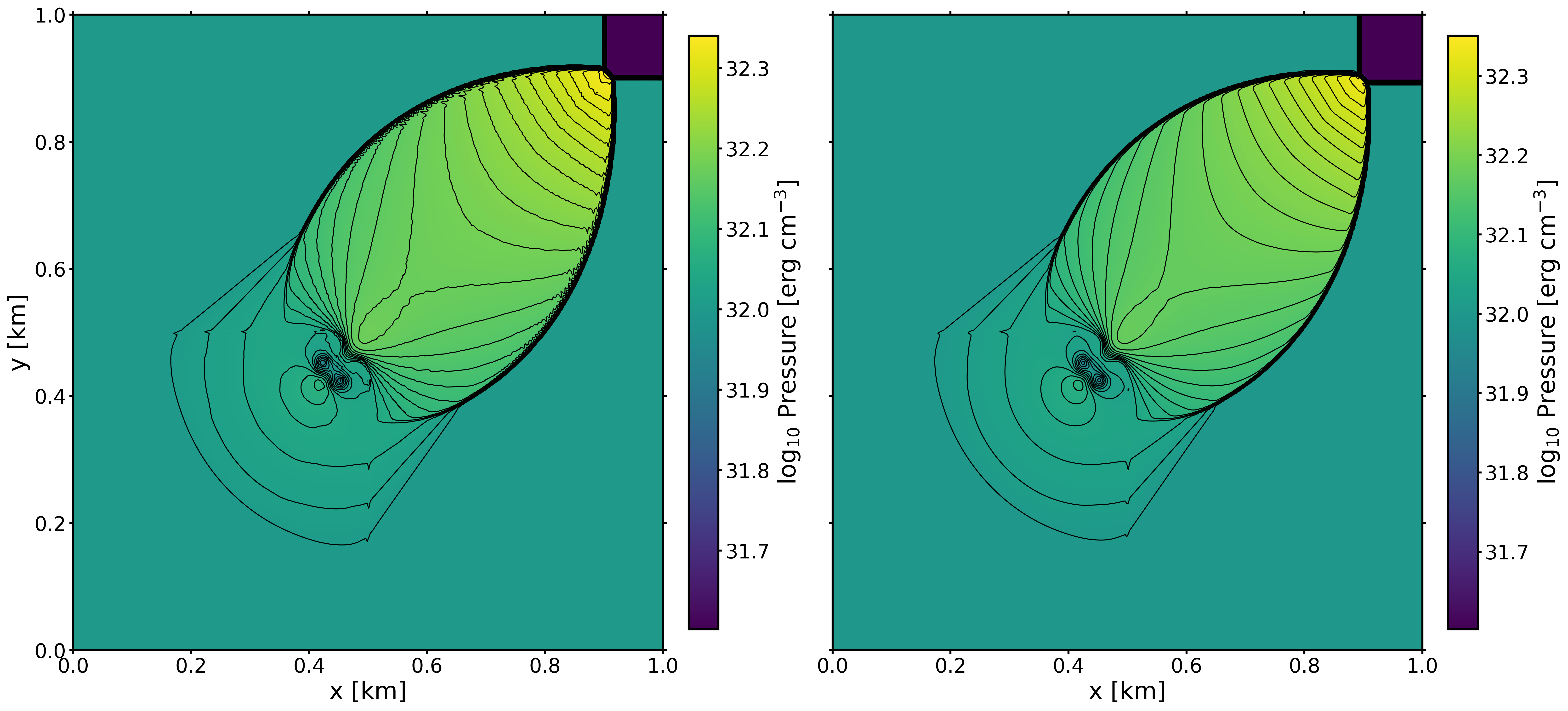}\hfill
  \caption{Numerical solution of a 2D Riemann problem (adopted from ``Configuration~12'' of \citet{lax:1998}) with a nuclear EoS at $t = 0.0025$~ms using 400$^2$ elements and third order spatial and temporal discretization for density (top panels) and pressure (bottom panels).
  We compare results obtained with component-wise (left panels) and characteristic (right panels) limiting.  
  Black lines on each plot show logarithmically spaced contours to highlight structures in the solutions.}
\end{figure}

Figure~\ref{fig:2d_riemann} shows the density (top panels) and pressure (bottom panels) at $t = 0.0025$~ms, from a run with component-wise limiting (left panels) and a run with characteristic limiting (right panels).  
Black lines on each plot show logarithmically spaced contours to highlight solution features.  
Overall, the morphology of the solutions obtained with \thornado\ --- using a nuclear EoS --- agree well with the results displayed by \citet{lax:1998}.  
Moreover, the use of characteristic limiting presents a tremendous improvement over component-wise limiting, particularly as the higher dimensionality of the problem allows for more complex flow patterns and discontinuity geometries.  
Notably, the density and pressure contours in the component-wise limiting case reveal more oscillations and deformities.  
These oscillations are particularly prominent near the boundary of the curved shock surface.  
There appears to be no oscillations present in the run performed with characteristic limiting.  
Similarly, the jet-like feature seen in the southwest quadrant of the density plots appear less resolved and are somewhat asymmetric in the component-wise limiting case.  

\subsection{Poisson Solver Test}
\label{sec:gravpotential}

The accuracy of the finite element method used by Poseidon to solve Equation~\eqref{eq:poissonEquation} is determined by the total number of degrees freedom used to solve the system.  
The number of degrees of freedom can be changed by either the $p$-method or the $h$-method.  
The $p$-method varies the degree $k$ of the polynomials used in the approximation of the solution and requires $k+1$ nodes per element.  
The $h$-method increases the number of elements $N_{e}$ used to discretize the system.  
These two methods are used together in the $hp$-method where both the refinement of the mesh and the degree of the approximation polynomials can be varied.  
In the $hp$-method, the number of degrees of freedom is given by $n_{\mathrm{DOF}} = (k+1) \times N_{e}$.  
The accuracy of the $hp$-method increases with increasing $n_{\mathrm{DOF}}$, and the error should decrease with increasing $n_{\mathrm{DOF}}$ as $1/n_{\mathrm{DOF}}^{k+1}$. 

We test the accuracy of Poseidon's Poisson solver using the density profile of a centrally condensed sphere of radius $R$.  
This test, from \cite{StoneNorman:1992a}, was chosen because it has a non-polynomial analytic solution, thus allowing us to better explore the convergence properties of the solver.  
(Problems with polynomial solutions are solved exactly for sufficiently high $k$.)  
The density profile and analytic solution for the test are given by
\beq
\rho(r) = 
	\begin{cases}
		\frac{\rho_{\rm{c}}}{1 + \left(\frac{r}{r_{\rm{c}}}\right)^2}		&	\text{if} \hspace{5pt} r \le R 	\\
		0										&	\text{if} \hspace{5pt} r > R
	\end{cases}	\label{eq:Poisson_DensityProfile}
\eeq
and
\beq
\Phi(r) = 
	\begin{cases}
		-4 \pi G \rho_{\rm{c}}r_{\rm{c}}^{2}\left[ 1 - \frac{ \arctan\left({\frac{r}{r_{\rm{c}}}}\right)}{\frac{r}{r_{\rm{c}}}} - \frac{1}{2}\left(\frac{1+ \left(\frac{r}{r_{\rm{c}}}\right)^2}{1 + \left(\frac{R}{r_{\rm{c}}}\right)^2} \right)  \right]	 & \text{if} \hspace{5pt} r \le R 	\\
		-4 \pi G \rho_{\rm{c}} \frac{r_{\rm{c}}^{3}}{r} \left[ \frac{R}{r_{\rm{c}}} - \arctan{\left( \frac{R}{r_{\rm{c}}} \right)} \right]	&	\text{if} \hspace{5pt} r > R,
	\end{cases}
\eeq
respectively, where $\rho_{\rm{c}}$ and $r_{\rm{c}}$ are the central density and core radius, respectively.  
For this test, we choose $\rho_{\rm{c}} = 150 \, \mathrm{g}~\mathrm{cm}^{-3}$, $r_{\rm{c}} = 0.2~R_{\odot}$, and $R = R_{\odot}$, and perform the calculations over the 1D computational domain $D = [0, 2~R_{\odot}]$.  
We compute the $L_{1}$ and $L_{\infty}$ error norms as 
\beq
	L_{1} \equiv \sum_{j=1}^{n_{\mathrm{DOF}}} | \Phi(r_{j}) - \Phi_{h}(r_{j}) |,
\eeq
and
\beq
	L_{\infty} \equiv \max_{j} \left( | \Phi(r_{j}) - \Phi_{h}(r_{j}) | \right)\, \quad\mathrm{for}\quad j \in \{1,\ldots,n_{\mathrm{DOF}}\}.
\eeq

\begin{figure}[h]
  \centering
  \includegraphics[width=1.0\textwidth]{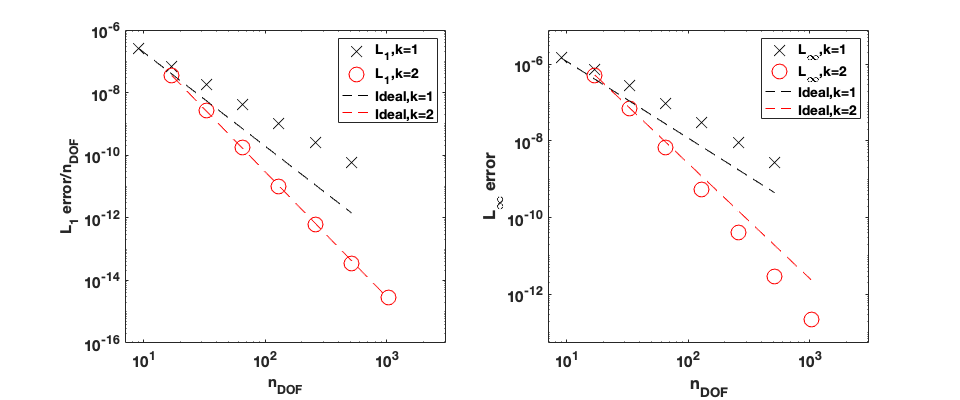}
  \caption{ $L_{1}$ (left panel) and $L_{\infty}$ (right panel) errors between the analytic and numerical solution calculated by the Poseidon solver for the case of a centrally condensed sphere.  The $L_{1}$ error norms are scaled by the number of degrees of freedom to obtain an average error per node.  The dashed lines are proportional to $1/n_{\mathrm{DOF}}^{k+1}$ and serve as references for the convergence rates of the numerical solutions.}
  \label{fig:Poisson_Results}
\end{figure}

In Figure~\ref{fig:Poisson_Results}, we plot the $L_{1}$ error norm (scaled by $n_{\mathrm{DOF}}$; left panel) and the $L_{\infty}$ error norm (right panel) versus $n_{\mathrm{DOF}}$.  
The numerical solutions were obtained using $k = 1$ (black symbols) and $k = 2$ (red symbols).  
For each value of the polynomial degree $k$, seven values of $N_{e}$ (8, 16, 32, 64, 128, 256, and 512) were used to create uniform grids.  
From these plots we see that for a specific value of $N_{e}$ the higher order method always provides a more accurate solution.  
The rate of convergence observed for the third-order method is as expected (or better) in both error norms (cf. red, dashed reference lines).  
The second-order method converges at a rate somewhat slower than expected when the error is measured in the $L_{1}$ error norm, but the $L_{\infty}$ error decreases roughly at the expected second-order rate (cf. black, dashed reference lines).  

\section{Adiabatic Collapse, Core-Bounce, and Shock Propagation}
\label{sec:collapse}

In this section we employ the DG method implemented in \thornado\ to evolve a non-rotating progenitor through adiabatic collapse, core bounce, and post-bounce shock propagation. 
The initial conditions are provided by a $15$~\Msun\ progenitor model from \citet{Woosley:2007}. 
Overall, this section will cover the chronological evolution of the stellar collapse model in three stages: (1) adiabatic collapse of the core, (2) core rebound and the formation of the shock shortly after nuclear saturation, and (3) the propagation of the shock through the outer core thereafter.  
In total, the evolution covers about 800~ms of physical time, which is divided into about 300~ms for collapse, and almost 500~ms of post-bounce evolution, until the bounce shock reaches the outer boundary.  

The following subsections will first discuss the physical conditions of the adiabatic collapse application that challenge any hydrodynamics method used for CCSN simulations.  
Then we focus on various features of the DG method in \thornado, such as (1) the performance of the bound-enforcing limiter during bounce and shock formation, (2) the response of the numerical solution to adjusting the troubled-cell indicator threshold parameter $C_{\mathrm{TCI}}$, (3) resolution dependence in the inner core, (4) the challenge of maintaining energy conservation when applying limiters, and (5) difficulties associated with employing characteristic limiting in the vicinity of the phase transition.  
Of course, being spherically symmetric and without neutrino transport, this adiabatic model does not describe a realistic evolutionary trajectory for a CCSN progenitor.  
However, this test does subject the numerical method to some of the physical conditions encountered, and we deem it a necessary step towards more realistic models.  

Using spherical-polar coordinates, the domain $D = [0, \, 8000]$~km is divided into $N=512$ elements.  
In the interest of capturing important physical characteristics while maintaining computational efficiency, this application implements a geometrically progressing grid that uses a finer spatial resolution in the inner core, which becomes progressively coarser according to
\begin{equation} \label{eq:geometricGridCellWidth}
	\Delta r_{i} = z\times\Delta r_{i-1},\quad i=2,\ldots,N,
\end{equation}
where $z>1$ is the `zoom factor'.  
This emphasizes the inner core, where most of the mass is concentrated after collapse, while deemphasizing the outer regions. 
To begin constructing the grid, the innermost cell width $\Delta r_{1}=\Delta r_{\min}$, the length $|D|$ of the spatial domain, and the number of elements $N$ are defined.  
Then, the zoom factor is obtained by solving
\begin{equation} \label{eq:zoomBisection}
    \eta \times \left( z^{N} - 1 \right) - \left( z - 1 \right) = 0,
\end{equation}
where $\eta = \Delta r_{\min} / |D|$.  
The fiducial run in this section uses an inner cell width of $\Delta r_{\min}=0.5$~km.  
Then, with $|D|=8000$~km and $N=512$, this results in a zoom factor (in double precision) of $z = 1.009967685243838$, and an outer cell width of $\Delta r_{N} = 79.45$~km.  
Also, for the fiducial run, we use second-order spatial ($k=1$) and temporal (SSP-RK2) discretization, combined with the component-wise limiting scheme discussed in Section~\ref{sec:slope}, $\beta_{\TVD}=1.75$, and $C_{\TCI}=0.0$.  
For all the runs, we use reflecting boundary conditions at the inner boundary and Dirichlet conditions (provided by the initial condition) at the outer boundary.  
The gravitational potential is obtained with a second-order accurate finite element method as discussed in section \ref{sec:poisson}.  


\subsection{Stage 1: Collapse}

Figure~\ref{fig:Gravitational Collapse, Collapse Phase} illustrates the collapse phase prior to core bounce.  
We scale such that bounce occurs at $t-t_{\rm{b}}=0$~ms with $t_{\rm{b}}=302.9$~ms for this model, which is defined as the time when the central density, $\rho_{\rm{c}}$, reaches its maximum.  
We plot the mass density (upper left panel), velocity (upper right panel), electron fraction (lower left panel), and entropy per baryon (lower right panel) versus radius for select times during collapse.  
We have chosen to display the collapse profiles at the times coinciding with each full decade in central density; i.e. $\rho_{\rm{c}}=10^{10,11,\ldots,14}$~g~cm$^{-3}$.  
The collapse dynamics is very similar to the self-similar solutions obtained by \citet{yahil:1983}, using a polytropic EoS.  
The central density increases with time and approaches nuclear densities ($10^{14}$~g~cm$^{-3}$) at $t-t_{\rm{b}}=-1$ ms while the outer region rarefies as indicated by the steeper slope in density outside the innermost core. 
Meanwhile, the infall velocity increases linearly with radius in the inner core (consistent with homologous collapse), and approaches free-fall beyond the maximum infall velocity, where it eventually falls off roughly as $r^{-1/2}$.   
The maximum infall velocity reaches $11-12 \%$ of the speed of light just before bounce. 
The electron fraction, $\ye$, is a monotonically increasing function of radius and its inner profile shifts inward --- in an approximately self-similar fashion --- with the decreasing core radius during collapse.  
Because this test models adiabatic flows (i.e., no neutrino physics is included), the electron fraction remains constant in the core.  
Before core bounce and shock formation the entropy profile shifts inward due to the collapsing core.  
In fact, both the electron fraction and entropy profiles remain constant in the core throughout collapse, bounce, and shock propagation, which we quantify further in Section~\ref{sec:collapseResolutionDependence}.

\begin{figure}[h]
    \centering
    \includegraphics[width=.50\textwidth]{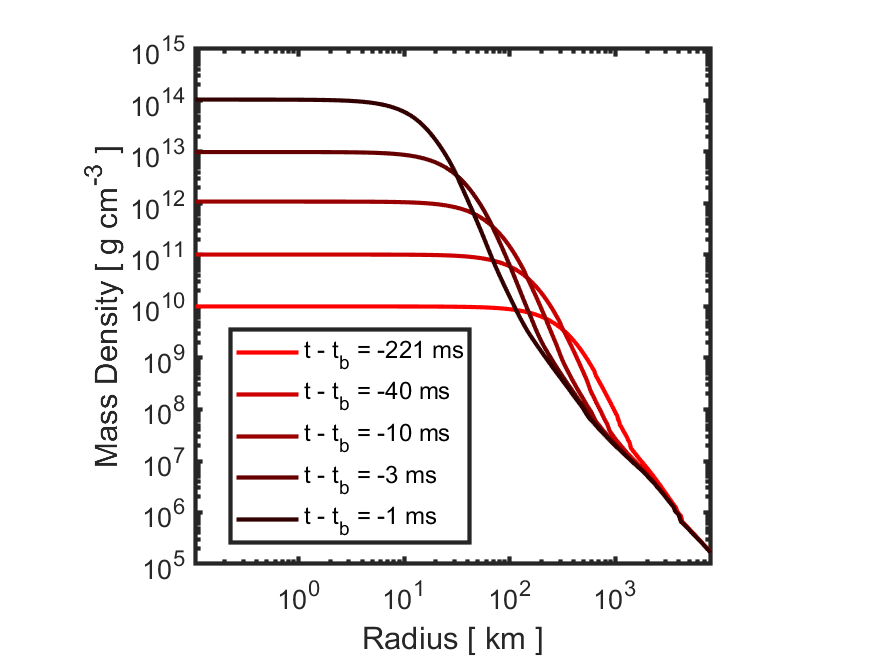}\hfill
    \includegraphics[width=.50\textwidth]{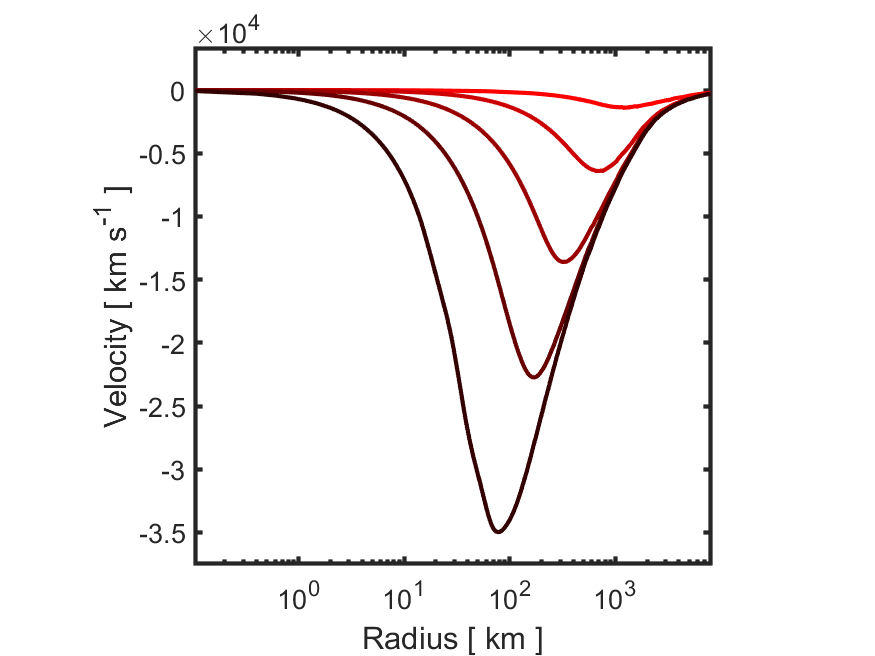}\hfill
    \includegraphics[width=.50\textwidth]{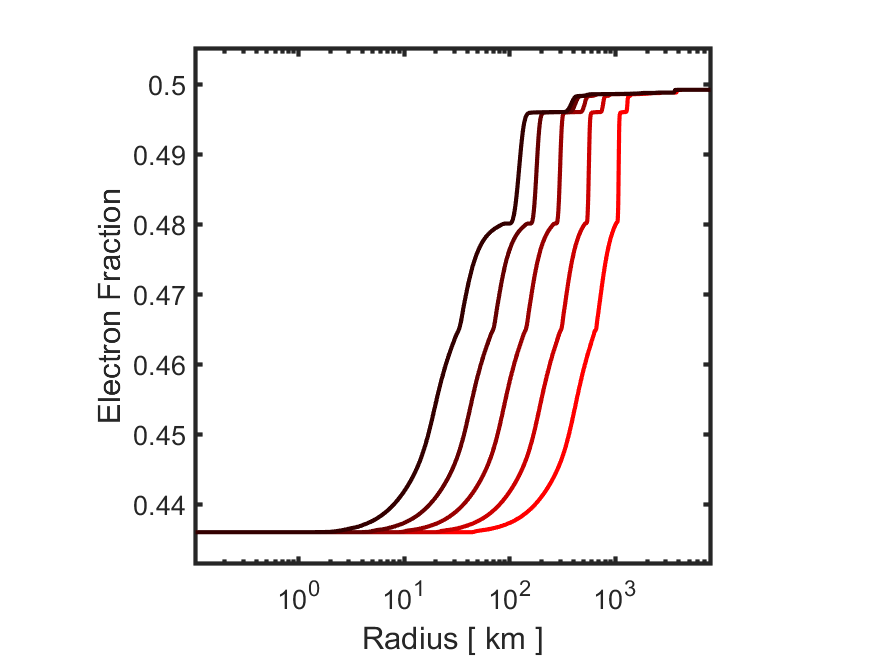}\hfill
    \includegraphics[width=.50\textwidth]{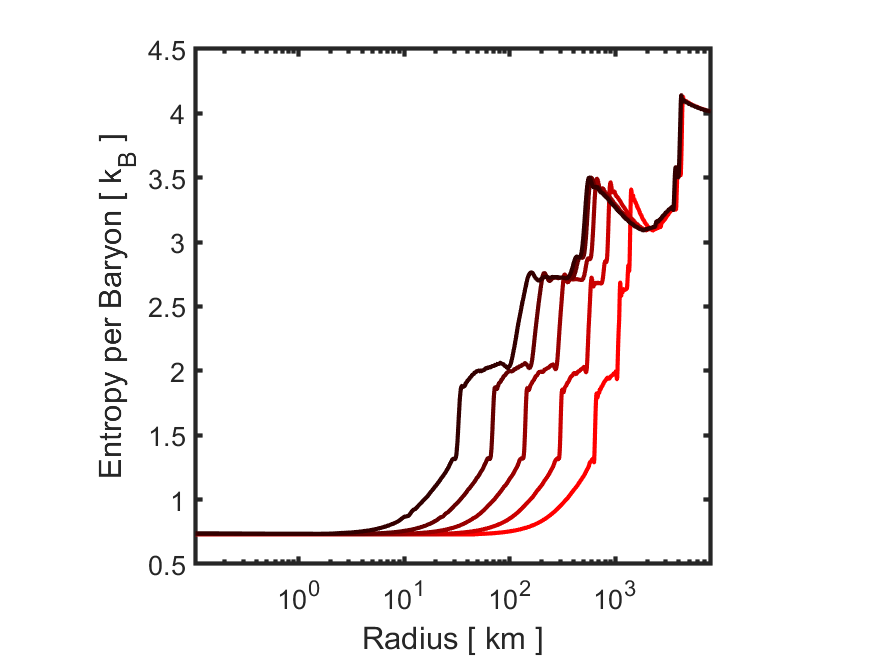}
    \caption{Numerical solutions for the adiabatic collapse of a $15~M_{\odot}$ progenitor from \citet{Woosley:2007}, obtained with \thornado\ using $512$ elements and a second-order DG scheme with component-wise limiting.  Plotted versus radius are mass density (upper left), velocity (upper right), electron fraction (lower left), and entropy per baryon (lower right) during collapse.  The time slices were chosen to depict the central mass density increasing by factors of 10.}
    \label{fig:Gravitational Collapse, Collapse Phase}
\end{figure}

\subsection{Stage 2: Core-Bounce}

Figure~\ref{fig:Gravitational Collapse, Bounce Phase} captures core-bounce and shock formation in the inner core ($r\in[0, \, 500]$~km).  
We plot the adiabatic index $\Gamma\equiv\big(\pderiv{\ln p}{\ln\rho}\big)$ (upper left), velocity (upper right), electron fraction (lower left), and entropy per baryon (lower right) versus radius.  
In each panel, blue curves illustrate the dynamics immediately before bounce (leading up to maximum $\rho_{\rm{c}}$), while red curves illustrate the dynamics immediately after bounce (see color maps to the right of each panel).  
The bounce dynamics is in response to the stiffening of the EoS, which is illustrated by the evolution of the adiabatic index during the transition to nuclear matter in the inner core.  
In the upper left panel, the adiabatic index is $ \Gamma \approx 4/3$ at $t-t_{\rm{b}}=-0.8$~ms.  
Once the core reaches nuclear densities and undergoes a phase transition to bulk nuclear matter, the EoS stiffens and the repulsive nuclear forces between the tightly packed nucleons results in a jump in $\Gamma$ to around $2.5$ at the inner boundary.  
After bounce, the inner core, $r\lesssim10$~km is characterized by $\Gamma\approx 2.5$, while $\Gamma\lesssim4/3$ at larger radii.  
Notice the sharp transition occurring around $r=10$~km, which we refer to as the \emph{phase transition}.  
The velocity profiles provide a clear demonstration of the genesis and evolution of the shock resulting immediately after bounce.  
When the EoS stiffens, collapse is halted, and a shockwave is formed in the region $r\in[10,20]$~km.  
Once formed, the shock must push through the supersonically collapsing outer core.  
In this adiabatic simulation, without neutrinos, the shockwave propagates relatively unencumbered through the outer core, and eventually reaches the outer boundary.  
The constant value in electron fraction in the very inner core is preserved through bounce and shock formation, meanwhile the profile in the outer region (around $10$ km) shifts as the shock travels through.  
There is no noticeable change in central entropy during bounce, but, as the shock forms, there is a large increase in the entropy across the shock, as expected.  

\begin{figure}[h]
    \centering
    \includegraphics[width=.50\textwidth]{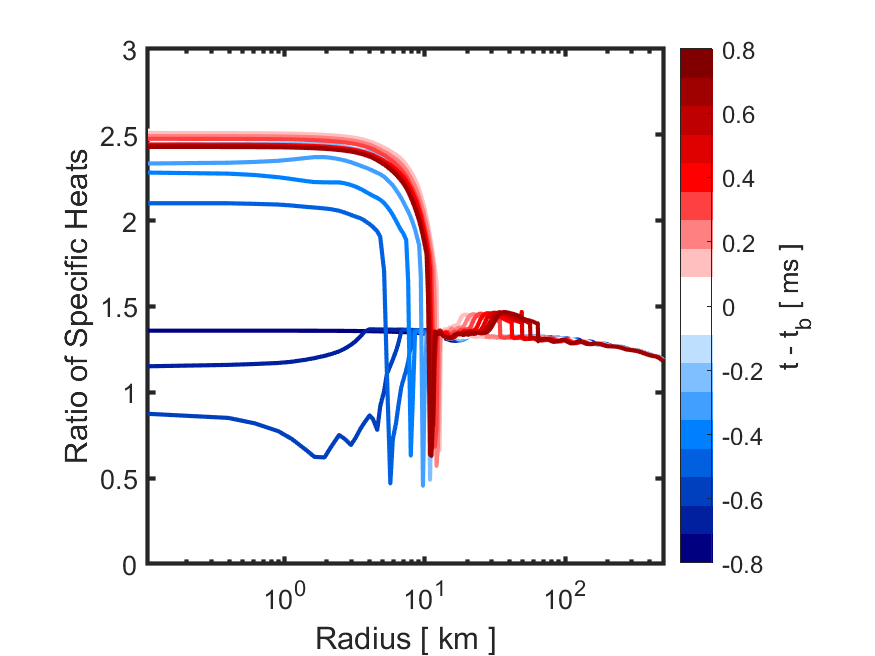}\hfill
    \includegraphics[width=.50\textwidth]{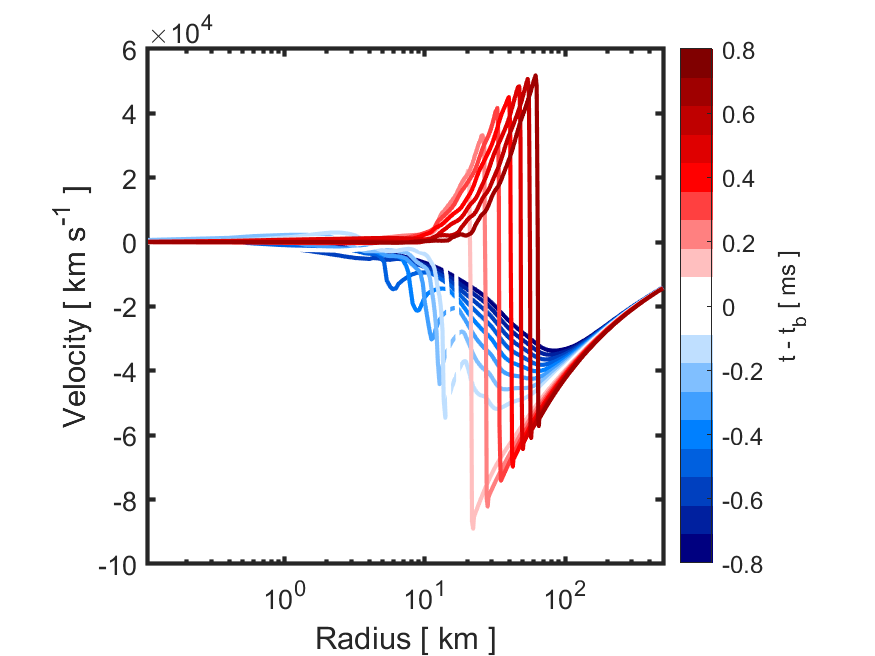}\hfill
    \includegraphics[width=.50\textwidth]{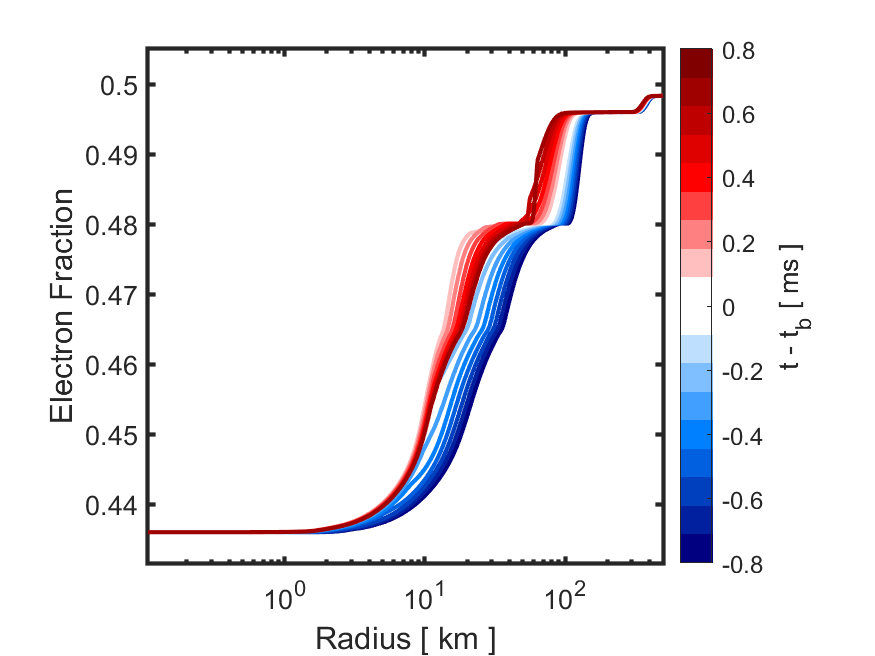}\hfill
    \includegraphics[width=.50\textwidth]{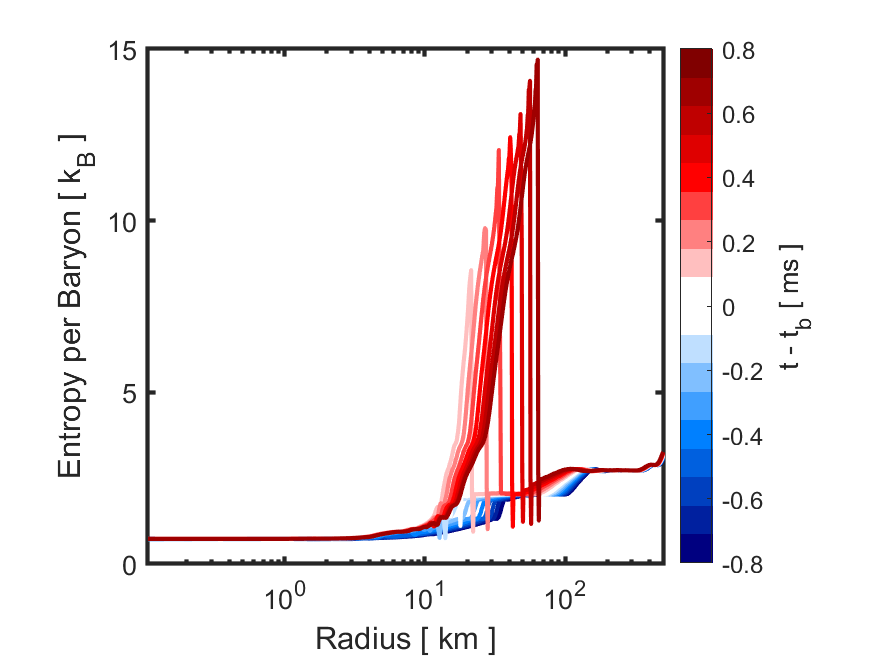}
    \caption{Numerical solutions of select quantities versus radius for adiabatic collapse evolved through bounce: adiabatic index $\Gamma\equiv\big(\pderiv{\ln p}{\ln\rho}\big)$ (upper left), velocity (upper right), electron fraction (lower left), and entropy per baryon (lower right). A finer time resolution is used here to exhibit the characteristics of bounce and shock formation, and the color map on the right of each panel is used to distinguish pre- and post-bounce profiles; blue and red, repsectively.}
    \label{fig:Gravitational Collapse, Bounce Phase}
\end{figure}

\subsection{Stage 3: Shock Propagation}

Figure~\ref{fig:Gravitational Collapse, Post-Bounce Phase} shows the shock's trajectory through the outer core on its way towards the outer boundary.  
In this figure, we plot the mass density (upper left), velocity (upper right), electron fraction (lower left), and temperature (lower right) versus radius for select times after bounce.  
As can be seen by inspecting all panels, the inner core (inside about 50~km) settles into an approximate hydrostatic equilibrium once the bounce shock has cleared.  
Inside this region, the velocity is small (compared with the sound speed), and the mass density, electron fraction, and temperature profiles remain practically unchanged for hundreds of milliseconds.  
This suggests that the DG method is quite capable of capturing the adiabatic nature of the flow (this is further supported by the results shown in the left panel in Figure~\ref{fig:Gravitational Collapse, Resolution Study}).  
In the velocity figure, the shock is seen to reduce in amplitude as it propagates towards the outer boundary.  
Early on, one can also observe secondary shocks, produced by the ring-down of the core as it settles into hydrostatic equilibrium, which later catch up with the main shock.  
Rarefaction of the gas occurs in the outer core (beyond 100~km) as the shock pushes through the infalling matter, notably at $t-t_{\rm{b}}=362$~ms in the mass density profile.  
The thermal energy behind the shock is partially used to dissociate heavy nuclei and alpha particles in the supersonically infalling outer core, causing the shock to lose energy while leaving behind free nucleons in its wake.  
As the shock travels outward, the electron fraction profile in the outer core is advected with the flow; cf. the sharp gradient located around 100~km at $t-t_{\rm{b}}=2$, which has moved to about 1000~km when $t-t_{\rm{b}}=362$~ms.  
The temperature inherently rises across the shock, and a sharp rise in temperature that traces the path of the shock is seen in the lower right panel.  

\begin{figure}[h]
    \centering
    \includegraphics[width=.50\textwidth]{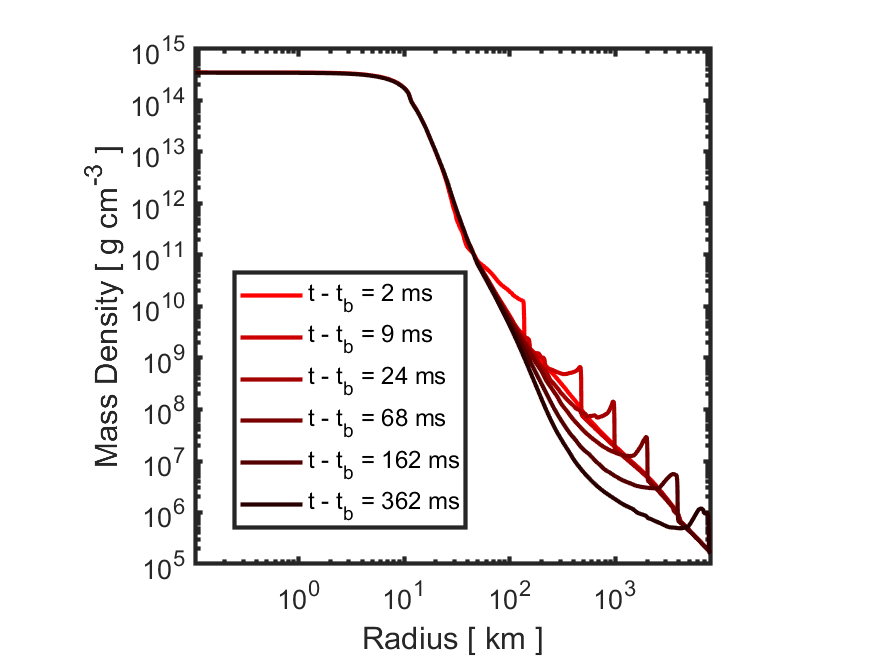}\hfill
    \includegraphics[width=.50\textwidth]{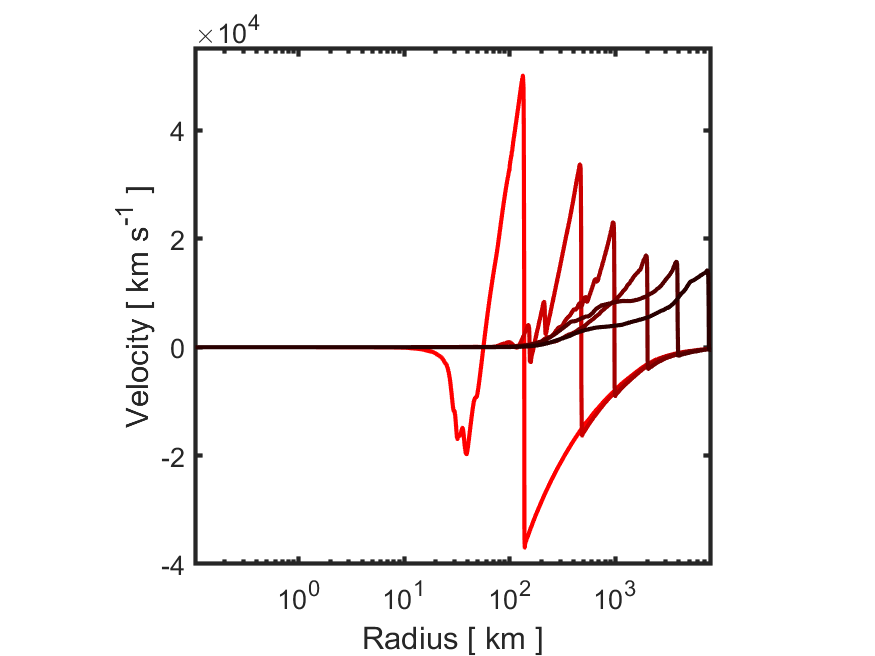}\hfill
    \includegraphics[width=.50\textwidth]{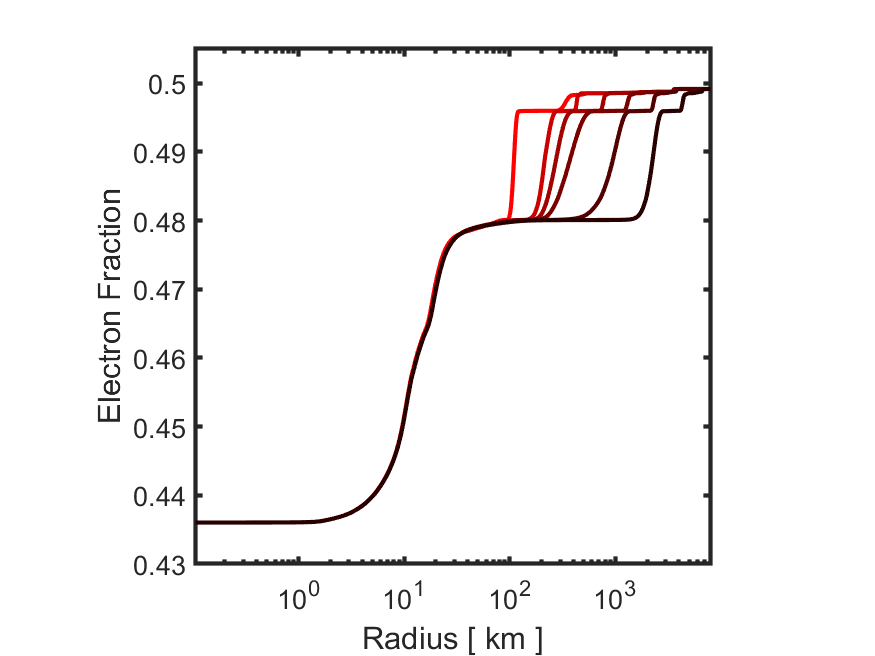}\hfill
    \includegraphics[width=.50\textwidth]{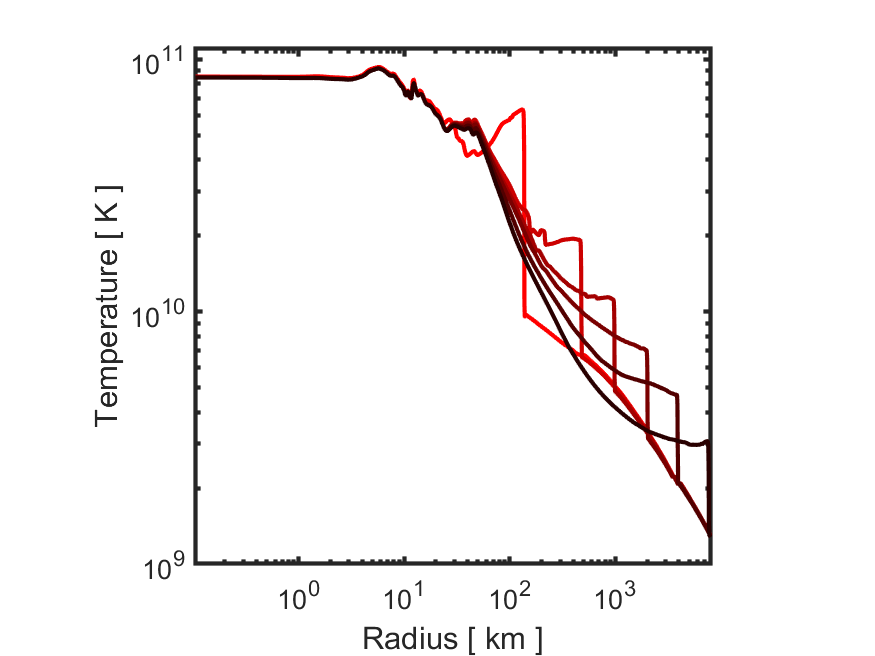}
    \caption{Numerical solutions for mass density (upper left), velocity (upper right), electron fraction (lower left), and temperature (lower right) versus radius at select times for the adiabatic collapse simulation evolved for several hundred milliseconds post bounce.  This time domain partially captures the structure of the core as the shock propagates from its origin to the outer boundary.}
    \label{fig:Gravitational Collapse, Post-Bounce Phase}
\end{figure}

\subsection{Bound-Enforcing Limiter}

The microphysical conditions encountered in this test are constrained by the nuclear EoS.  
However, some extreme conditions encountered are difficult to resolve numerically, and thus may push the solutions beyond the boundaries of the admissible state set.  
For example, when the core bounces and launches the bounce shock, the discontinuity can generate oscillations in the numerical solution.  
These oscillations are to a certain degree suppressed by the slope limiting procedure described in Section~\ref{sec:slope}, but the solution can still exceed the limits of the tabulated EoS.  
Thus, the bound-enforcing limiting procedure from Section~\ref{sec:boundEnforcing} is required to ensure that the numerical solution remains physically valid, mostly at bounce and shock formation.  
When necessary, the bound-enforcing limiter acts to constrain the mass density, electron fraction, and specific internal energy.  
However, for the conditions encountered in the adiabatic collapse simulations discussed in this section, only violations of the bounds on the specific internal energy trigger limiting (cf. Step~3 in Section~\ref{sec:boundEnforcing}), namely during the early stages of shock formation.  
We note that, without the bound-enforcing limiter, the specific internal energy falls below the minimum possible value at certain locations, which then implies that a valid temperature --- required, e.g., to compute the pressure --- cannot be found, and the algorithm fails.  
Therefore, the bound-enforcing limiter is a critical component of the DG algorithm in \thornado.

Figure~\ref{fig:Gravitational Collapse, Bound Enforcing Limiter Activation Sites} illustrates the action of the bound-enforcing limiter during bounce in the fiducial run discussed in the previous subsections.  
The left panel is a space-time plot of the limiter parameter $\vartheta_{3}\in[0,1]$ (cf. Equation~\eqref{eq:limitedPolynomialStep3}), and shows the activation sites of the bound-enforcing limiter acting to constrain the specific internal energy $\epsilon$.  
Values of $\vartheta_{3}<1$ imply some amount of limiting.  
The region displayed in the figure captures the brief moment around shock formation where $\epsilon$ drops below the minimum value, but is corrected by shifting the DG solution toward the cell average by an amount determined by $\vartheta_{3}$.  
The darker regions indicate more aggressive limiting, and we find that $\vartheta_{3}$ can become as small as 0.4 in this case.  
In the right panel in Figure~\ref{fig:Gravitational Collapse, Bound Enforcing Limiter Activation Sites}, the specific internal energy is plotted versus radius for select times during the initial shock propagation (black lines).  
We also plot the minimum specific internal energy $\epsilon_{\min}(\rho,\ye)$, using the corresponding numerical solutions for $\rho$ and $\ye$ (red lines).  
This figure captures $\epsilon$ being very close to, but above, $\epsilon_{\min}$ --- especially around the shock, which is located roughly $r=20$, $40$, and $70$~km for the times displayed.  

\begin{figure}[h]
    \centering
    \includegraphics[width=0.5\textwidth]{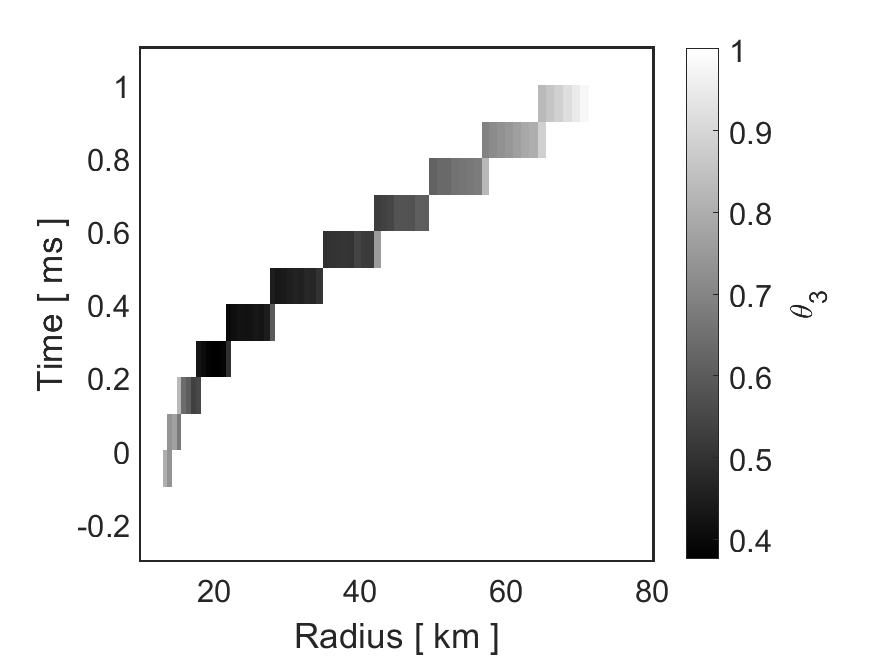}\hfill
    \includegraphics[width=0.5\textwidth]{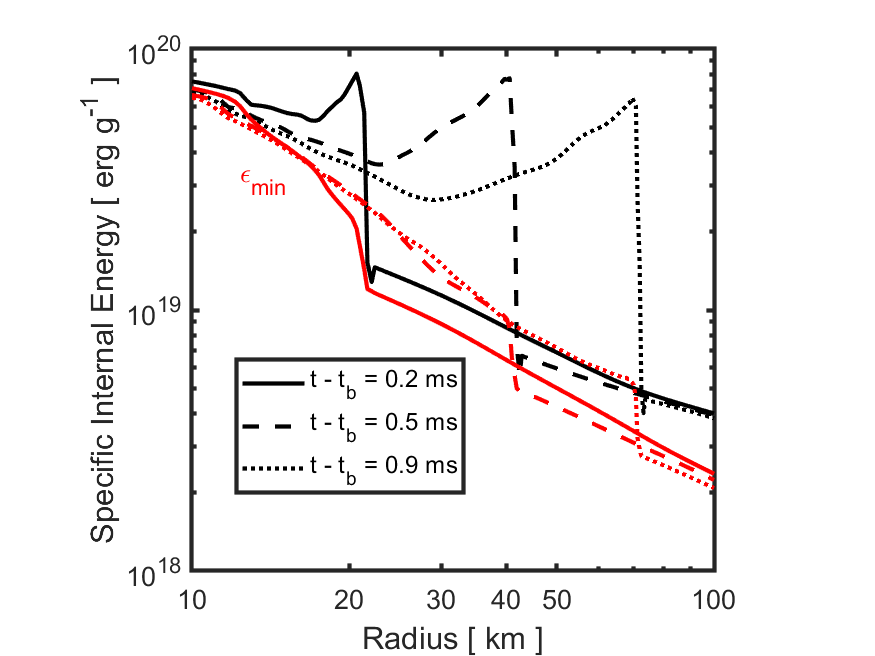}
    \caption{Activation of the bound-enforcing limiter in the fiducial adiabatic collapse simulation.  The left panel shows the value of the limiter parameter $\vartheta_{3}$ from Equation~\eqref{eq:limitedPolynomialStep3} in space and time. In the right panel we plot the solution for $\epsilon$ (black) and the minimum $\epsilon_{\min}$ (red), described in Section~\ref{sec:boundEnforcing}. Each profile captures a moment in time briefly after bounce, when the bound-enforcing limiter is required to maintain $\epsilon>\epsilon_{\mathrm{min}}$.}
    \label{fig:Gravitational Collapse, Bound Enforcing Limiter Activation Sites}
\end{figure}

Figure~\ref{fig: Gravitational Collapse, Activation Sites over EoS} shows activation sites of the bound-enforcing limiter in the $\rho\ye$-plane (white dots).  
The majority of the activation sites are seen at higher mass densities, and correspond to the formation of the shock.  
These points appear to occupy a locally convex region of $\epsilon_{\min}(\rho,\ye)$.  
However, some points also appear at a low density and higher electron fraction.  
These points correspond to a moment toward the end of the simulation, specifically when the shock passes through the outer boundary.  
This portion of the EoS table may also be locally convex, thus the limiting scheme is expected to operate robustly in that region as well.  
Future work will involve an investigation of the EoS surface at minimum temperature to further challenge the robustness of our bound-enforcing limiter.  
This work, however, will need to be carried out in the context of neutrino radiation-hydrodynamics simulations of CCSNe, which access different and/or larger regions of the $\rho\ye$-plane.  

\begin{figure}[h]
  \centering
  \includegraphics[width = 0.75\textwidth]{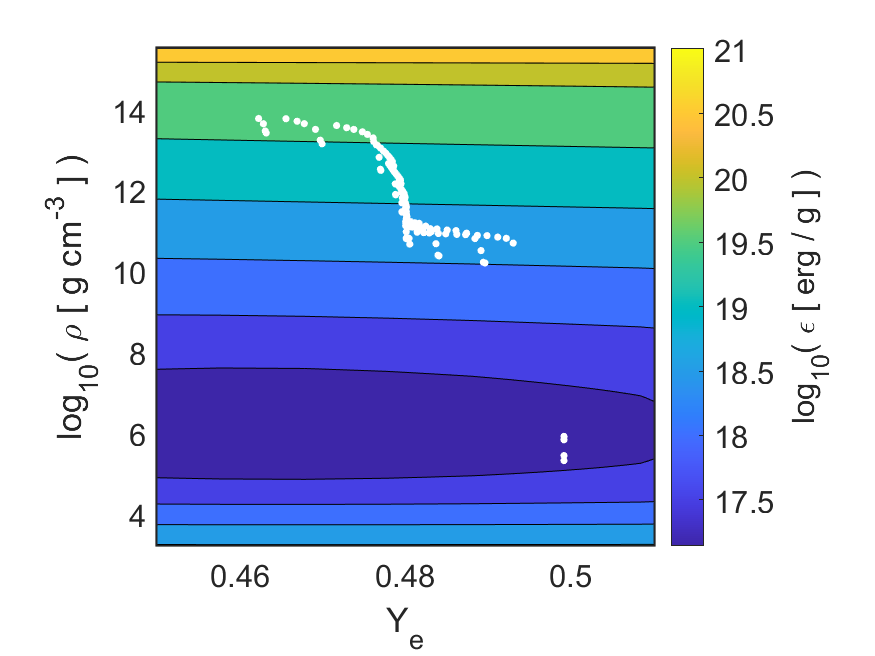}
  \caption{Activation sites (white dots) of the bound-enforcing limiter during the adiabatic collapse simulations in the $\rho\ye$-plane, placed over a contour plot of the surface defined by $\epsilon_{\min}=\epsilon(\rho,T_{\min},\ye)$. The points in the mass density range $\log_{10}\left(\rho\right) \in \left[10, \, 14\right]$ show the limiter being activated during bounce and shock formation. The limiter is again briefly applied in the low density region, $\log_{10}\left(\rho\right) \in[5,6]$. This corresponds to when the shock momentarily forces the solution below the lower EoS table boundary as the shock passes through the outer boundary of the spatial domain.}
  \label{fig: Gravitational Collapse, Activation Sites over EoS}
\end{figure}

\subsection{Troubled-Cell Indicator Threshold Dependence}
\label{sec:collapseTCI}

In this section, we investigate the effect of varying the troubled-cell indicator threshold $C_{\TCI}$ on the adiabatic collapse simulations.  
The numerical results discussed in the previous subsections applied the slope limiter everywhere; i.e. the TCI threshold $C_{\TCI}$ was set to zero such that all elements are flagged for (component-wise) limiting.  
As seen in Section~\ref{sec:riemannShuOsher}, increasing the value of $C_{\TCI}$ prevents limiting at smooth extrema and preserves the accuracy of the solution.  
However, in contrast to the shock tube problem, the solutions for the adiabatic collapse problem exhibit nonzero slopes almost everywhere.  
This leads to more areas that may require limiting, and it becomes more difficult to find an optimal value for $C_{\TCI}$.  
Moreover, various quantities vary by many orders of magnitude across the computational domain, and it is not clear which variables are optimal for detecting troubled cells.  
When using the mass density, the total fluid energy density, and the electron fraction as the variables to sense troubled cells, we find that if $C_{\TCI}$ is set too high, some areas that may require limiting are not flagged, and oscillations can start to develop.  

In general, we have found that thermodynamic quantities such as the temperature and entropy per baryon demonstrate a higher sensitivity to $C_{\TCI}$ than the evolved quantities $\vect{U}$.  
Thus, this section will focus on the solution for the temperature and its sensitivity to $C_{\TCI}$.  
Figure~\ref{fig:Gravitational Collapse, Troubled Cell Indicator Study} shows the evolution of the troubled-cell indicator $I_{\vect{K}}$ (cf. Equation~\eqref{eq:tci}) versus radius for adiabatic collapse simulations with various values of $C_{\TCI}$: 0.01 (upper left panel), 0.03 (upper right panel), and 0.05 (lower left panel).  
The plotted quantities are derived from the maximum value across all fields in each element; i.e., 
\begin{equation}
  I_{\vect{K}}=\max_{G\in\vect{G}}I_{\vect{K}}(G),\quad\text{where}\quad \vect{G}=(\rho,E,\ye)^{\rm{T}}.  
\end{equation}
In each panel, the red curve represents the time-averaged value (from $t_{\rm{b}}$ to $t_{\rm{end}}-t_{\rm{b}}=497.1$ ms), while the maximum and minimum values are given by the boundaries of the light gray-shaded region, and the positive standard deviations (i.e., average plus one $\sigma$) are given by the upper boundary of the dark gray-shaded region.  
We also plot $C_{\TCI}$ in each panel (dashed horizontal line).  
Recall that an element is flagged for limiting whenever $I_{\vect{K}}>C_{\TCI}$.  
In the lower right panel of Figure~\ref{fig:Gravitational Collapse, Troubled Cell Indicator Study}, we plot the temperature versus radius at the end of each simulation with a different value of $C_{\TCI}$.  
For comparison, we also plot the temperature for the simulation from the previous subsections, with $C_{\TCI}=0$, which applies limiting everywhere.  

As can be seen in the lower right panel in Figure~\ref{fig:Gravitational Collapse, Troubled Cell Indicator Study}, the temperature profiles from all four runs display the same general trend, and fall on top of each other outside $r=50$~km.  
The simulations with $C_{\TCI}=0$ and $C_{\TCI}=0.01$ (magenta and black lines, respectively) are practically indistinguishable everywhere.  
However, inside $r=50$~km, the simulations with the larger values of $C_{\TCI}$ (0.03 and 0.05) exhibit some oscillations about the temperature profile from the fiducial run with $C_{\TCI}=0$, and the amplitude appears to increase with increasing $C_{\TCI}$.  
The TCI maxima (upper boundary of the light gray-shaded region) are generally above the threshold in all cases, which implies that limiting has been applied at least once in most of the domain displayed.  
However, the average and the one sigma values serve as better indicators for where limiting occurs.  
Although the $I_{\vect{K}}$ values tend to be above the threshold inside the first $100$~km in the $C_{\TCI}=0.01$ case, the solution is limited most frequently inside $r=50$~km, which corresponds to the region where the temperature displays oscillatory behavior in the runs with larger values of $C_{\TCI}$.  
As the threshold is increased, less of this region receives limiting.  
And in particular, for the $0.03$ and $0.05$ threshold cases, oscillations have developed in this region.  
Ideally, the limiting procedure should both preserve the original order of accuracy and prevent the development of spurious oscillations.  
However, for the adiabatic collapse simulations, inside $r=50$~km, there seems to be a trade-off between these two features which leaves little flexibility for selecting a large value for $C_{\TCI}$.


\begin{figure}[h]
    \centering
    \includegraphics[width=.50\textwidth]{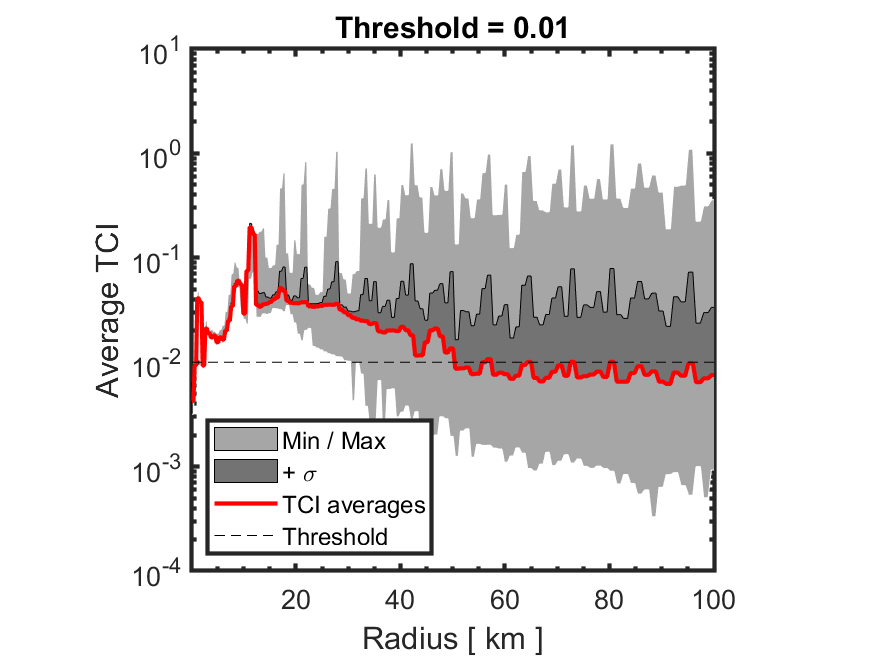}\hfill
    \includegraphics[width=.50\textwidth]{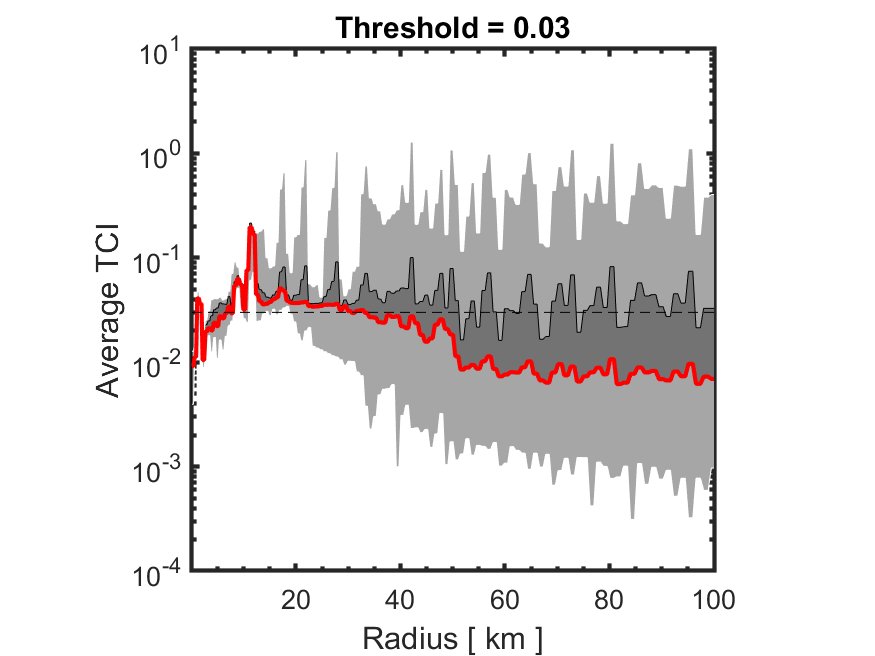}\hfill
    \includegraphics[width=.50\textwidth]{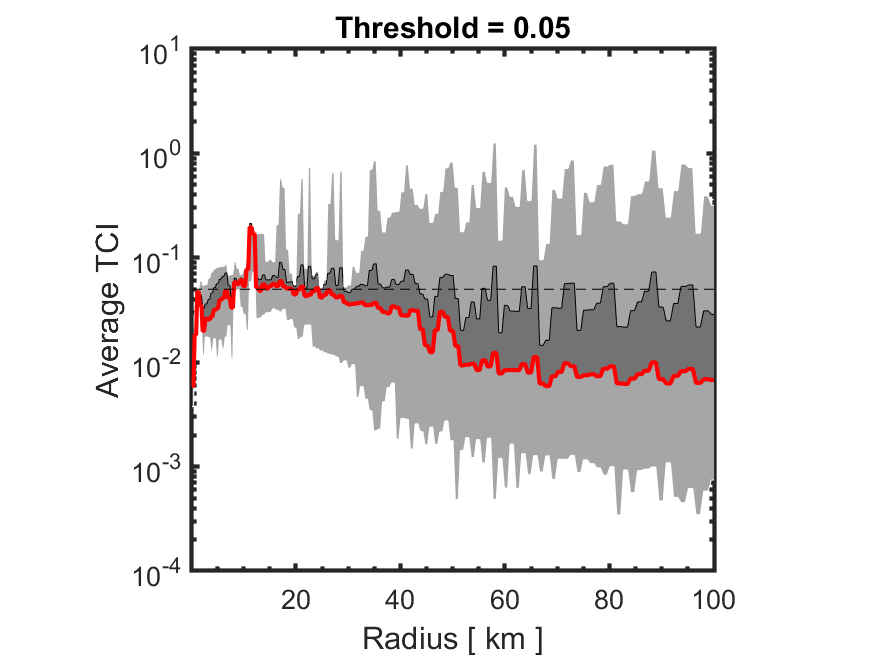}\hfill
    \includegraphics[width=.50\textwidth]{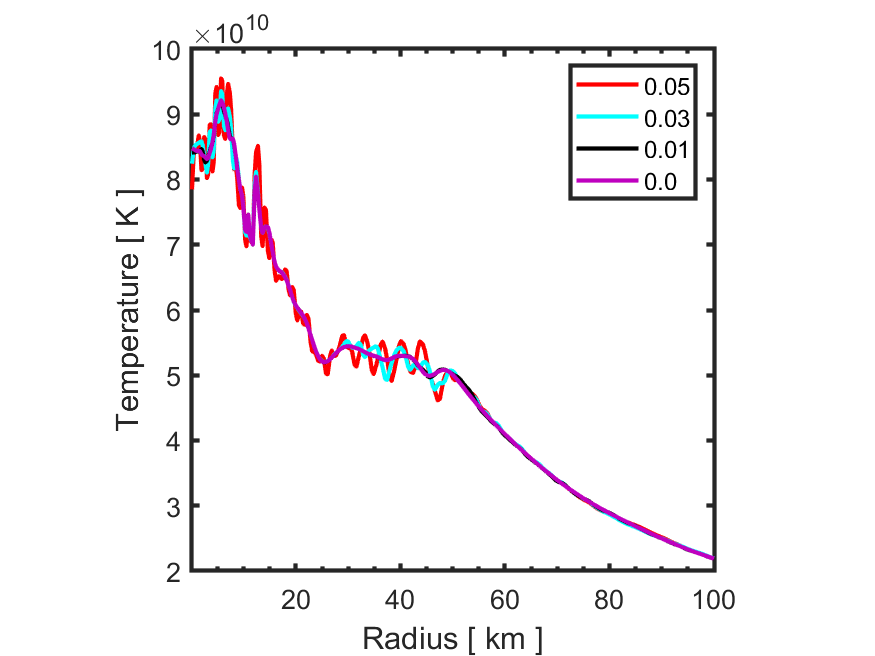}
    \caption{Troubled-cell indicator values for adiabatic collapse simulations with various thresholds $C_{\TCI}$ (upper panels and lower left panel).  In each panel, the red curve represents the time-averaged troubled-cell indicator value (averaged from $t_{\rm{b}}$ to $t_{\rm{end}}-t_{\rm{b}}=497.1$ ms of the simulation). The lighter gray-shaded regions represent the extreme TCI values in each element (taken over all post-bounce times).  The darker shaded region represents the positive TCI standard deviation.  In the lower right panel, the temperature at the end of each simulation with different $C_{\TCI}$ is plotted versus radius. A higher threshold results in less slope limiting, which allows for some oscillations to develop in the temperature profile.}
    \label{fig:Gravitational Collapse, Troubled Cell Indicator Study}
\end{figure}

\subsection{Resolution Dependence}
\label{sec:collapseResolutionDependence}

In this section we investigate the effect of varying the spatial resolution in the adiabatic collapse simulation.  
To do this, we keep the number of elements fixed to $N=512$, and vary the innermost cell width $\Delta r_{1}$ from $0.125$~km to $1.0$~km.  
Table~\ref{tab:ResolutionStudy} lists the inner- and outer-most cell widths along with the cell widths at $r=10~$km and $r=100$~km, and the corresponding zoom factors $z$, in Equation~\eqref{eq:geometricGridCellWidth}.  
Since we keep the number of elements fixed, the the zoom factor increases with decreasing $\Delta r_{1}$, which also results in coarser resolution in the outer regions of the computational domain.  
We find that the general features of the solution --- e.g., density and velocity profiles --- are rather insensitive to the numerical resolution.  
Instead, we focus on the long term evolution (i.e., hundreds of milliseconds) of the central density, electron fraction, and entropy per baryon.  
After bounce, when the inner core settles into hydrostatic equilibrium, the central density should remain relatively constant with time.  
Similarly, since we do not include neutrinos and the evolution is adiabatic, the central electron fraction and entropy per baryon should also remain constant throughout the simulation.  
Figure \ref{fig:Gravitational Collapse, Resolution Study} shows results from varying the inner cell width.  
In the left panel, we plot the central density $\rho _{\rm{c}}$ versus time after bounce; i.e. the time when maximum central density is achieved.  
(To better visualize with a logarithmic abscissa, we have applied an arbitrary shift of $0.6$~ms.)  
The right panel displays the evolution of the central entropy per baryon $S_{\rm{c}}$ (top) and electron fraction $Y_{\rm{e,c}}$.  
During collapse, these quantities are plotted versus central density, while after bounce they are plotted versus time.  
There is some spread in the central density curves before bounce, but they all reach about the same maximum, $\rho_{\rm{c}}\approx4.2\times10^{14}$~g~cm$^{-3}$, and, after the core stabilizes after bounce, $\rho_{\rm{c}}$ remains constant with time for all resolutions.  
For the coarsest resolution run ($\Delta r_{1}=1$~km), the central density settles down to about $3.425\times10^{14}$~g~cm$^{-3}$, while in the finer resolution models it settles down to about $3.475\times10^{14}$~g~cm$^{-3}$.  
Because the collapse is adiabatic and the profiles are constant with radius in the very inner core (cf. lower panels in Figure~\ref{fig:Gravitational Collapse, Collapse Phase}), $S_{\rm{c}}$ and $Y_{\rm{e,c}}$ should remain constant throughout the evolution.  
All the simulations exhibit this behavior before $\rho _{\rm{c}}\approx10^{13}$~g~cm$^{-3}$; i.e., before the phase transition into nuclear densities.  
(There is a slow increase in $S_{\rm{c}}$, from 0.73 to 0.74, during collapse.)  
Just before core bounce, the profiles deviate somewhat from their constant values, and the lower resolutions exhibit larger deviations.  
For both central entropy and electron fraction, the profiles for the runs with $\Delta r_{1}=0.75$~km and $\Delta r_{1}=1.0$~km undergo notably larger changes than the higher resolution profiles.  
$Y_{\mathrm{e,c}}$ remains nearly constant through bounce for the $0.125$~km, $0.25$~km, and $0.5$~km simulations.  
For both $Y_{\mathrm{e,c}}$ and $S_{\rm{c}}$, the two lowest resolution profiles drop further down before maximum central density, and then exhibit a slight drift with time after bounce.  
However, both of these quantities remain relatively constant with time after bounce in the higher resolution cases.  
Thus, a threshold resolution seems to be required to accurately capture the physical behavior in the inner core.  
Considering the balance between computational cost and physical fidelity, an inner cell width of $0.5$~km (as in the fiducial run) appears to be close to the optimal choice among the tested resolutions.  
For example, the central density for this run remains constant after bounce, as desired.  
It also maintains approximately constant central entropy and electron fraction through bounce.  
The central entropy deviates by no more than 0.02~$k_{\mathrm{B}}$, while the electron fraction changes by no more than about $10^{-6}$.  

\begin{table}[h]
  \caption{Inner, $r=10$~km, $r=100$~km, outer cell widths, and zoom factors for geometrically progressing grids with $N=512$ elements.\label{tab:ResolutionStudy}}
  \small
  \vspace{-6pt}
  \begin{center}
  \begin{tabular}{llllc}
    \midrule
    $\Delta r_{1}$  [ km ] & $\Delta r_{10 \, \rm{km}} $ [ km ] & $\Delta r_{100 \, \rm{km}}$ [ km ] & $\Delta r_{N}$ [ km ] & Zoom Factor \\
    \midrule
    \midrule
    $0.125$ & $0.258$ & $1.430$ & $1.048 \times 10^{2}$ & $1.013260722382225$  \\
    $0.25$  & $0.366$ & $1.401$ & $9.225 \times 10^{1}$ & $1.011634298318296$ \\
    $0.5$ & $0.598$ & $1.489$  & $7.945 \times 10^{1}$ & $1.009967685243838$ \\
    $0.75$ & $0.835$ & $1.630$  & $7.185 \times 10^{1}$ & $1.008968091682754$ \\
    $1.0$ & $1.077$ & $1.821$  & $6.641 \times 10^{1}$ & $1.008244905346311$ \\
    \midrule
    \midrule
  \end{tabular}
  \end{center}
\end{table}

\begin{figure}[h]
    \centering
    \includegraphics[width=.50\textwidth]{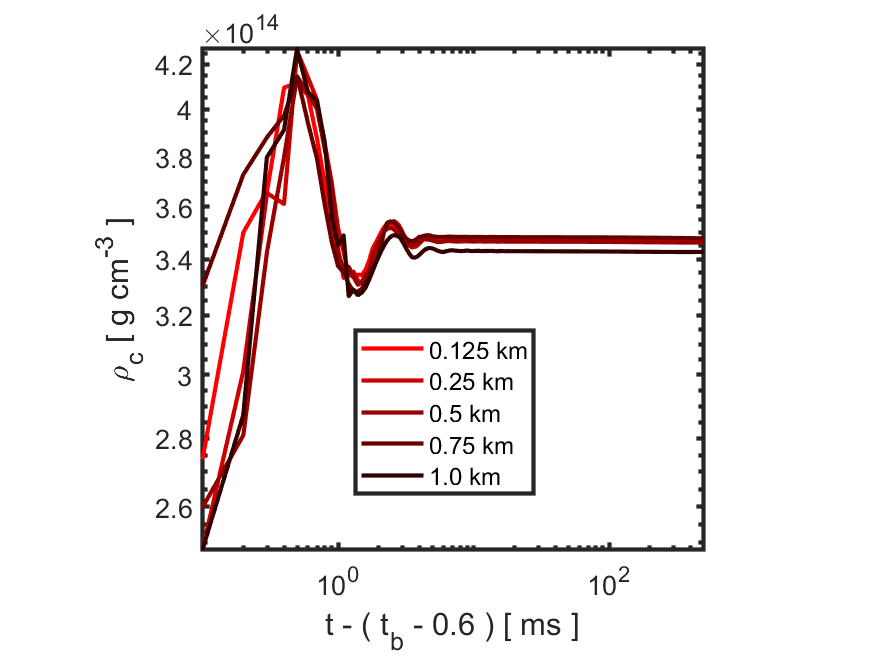}\hfill
    \includegraphics[width=.50\textwidth]{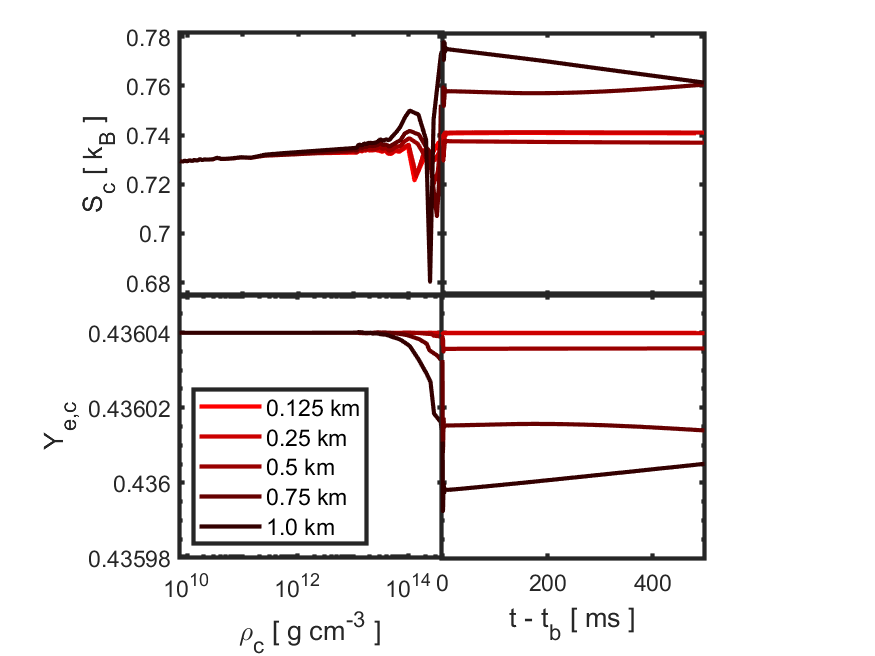}\hfill
    \caption{Results from adiabatic collapse simulations where the innermost cell width has been varied.  The left panel shows the central density as a function of time for various $\Delta r_{1}$. The right panel shows the central entropy (top) and central electron fraction (bottom) versus central density (up to its maximum value). Beyond the maximum central density, the entropy and electron fraction are plotted versus time.}
    \label{fig:Gravitational Collapse, Resolution Study}
\end{figure}

\subsection{Energy Conservation}
\label{sec:collapseEnergyConservation}

In this section we investigate total energy conservation with the DG method in \thornado\ in the context of the adiabatic collapse simulations.  
Exact conservation of total energy is nontrivial to achieve in simulations of self-gravitating flows because of the adopted formulation of the fluid energy equation given by Equation~\eqref{eq:energyConservation}, which is in non-conservative form due to the gravitational source term on the right-hand side.  
For simplicity, we limit the discussion to the present context of spherical-polar coordinates with spherical symmetry imposed.  
Then, by combining Equations~\eqref{eq:massConservation}, \eqref{eq:energyConservation}, and \eqref{eq:poissonEquation}, it is possible to formulate a conservation law for the total energy
\begin{equation}
  \pd{\mathcal{E}}{t} + \f{1}{r^{2}}\pd{}{r}\big(\,r^{2}\,\mathcal{F}\,\big) = 0,
  \label{eq:totalEnergyConservation}
\end{equation}
where
\begin{equation}
  \mathcal{E} = \rho\,\big(\,\epsilon+\f{1}{2}\,v^{2}+\f{1}{2}\,\Phi\,\big)
  \quad\text{and}\quad
  \mathcal{F} = \big(\,E+p+\rho\,\Phi\,\big)\,v + \f{1}{8\pi G}\Big(\,\Phi\,\pderiv{\dot{\Phi}}{r}-\dot{\Phi}\,\pderiv{\Phi}{r}\,\Big)
\end{equation}
are the total energy density and total energy flux density, respectively, $v$ is the radial component of the fluid three-velocity, and $\dot{\Phi}=\pd{\Phi}{t}$.  
Because the corresponding RKDG discretization of Equations~\eqref{eq:massConservation} and \eqref{eq:energyConservation}, and the finite element discretization of Equation~\eqref{eq:poissonEquation}, do not combine exactly to form a discrete equivalent to Equation~\eqref{eq:totalEnergyConservation}, the conservation of total energy is not expected to be exact in the adiabatic collapse simulations.  
Although we find that the combination of RKDG and finite element discretization exhibits surprisingly good energy conservation properties, we find evidence that the application of the slope and bound-enforcing limiters, mainly around core bounce, compromise the conservation of total energy.  
As seen in Figure~\ref{fig:Pochik Shock Tube} for the Riemann problem invoking the bound-enforcing limiter, in the absence of gravity, the total fluid energy (i.e., internal plus kinetic) is by construction conserved to machine precision.  
The slope limiter is also conservative with respect to the total fluid energy.  
Conservation of total energy is more difficult to achieve for self-gravitating flows such as in the adiabatic collapse problem.  

By integrating Equation~\eqref{eq:totalEnergyConservation} over the computational domain $D=[0,R]$, and from $t_{0}$ to $t$, the total energy in the system is given by
\begin{equation}
  E_{\rm{total}}(t) = E_{\rm{total,0}} - 4\pi R^{2}\int_{t_{0}}^{t}\mathcal{F}(R,\tau)\,d\tau,
  \label{eq:totalEnergyVersusTime}
\end{equation}
where
\begin{align}
  E_{\rm{total}}
  =\int_{D}\rho\,\epsilon\,dV + \f{1}{2}\int_{D}\rho\,v^{2}\,dV + \f{1}{2}\int_{D}\rho\,\Phi\,dV
  \equiv E_{\rm{i}} + E_{\rm{k}} + E_{\rm{g}},
  \label{eq:totalEnergyComponents}
\end{align}
and $dV=4\pi r^{2}dr$.  
Figure~\ref{fig:Gravitational Collapse, Energy Conservation} shows energy conservation results from adiabatic collapse simulations.  
In the left panel, we plot the kinetic, gravitational, internal, and total energy versus time for the fiducial run with $\Delta r_{1}=0.5$~km.  
Approaching core-bounce, the internal energy $E_{\rm{i}}$ and the gravitational energy $E_{\rm{g}}$ grow rapidly in concert (with opposite signs), before stabilizing after bounce with $E_{\rm{i}}\approx157$~B and $E_{\rm{g}}\approx-158$~B, where 1~B = 10$^51$ erg.  
The kinetic energy $E_{\rm{k}}$ peaks at about $10$~B at bounce before decreasing again, and is down to $1$~B, when $t-t_{\rm{b}}=5.5$~ms.  
For $t-t_{\rm{b}}\gtrsim40$~ms, the kinetic energy starts slowly increasing again, and is back up at $1$~B for $t-t_{\rm{b}}=200$~ms.  

The change in the total energy versus time, $E_{\rm{total}}-E_{\rm{total},0}$, is plotted in the middle panel of Figure~\ref{fig:Gravitational Collapse, Energy Conservation} for the various spatial resolutions investigated in Section~\ref{sec:collapseResolutionDependence}.  
As can be seen, the total energy remains relatively constant during collapse, makes an almost discontinuous jump around bounce, before remaining relatively constant again after bounce.  
(At $t-t_{\rm{b}}\approx375$~ms, the bounce shock reaches the outer boundary, and the total energy starts to decrease due to the energy flux through the boundary; cf. the second term on the right-hand side of Equation~\eqref{eq:totalEnergyVersusTime}, which has not been accounted for in the figure.)  
The magnitude of the jump in total energy at bounce decreases with increasing resolution in the core.  
Around $t=t_{\rm{b}}$, the total energy in the fiducial run ($\Delta r_{1}=0.5$~km) increases by less than $0.5$~B, as is seen from the middle curve (after bounce) in the middle panel in Figure \ref{fig:Gravitational Collapse, Energy Conservation}.  
The difference $E_{\mathrm{total}}-E_{\mathrm{total},0}$ ought to remain zero throughout the simulation, but the extreme conditions during core-bounce --- due to short time and length scales, and the necessity of applying limiters around the region of shock formation, which occurs at high energy densities --- result in energy conservation violations.  
The change in the total energy in the fiducial run is less than $0.5$\% of the gravitational energy at bounce, and about $5$\% of the kinetic energy at bounce.  
Without accounting for the energy flowing through the outer boundary, the total energy changes by less than $1.5\times10^{-3}$~B during collapse, until $t-t_{\rm{b}}=-0.6$~ms, when the central density is about $1.5\times10^{14}$~g~cm$^{-3}$.  
Then, after bounce, from $t-t_{\rm{b}}\approx50$~ms to $t-t_{\rm{b}}\approx350$~ms, the total energy changes by less than $2.5\times10^{-3}$~B, which is small compared to any of the individual components of the total energy.  

We have found that the slope and bound-enforcing limiters contribute to the violation of total energy conservation at bounce.  
To investigate the impact of limiters on total energy conservation, we restarted the fiducial run, which employs slope and bound-enforcing limiters, at $t-t_{\rm{b}}=1$~ms, and ran one model with the slope limiter turned off, and one model with both the slope and bound-enforcing limiters turned off.  
The right panel in Figure \ref{fig:Gravitational Collapse, Energy Conservation} shows the total energy conservation versus time for these models.  
The largest violation of total energy conservation is observed in the fiducial run (red line).  
For the model where the slope limiter is turned off, but the bound-enforcing limiter is still active, the change in the total energy is noticeably reduced (black line).  
For example, the black line demonstrates no noticeable change in the total energy briefly before bounce, while the red line shows a minor increase starting at $t-t_{\rm{b}}=-0.5$ ms.  
Thus, the slope limiter begins adding energy to the system shortly before bounce.  
Meanwhile, the bound-enforcing limiter remains inactive until about $0.2$~ms before bounce. 
Once activated, the bound-enforcing limiter breaks total energy conservation, but to a lesser extent than when both limiters are active.  
The reason the limiters contribute to total energy violation is the gravitational potential energy, the third integral on the right-hand side of Equation~\eqref{eq:totalEnergyComponents}.  
While both limiters preserve the cell-averaged fluid energy, and thus leave the first two integrals on the right-hand side of Equation~\eqref{eq:totalEnergyComponents} unchanged, the cell-averaged gravitational potential energy density is defined as a higher moment of the mass density ($\Phi$ depends on position), which is not preserved by any of the limiters.  
It is interesting to note that the DG method manages to model core bounce and shock formation without the slope limiter activated.  
When both limiters are turned off, the run fails at bounce because $\epsilon$ may fall below the minimum value required by the EoS 
Until then, the DG method maintains total energy conservation well.  
For example, we find $E_{\mathrm{total}}-E_{\mathrm{total}}(t_{\rm{b}}-1~\text{ms})=8.6 \times 10^{-6}$~B at the time when the run crashes, which occurs when $\rho_{\rm{c}}=3.65 \times 10^{14}$~g~cm$^{-3}$.  
In the future, we will investigate ways of improving the conservation of total energy while applying both limiters through bounce.  

\begin{figure}[h]
  \centering
  \includegraphics[width = 0.31\textwidth]{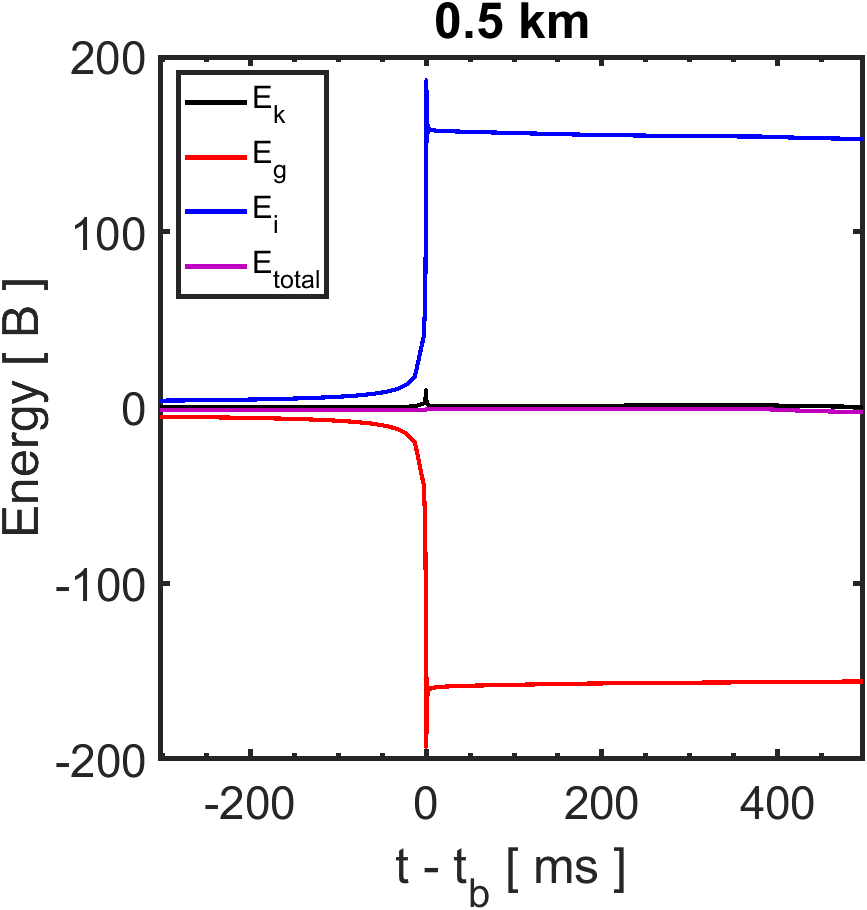}
  \hfill
  \centering
  \includegraphics[width = 0.31\textwidth]{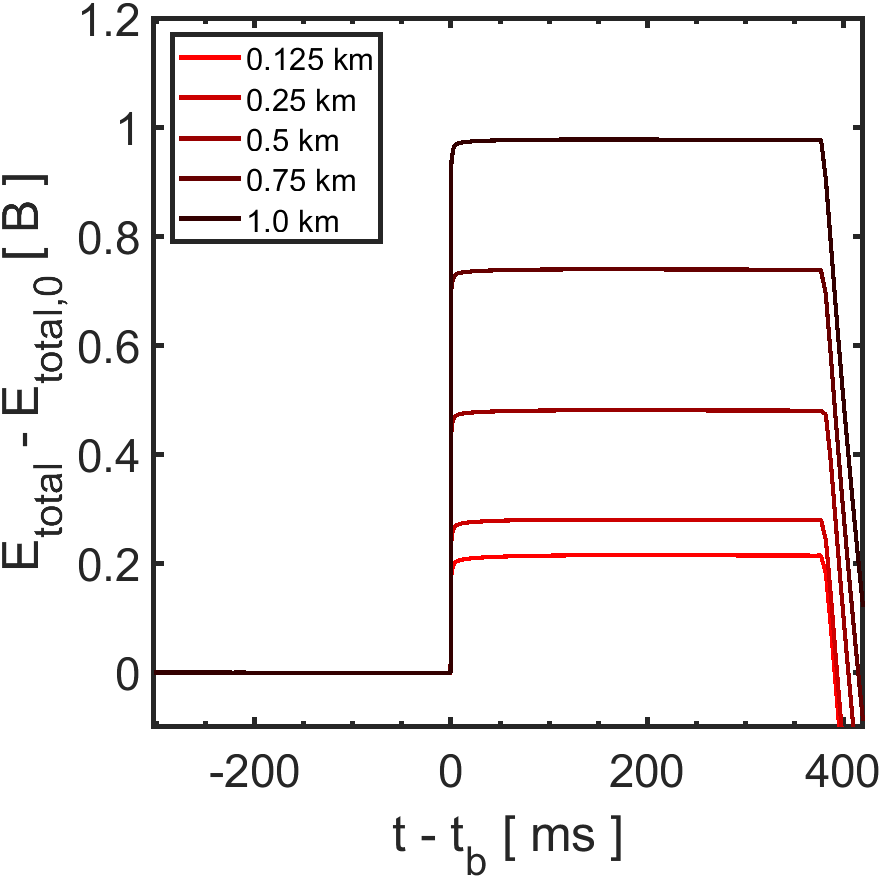}
  \hfill
  \centering
  \includegraphics[width = 0.32\textwidth]{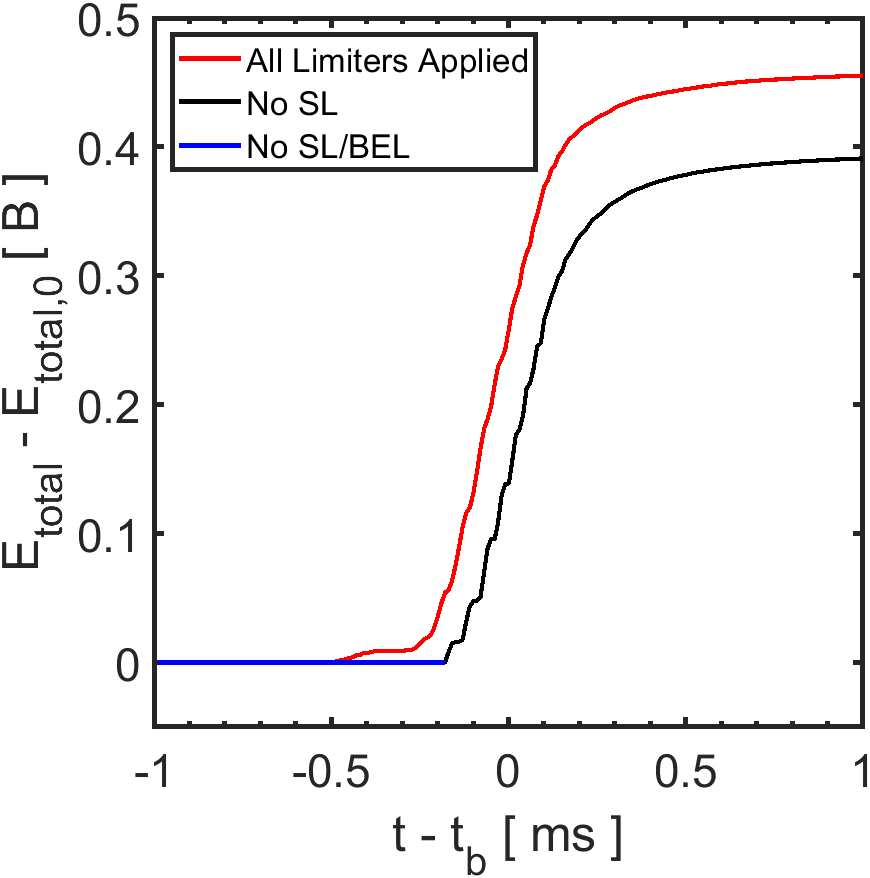}
  \caption{Energy conservation by the RKDG method in \thornado\ for adiabatic collapse simulations.  The left panel shows gravitational (red), kinetic (black), internal (blue), and total (magenta) energy versus time for the fiducial run with $\Delta r_{1} = 0.5$~km. Due to the relative magnitude of $E_{\rm{i}}$ and $E_{\rm{g}}$, the details in the kinetic and total energies are obscured.  The middle panel shows the change in total energy versus time for all the resolutions considered in Section~\ref{sec:collapseResolutionDependence}.  The decrease in the total energies around $t-t_{\rm{b}}\approx375$~ms is due to the bounce shock reaching the outer boundary of the domain.  The right panel shows the total energy versus time for models with various combinations of limiters enabled for the fiducial run.  The red line represents the total energy when applying both the slope limiter and the bound-enforcing limiter. The black line shows this quantity when only applying the bound enforcing limiter. The blue line represents a model with both limiters off, which eventually crashed at bounce.}
  \label{fig:Gravitational Collapse, Energy Conservation}
\end{figure}

\subsection{Characteristic Limiting}
\label{sec:collapseCharacteristicLimiting}

In contrast to the Riemann problems discussed in Section~\ref{sec:riemann}, the adiabatic collapse application does not currently benefit from characteristic limiting.  
As discussed earlier, toward the end of collapse, the core undergoes a phase transition from atomic nuclei and nucleons to bulk nuclear matter.  
However, the tabulated nuclear matter EoS appears to not be sufficiently smooth around this transition to enable robust construction of the characteristic fields, which depends on thermodynamic derivatives from the EoS (see Appendix~\ref{appendix:deriv}).  
Moreover, the interpolation scheme discussed in Section~\ref{sec:tableInterpolation} is only $C^{0}$ continuous, which implies that derivatives are discontinuous across adjacent cubes in the table.  
As a result, the thermodynamic derivatives are not smooth around the phase transition, which appears to give rise to unphysical perturbations.  
These perturbations manifest as acoustic noise, in the characteristic and, eventually, the conserved fields, and this is clearly evident by considering the pressure.  
Figure~\ref{fig:Gravitational Collapse, Characteristic Limiting Study} displays space-time plots of the logarithmic pressure gradient, $(\ln p/\ln r)$, shortly after bounce from two simulations --- one employing component-wise limiting (left panel), and one employing characteristic limiting (right panel).  
In both panels, the dashed black line corresponds to the minimum in the adiabatic index (cf. upper left panel in Figure~\ref{fig:Gravitational Collapse, Bounce Phase}), which we refer to as the phase transition.  
The formation of the bounce shock and subsequent acoustic waves during the ring-down phase after bounce are clearly seen in the lower part of both panels, which qualitatively agree.  
However, about $5$~ms after bounce, acoustic waves are seen to continuously emanate from the core in the model with characteristic limiting.  
These waves are absent in the model with component-wise limiting.  
The accoustic waves in the model with characteristic limiting appear to emanate from the vicinity of the phase transition; i.e. originate around the vertical dashed black line (see also Figure~\ref{fig:Gravitational Collapse, Characteristic Limiting Zoomed-In}).  
Another difference in the results obtained with the two limiters is the behavior of the maximum logarithmic pressure gradient.  
In the component-wise case, the peak in the pressure gradient remains fixed at approximately $30$~km for the duration of the run after bounce.  
With characteristic limiting, this peak has a slow trajectory, starting at about $r=15$~km and ending at $r=30$~km.  

\begin{figure}[h]
    \centering
    \includegraphics[width=.50\textwidth]{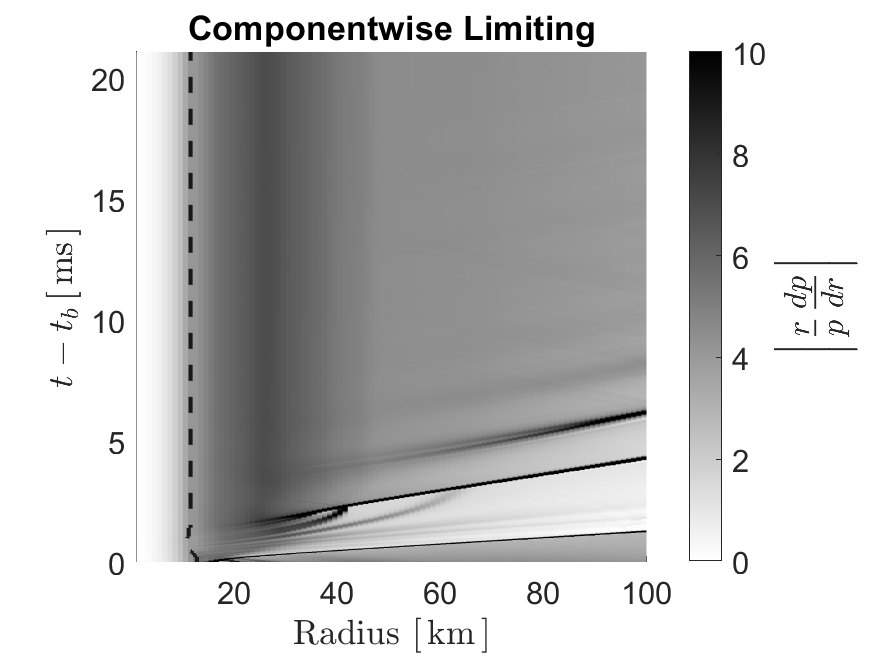}\hfill
    \includegraphics[width=.50\textwidth]{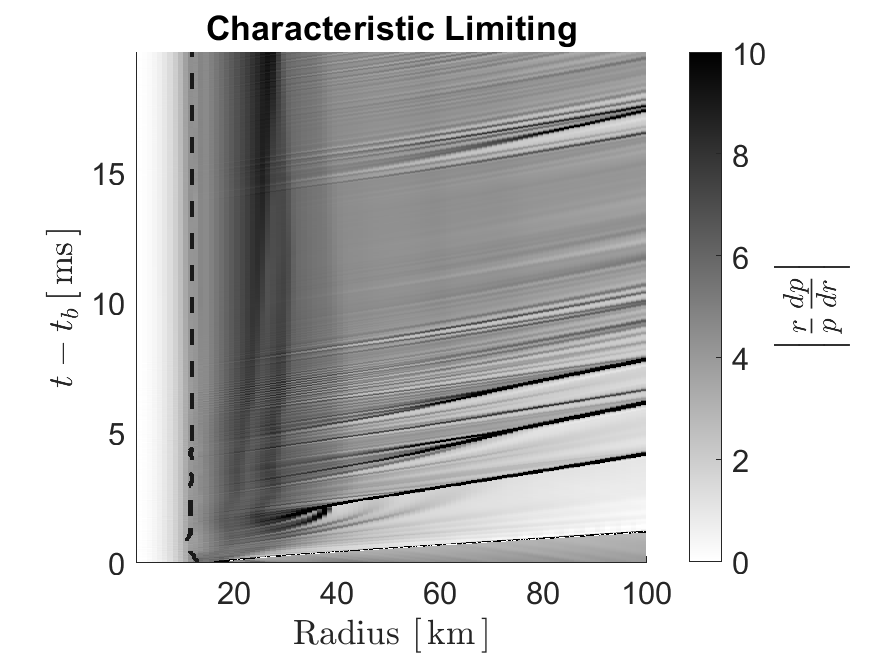}\hfill
   \caption{Space-time plots of the absolute value of the logarithmic pressure gradient for simulations employing component-wise (left) and characteristic (right) limiting.  
   The time domain extends over a brief period after bounce $t \in \left[t_{\rm{b}}, \, \, t_{\rm{b}}+20 \, \mathrm{ms}\right]$. 
   The vertical dashed line around $10$~km, which traces the minimum of the adiabatic index $\Gamma _{\mathrm{min}}$, represents the approximate position of the phase transition.  Near the bottom of each figure, the black line extending from approximately $20$~km to $100$~km for a duration of $1$~ms traces the bounce shock. The lines which form after this are traces of secondary or tertiary ``ripples'' that propagate outward from the inner core, and follow the shock shortly after bounce.  With component-wise limiting, the pressure gradient is relatively smooth at about $10$~ms after bounce and thereafter.  With characteristic limiting, perturbations continue to develop in the solution after bounce, which is visible as noise in the pressure gradient.  A high time resolution was used in this case to better capture details of the dynamics around the phase transition, such as the formation of acoustic waves and their reflections off the inner boundary. }
   \label{fig:Gravitational Collapse, Characteristic Limiting Study}
\end{figure}

Figure~\ref{fig:Gravitational Collapse, Characteristic Limiting Zoomed-In} shows a zoomed-in portion of the logarithmic pressure gradient for the characteristic limiting case displayed in the right panel in Figure~\ref{fig:Gravitational Collapse, Characteristic Limiting Study}.  
Spurious pressure waves appear to be generated around the phase transition (or slightly ahead of the dashed black line), which then propagate across the entire domain.  
Prominent examples of this are seen around $t-t_{\rm{b}}=0.5$~ms, $1.65$~ms, and $2.45$~ms, where pairs of left- and right-propagating waves emanate from the phase transition.  
The left-going waves propagate toward the inner boundary and are then reflected back out.  
These waves lead to the noisy pattern seen in the right panel in Figure~\ref{fig:Gravitational Collapse, Characteristic Limiting Study}.  

\begin{figure}[h]
  \centering
  \includegraphics[width = 0.85\textwidth]{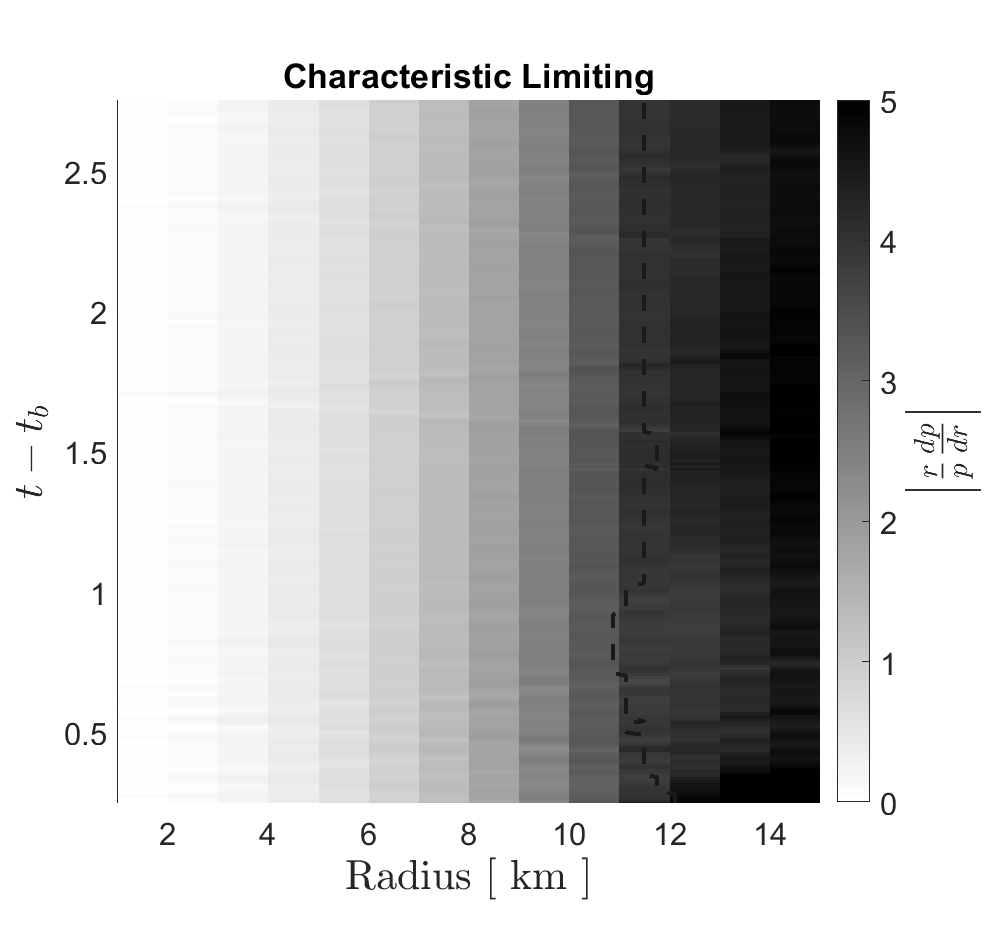}
  \caption{Zoomed-in view of the logarithmic pressure gradient (to better capture details of the dynamics around the phase transition) for the simulation employing characteristic limiting shown in the right panel in Figure~\ref{fig:Gravitational Collapse, Characteristic Limiting Study}.}
  \label{fig:Gravitational Collapse, Characteristic Limiting Zoomed-In}
\end{figure}

As discussed in Section~\ref{sec:slope}, characteristic limiting relies on transforming the set of conserved variables to the set of characteristic variables by applying the matrix of left eigenvectors from the eigendecomposition of the flux Jacobian.  
The construction of this matrix involves thermodynamic derivatives of the pressure and other quantities which, in the case of the tabulated EoS, do not have analytic expressions.  
Instead, these derivatives are obtained by differentiating the trilinear interpolation formula used to obtain quantities from the EoS table, and are not necessarily smooth --- especially across the phase transition.  
The result is a discontinuity in every characteristic variable around the location of the phase transition.  
Because of this, it appears to no longer be beneficial to employ characteristic limiting, as it results in the waves seen in Figures~\ref{fig:Gravitational Collapse, Characteristic Limiting Study}~and~\ref{fig:Gravitational Collapse, Characteristic Limiting Zoomed-In}, and destroys the accuracy gained from characteristic limiting observed in Section~\ref{sec:riemann}.  
Moreover, the expression for the sound speed given in Appendix~\ref{appendix:characteristic}, obtained from the eigendecomposition of the flux Jacobian, may transiently become imaginary due to variations in the derivatives.  
In this case we default to constructing the sound speed as provided by the EoS table.  
Future work will include improving the fidelity of \thornado's interface with the EoS table, especially around the phase transition, in order to circumvent these problems.  

\section{Summary, Conclusions, and Outlook}
\label{sec:conclusions}

\subsection{Summary}

We have extended the Runge--Kutta discontinuous Galerkin (RKDG) method for the Euler equations to accommodate an equation of state for dense nuclear matter, to solve problems in Cartesian, spherical-polar, and cylindrical coordinate systems in a three-covariant framework, and to simulate adiabatic, spherically-symmetric stellar collapse with self-gravity.  
More specifically, we have implemented a spectral-type nodal collocation DG approximation, which leads to simplifications in the semi-discrete equations --- especially for problems that make use of curvilinear coordinates.  
In making these extensions to the RKDG method, we extended various limiters to maintain physically sound solutions:
\begin{itemize}
    \item We have supplemented the RKDG method with a standard total variation diminishing slope limiter, combined with a troubled-cell indicator, to maintain time-integration stability and to reduce spurious oscillations around discontinuities. 
    For our purposes, this involved a non-trivial adaptation of the limiter to nuclear equations of state, specifically when limiting the characteristic fields, and we have provided the necessary characteristic decomposition to achieve this in Appendix~\ref{appendix:characteristic}.  
    \item We have designed a bound-enforcing limiter to prevent the numerical solutions from becoming physically inadmissible; i.e. exceeding bounds imposed by the tabulated EoS.  
    The tabulated EoS is supplied with strict boundaries in which the solution must be confined. 
    However, critical thermodynamic quantities provided by the EoS are not necessarily globally convex, and this complicates the design of the bound-enforcing limiter, which currently operates under the assumption of a convex EoS.
\end{itemize}

We have developed \thornado\ based on this extended RKDG method.  
\thornado\ is written in modern Fortran, which is a general purpose programming language for high-performance scientific computing. 
Moreover, \thornado\ is intended for multiphysics CCSN simulations with high-order methods, and to this end the RKDG method for hydrodynamics has been chosen, in part, for its ability to faithfully capture discontinuities and its ability to maintain high-order accuracy in smooth flows with a compact computational stencil.  
Distributed parallel computing capabilities with MPI are enabled through an interface with AMReX~\citep{zhang:2019}.  
(The incorporation of AMReX's adaptive mesh refinement is deferred to future work.)  
We also mention that, in addition to distributed parallelism with MPI, \thornado\ has been partially ported to utilize graphics processing units (GPUs) through the OpenACC\footnote{\url{https://www.openacc.org}} and OpenMP\footnote{\url{https://www.openmp.org}} standards, which will allow \thornado\ to utilize heterogeneous architectures.  
Details on this progress will be reported in a future publication.  

We have tested 
\thornado\ against a suite of diverse and challenging problems incorporating a tabulated nuclear EoS in one and two spatial dimensions (see \citet{endeve:2019} for further tests in the ideal EoS case):
\begin{itemize}
    \item To test the formal order of accuracy of the RKDG method with a nuclear EoS we performed an advection test with a smooth mass density profile using second- and third-order methods and various degrees of freedom to determine the rate of convergence.  
    It was found that the third-order method is significantly more accurate than the second-order method, but the rate of convergence for the third-order method deteriorates to second-order at higher resolution, possibly due to the use of trilinear EoS interpolation.  
    To further examine the efficacy of the high-order RKDG method, a discontinuous multi-shaped mass density profile was advected using characteristic limiting, and the initial condition was compared with the numerical solution after one and ten periods. 
    We compared results obtained with second- and third-order methods using the same total number of degrees of freedom, by adjusting the number of cells.  
    The third-order method was found to be superior to the second-order method in this case as well.  
    \item We conducted several well-known Riemann problem tests --- adapted to the nuclear EoS case --- in Cartesian, spherical-polar, and cylindrical coordinates, and one and two spatial dimensions, to examine \thornado's ability to resolve discontinuities with high-order RKDG methods, without introducing spurious oscillations. 
    It was demonstrated that results obtained with characteristic limiting are far superior to corresponding results obtained with component-wise limiting.  
    Finally, a special version of the Sod shock tube test was constructed to examine the efficacy of the bound-enforcing limiter.  
    In this case, it was demonstrated that the bound-enforcing limiter maintains physically admissible solutions, while at the same time preserving the inherent conservation properties of the RKDG method.
\end{itemize}

We have applied 
\thornado\ to the problem of adiabatic stellar core collapse of a realistic non-rotating progenitor in spherical symmetry:
\begin{itemize}
    \item We modeled the critical phases of collapse, through nuclear densities, the phase transition to bulk nuclear matter, core bounce, shock formation, and the propagation of the shock through the outer stellar layers.  
    \item The complexity of this application necessitated additional investigations to probe the features of the RKDG method for hydrodynamics in \thornado, such as the role of limiting and how it contributes to improved robustness of these simulations, the dependence of the solution on the troubled-cell indicator threshold and spatial resolution, and the conservation of energy through challenging stages in the simulation, such as stellar core bounce.
\end{itemize}

\subsection{Conclusions and Outlook}

\begin{itemize}
    \item We successfully evolved a non-rotating, spherically symmetric, 15~\Msun\ progenitor with self-gravity through adiabatic collapse, bounce, and several hundred milliseconds of shock propagation past bounce, while maintaining adiabaticity (e.g. the entropy and electron fraction profiles remained constant in the core). 
    The success of this application marks an important step toward applying DG methods to more realistic CCSN simulations, and given the results obtained, we are in a position to develop \thornado\ further towards more physically complete CCSN simulations; e.g., by incorporating neutrino transport. 

    \item In the adiabatic collapse application, the bound-enforcing limiter is critical in allowing the solution to evolve through bounce.  
    Without this limiter, the solution exceeds the limits of the EoS and the algorithm fails.  
    The bound-enforcing limiter is required to maintain a physically valid solution for this application, but it, along with the slope limiter, interferes with the inherently good energy conservation properties of the RKDG scheme.  
    Before and after bounce, the change in total energy is relatively low.  
    However, when limiters are applied through bounce, an artificial jump in the total energy compromises the energy conservation.  
    The change in total energy is less than $0.5$~B for the fiducial run with inner cell width of $0.5$~km, and decreases with increasing spatial resolution.  
    While the change in total energy during bounce is relatively small, when compared to any of the individual energy components, future work focusing on reducing this unphysical change in total energy is warranted.

    \item For standard hydrodynamics tests with shocks, such as Riemann problems, we have shown that characteristic limiting is superior to component-wise limiting for resolving discontinuities while suppressing nonphysical, oscillatory features.  
    However, characteristic limiting depends on derivatives of thermodynamic quantities, which are estimated from the tabulated EoS, and may not be sufficiently smooth.
    In particular, for the adiabatic collapse application, we observed anomalous behavior in the form of acoustic noise, which appears to originate around the phase transition.  
    Thus, characteristic limiting currently does not provide the desired improvements for the adiabatic collapse application, or any problem that may involve a phase transition. 
    The issue associated with these thermodynamic derivatives must be further investigated and resolved before our numerical method can be extended more generally to employ more sophisticated limiters, such as moment limiters \citep{krivodonova:2007} or WENO-type limiters for DG \citep{Zhu:2020}, which also rely on limiting of characteristic fields.  
    This may involve an improved EoS interpolation scheme that enforces thermodynamic consistency.  

    \item As seen in the convergence tests, the RKDG method in \thornado\ gained accuracy over lower-order schemes by implementing high-order discretization ($N=3$) for a fixed number of degrees of freedom. 
    However, third-order methods diminished to second-order accuracy for higher degrees of freedom, and the interpolation of the tabulated EoS may have been an agent in this loss of accuracy, but further investigation is required to confirm this.  
    Moreover, for all the tests in Section \ref{sec:results}, our method reliably captured physical discontinuities and oscillations with high-order instantiations of the RKDG scheme in \thornado.
    However, for the adiabatic collapse application, we consistently employed a second-order accurate approach.  
    The main reason: transient spurious oscillations (or perturbations) developed when we employed third-order discretization.  
    Again, the interpolation of the EoS may be impacting the performance of the high-order scheme.  
    We emphasize that the results for the gravitational collapse application obtained with second-order methods and component-wise limiting are satisfactory, and provides the basis for incorporating neutrino transport algorithms also based on DG methods.  
    However, while the present paper represents a step towards our goal, further work is required to realize CCSN simulations with high-order DG methods.  

    \item All results presented here were obtained with the HLL Riemann solver \citep{Harten:1983}.  
    While we have also implemented the HLLC Riemann solver \citep{Toro:1994}, which is designed to account for contact discontinuities and has been shown to give superior results \citep[see, e.g.,][]{Genasis}, we decided not to use this Riemann solver here. 
    The known ``odd-even" instability \citep{quirk:1994}, which develops with the HLLC Riemann solver in some multidimensional settings, is the main reason for our decision.  
    Future work includes development of a hybrid solver, with the capability of applying the HLLC solver in regions of smooth flow while switching to the HLL solver in the vicinity of shocks by means of a shock detector.
    
    \item Because CCSNe are general relativistic in nature, we are extending the hydrodynamics
    in \thornado{} to accommodate general relativity under the conformally-flat approximation 
    \citep[see, e.g.,][]{wilson:1996},
    some details of which are given in \citet{dunham:2020}.
    
\end{itemize}

\acknowledgments{
Eirik Endeve and Anthony Mezzacappa, along with Jesse Buffaloe, Sam Dunham, David Pochik, and Nick Roberts, acknowledge support from the NSF Gravitational Physics Theory Program (NSF PHY 1505933 and 1806692).  
Brandon L. Barker is supported by the National Science Foundation Graduate Research Fellowship Program under grant number DGE-1848739.
Eirik Endeve was partially supported by the Exascale Computing Project (17-SC-20-SC), a collaborative effort of the U.S.\ Department of Energy Office of Science and the National Nuclear Security Administration.  
We thank Ann S. Almgren and Don E. Willcox for help with interfacing \thornado\ with AMReX.
}

\software{
  \href{https://matplotlib.org/}{Matplotlib} \citep{hunter:2007a},
  \href{http://www.numpy.org/}{NumPy} \citep{numpy},
  \href{https://www.scipy.org/}{SciPy} \citep{jones:2001},
  \href{http://yt-project.org/}{yt} \citep{turk:2011}
  \href{https://amrex-codes.github.io/amrex/}{AMReX} \citep{zhang:2019}
  }

\appendix

\section{Characteristic Decomposition}
\label{appendix:characteristic}
\input{characteristic}

\section{Thermodynamic Derivatives}
\label{appendix:deriv}
\input{derivatives}

\bibliography{ms}

\end{document}

%% file: characteristic.tex
In this appendix, we provide the characteristic decomposition of the flux Jacobians, which are needed for slope limiting in characteristic fields.  
Recall that for the characteristic slope limiting described in \ref{sec:slope}, we require the eigendecomposition of the flux Jacobian
\beq
  \pderiv{\mathbf{F}^{i}(\mathbf{U})}{\mathbf{U}} =
  \mathcal{R}^{i}\, \Lambda^{i}\, (\mathcal{R}^{i})^{-1}\quad(i=1,\ldots,d).
\eeq
In the following, we will express the pressure from the EoS as $p \,=\, p(\tau, \epsilon, D_{\rm{e}})$; i.e., with independent variables $\tau \,=\, \rho^{-1}$, $\epsilon \,=\, e/ \rho$, and $D_{\rm{e}} = \rho \, \ye$, instead of the usual function of $\rho$, $T$, and $\ye$.
This choice is arbitrary, but follows the approach outlined in \citet{colella:1985} for a general EoS without the addition of the conservation equation for electron number (cf. Equation~\eqref{eq:electronConservation}).
The necessary transformations of thermodynamic derivatives between these two sets of independent variables are given in Appendix~\ref{appendix:deriv}.  
From the state and flux vectors given in Equation~\eqref{eq:eulerSystem}, we can calculate the following flux Jacobian matrices in each direction:

\begin{equation}
  \centering
  \pderiv{\mathbf{F}^{1}(\mathbf{U})}{\mathbf{U}}
  = \left[
	\begin{array}{cccccc}
		0 & \gamma^{11} & 0 & 0 & 0 & 0 \\
		-v^{1}v_{1} -p_{\tau}\tau^{2} - p_{\epsilon}\tau\left(\epsilon - \frac{v^{i}v_{i}}{2}\right) & v^{1}(2-p_{\epsilon}\tau)  & -p_{\epsilon}v^{2}\tau & 
			-p_{\epsilon}v^{3}\tau  & p_{\epsilon}\tau  & p_{D_{\rm{e}}} \\
		-v^{1}v_{2} & \gamma^{11}v_{2} & v^{1} & 0 & 0 & 0 \\
		-v^{1}v_{3} & \gamma^{11}v_{3} & 0 & v^{1} & 0 & 0 \\
		v^{1}\left(-H - p_{\tau}\tau^{2} -p_{\epsilon}\tau\left(\epsilon - \frac{v^{i}v_{i}}{2}\right)\right) & \gamma^{11}H - p_{\epsilon}(v^{1})^{2}\tau  & 
			-p_{\epsilon}v^{1}v^{2}\tau & -p_{\epsilon}v^{1}v^{3}\tau  & v^{1}(1+p_{\epsilon}\tau) & v^{1} p_{D_{\rm{e}}} \\
		-v^{1} \ye & \gamma^{11} \ye & 0 & 0 & 0 & v^{1} \\
	\end{array}
    \right],
\end{equation}

\begin{equation}
  \centering
  \pderiv{\mathbf{F}^{2}(\mathbf{U})}{\mathbf{U}}
  = \left[
	\begin{array}{cccccc}
		0 & 0 & \gamma^{22} & 0 & 0 & 0 \\
		-v^{2}v_{1} & v^{2} & \gamma^{22}v_{1} & 0 & 0 & 0 \\
		-v^{2}v_{2} -p_{\tau}\tau^{2} - p_{\epsilon}\tau\left(\epsilon -\frac{v^{i}v_{i}}{2}\right) & -p_{\epsilon}v^{1}\tau &
			v^{2}(2-p_{\epsilon}\tau) &  -p_{\epsilon}v^{3}\tau & p_{\epsilon}\tau & p_{D_{\rm{e}}} \\
		-v^{2}v_{3} & 0 & \gamma^{22}v_{3} & v^{2} & 0 & 0 \\
		v^{2}\left(-H - p_{\tau}\tau^{2} -p_{\epsilon}\tau\left(\epsilon - \frac{v^{i}v_{i}}{2}\right)\right) &
			- p_{\epsilon}v^{2}v^{1}\tau & \gamma^{22}H - p_{\epsilon}(v^{2})^{2}\tau &
			- p_{\epsilon}v^{2}v^{3}\tau & v^{2}(1+p_{\epsilon}\tau) & v^{2}p_{D_{\rm{e}}} \\
		-v^{2}\ye & 0 & \gamma^{22}\ye & 0 & 0 & v^{2}\\
	\end{array}
  \right],
\end{equation}
and
\begin{equation}
  \centering
  \pderiv{\mathbf{F}^{3}(\mathbf{U})}{\mathbf{U}}
  = \left[
  \begin{array}{cccccc}
	0 & 0 & 0 & \gamma^{33} & 0 & 0 \\
	-v^{3}v_{1} & v^{3} & 0 & \gamma^{33}v_{1} & 0 & 0 \\
	-v^{3}v_{2} & 0 & v^{3} & \gamma^{33}v_{2} & 0 & 0 \\
	-v^{3}v_{3} -p_{\tau}\tau^{2} - p_{\epsilon}\tau\left(\epsilon - \frac{v^{i}v_{i}}{2}\right) & -p_{\epsilon}v^{1}\tau &
		- p_{\epsilon}v^{2}\tau &  v^{3}(2 - p_{\epsilon}\tau) & p_{\epsilon}\tau & p_{D_{\rm{e}}} \\
	v^{3}\left(-H - p_{\tau}\tau^{2} -p_{\epsilon}\tau\left(\epsilon - \frac{v^{i}v_{i}}{2}\right)\right) &
		- p_{\epsilon}v^{3}v^{1}\tau & -p_{\epsilon}v^{3}v^{2}\tau & \gamma^{33}H - p_{\epsilon}(v^{3})^{2}\tau &
	v^{3}(1+p_{\epsilon}\tau) & v^{3}p_{D_{\rm{e}}} \\
	-v^{3}\ye & 0 & 0 & \gamma^{33}\ye & 0 & v^{3} \\
  \end{array}
  \right],
\end{equation}
where we have defined the specific enthalpy of stagnation $H = \tau(E + p)$ and introduced the compact notation 
\beq
    p_{\epsilon}  = \left(\frac{\partial{p}}{\partial{\epsilon}}\right)_{\tau,D_{\rm{e}}}, \quad
    p_{D_e}  = \left(\frac{\partial{p}}{\partial{D_{\rm{e}}}}\right)_{\tau, \epsilon}, \quad
    p_{\tau}  = \left(\frac{\partial{p}}{\partial{\tau}}\right)_{\epsilon, D_{\rm{e}}}
\eeq
to express the necessary partial derivatives. 
The eigenvalues of the flux Jacobian are given by the diagonal matrix
\begin{align}
	\Lambda^{i} =
	\begin{bmatrix}
  		v^{i} - \CS\sqrt{\gamma^{ii}} & 0 & 0 & 0 & 0 & 0 \\
  		0 & v^{i} & 0 & 0 & 0 & 0      \\
  		0 & 0 & v^{i} & 0 & 0 & 0      \\
  		0 & 0 & 0 & v^{i} & 0 & 0      \\
  		0 & 0 & 0 & 0 & v^{i} & 0      \\
  		0 & 0 & 0 & 0 & 0 & v^{i} + \CS\sqrt{\gamma^{ii}}
	\end{bmatrix},
\end{align}
where $\CS = \sqrt{\Gamma p \tau}$, with
\beq
  \Gamma = \left(\tau (p p_{\epsilon} - p_{\tau}) + p_{D_e} \ye \tau^{-1}\right) p^{-1}, 
\eeq
is the local sound speed.  
In the less general case where we ignore the electron contribution (i.e. $p_{D_e}=0$), this reduces to the expression given by \citet{colella:1985}.
The right eigenvectors are then given by the column vectors of the following matrices
\begin{align}
  \mathcal{R}^{1} =
  \left[
  \begin{array}{cccccc}
    1 & 0 & 1 & 1 & 0 & 1 \\
    v_{1}-\CS\sqrt{\gamma_{11}} & 0 & v_{1} & v_{1} & 0 & v_{1}+\CS\sqrt{\gamma_{11}} \\
    v_{2} & 1 & 0 & 0 & 0 & v_{2} \\
    v_{3} & 0 & 0 & 0 & 1 & v_{3} \\
    H - \CS\sqrt{\gamma_{11}}v^{1} & v^{2} & \beta_{1} & 0 & v^{3} & H +\CS\sqrt{\gamma_{11}}v^{1} \\
    \ye  & 0 & 0 & \frac{\tau  \chi_{1} }{2 p_{D_{\rm{e}}}} & 0 & \ye  \\
  \end{array}
  \right],
\end{align}

\begin{align}
  \mathcal{R}^{2} =
  \left[
  \begin{array}{cccccc}
    1 & 0 & 1 & 1 & 0 & 1 \\
    v_{1} & 1 & 0 & 0 & 0 & v_{1} \\
    v_{2}-\CS\sqrt{\gamma_{22}}& 0 & v_{2} & v_{2} & 0 & v_{2} + \CS\sqrt{\gamma_{22}} \\
    v_{3} & 0 & 0 & 0 & 1 & v_{3} \\
    H - \CS\sqrt{\gamma_{22}}v^{2} & v^{1} & \beta_{2} & 0 & v^{3} & H + \CS\sqrt{\gamma_{22}}v^{2} \\
    \ye  & 0 & 0 & \frac{\tau  \chi_{2} }{2 p_{D_{\rm{e}}}} & 0 & \ye  \\
  \end{array}
  \right],
\end{align}
and
\begin{align}
  \mathcal{R}^{3} =
  \left[
  \begin{array}{cccccc}
    1 & 0 & 1 & 1 & 0 & 1 \\
    v_{1} & 1 & 0 & 0 & 0 & v_{1} \\
    v_{2} & 0 & 0 & 0 & 1 & v_{2} \\
    v_{3}-\CS\sqrt{\gamma_{33}} & 0 & v_{3} & v_{3} & 0 & v_{3}+\CS\sqrt{\gamma_{33}} \\
    H - \CS\sqrt{\gamma_{33}}v^{3} & v^{1} & \beta_{3} & 0 & v^{2} & H + \CS\sqrt{\gamma_{33}}v^{3} \\
    \ye  & 0 & 0 & \frac{\tau  \chi_{3} }{2 p_{D_{\rm{e}}}} & 0 & \ye  \\
  \end{array}
  \right],
\end{align}

\noindent where the following definitions have been used:

\begin{eqnarray*}
  &\Delta_{1} =& 2 v^{1}v_{1}-v^{i}v_{i}, \\
  &\Delta_{2} =& 2 v^{2}v_{2}-v^{i}v_{i}, \\
  &\Delta_{3} =& 2 v^{3}v_{3}-v^{i}v_{i}, \\
  &\chi_{i} =& p_{\epsilon} ( \Delta_{i} + 2\epsilon) + 2p_{\tau}\tau, \\
  &\beta_{i} =& \frac{1}{2} \left(\Delta_{i}+2 \epsilon +\frac{2 p_{\tau} \tau }{p_{\epsilon}}\right).
\end{eqnarray*}

\noindent The left eigenvectors are given by the row vectors of the inverse matrix $\mathcal{L}^{i} = (\mathcal{R}^{i})^{-1}$

\begin{align}
  (\mathcal{R}^{1})^{-1} = \frac{1}{\CS^{2}}
  \left[
  \begin{array}{cccccc}
   \frac{1}{4} (\omega + 2 \CS\sqrt{\gamma_{11}}v^{1}) &
   -\frac{1}{2} (\CS\sqrt{\gamma^{11}} + \phi^{1} )    &
   -\frac{1}{2} \phi^{2}                          &
   -\frac{1}{2} \phi^{3}                          &
   \frac{p_{\epsilon} \tau }{2} & \frac{p_{D_{\rm{e}}}}{2}
   \\
   -\frac{v_{2} \omega }{2}                 &
   \phi^{1} v_{2}  & \CS^{2}+\phi^{2} v_{2} &
   \phi^{3} v_{2}                           &
   -\phi_{2}                                &
   -p_{D_{\rm{e}}} v_{2}
   \\
   \frac{2 \chi_{1}  \CS^{2}+\alpha_{1}  \omega \tau^{-1} }{2 \chi_{1} } &
   -\frac{\phi^{1} \alpha_{1}  }{\chi_{1} \tau }                         &
   -\frac{\phi^{2} \alpha_{1}  }{\chi_{1} \tau }                         &
   -\frac{\phi^{3} \alpha_{1}  }{\chi_{1} \tau }                         &
   \frac{p_{\epsilon} \alpha_{1} }{\chi_{1} }                            &
   \frac{p_{D_{\rm{e}}} \left(\alpha_{1} -2 \CS^{2}\right)}{\tau \chi_{1} }
   \\
   -\frac{\ye p_{D_{\rm{e}}} \omega }{\chi_{1} \tau }    &
   \frac{2 \ye p_{D_{\rm{e}}} \phi^{1} }{\chi_{1} \tau } &
   \frac{2 \ye p_{D_{\rm{e}}} \phi^{2} }{\chi_{1} \tau}  &
   \frac{2 \ye p_{D_{\rm{e}}} \phi^{3} }{\chi_{1} \tau}  &
   -\frac{2 \ye p_{D_{\rm{e}}} p_{\epsilon} }{\chi_{1} } &
   \frac{2 p_{D_{\rm{e}}}\left(\CS^{2}-\ye p_{D_{\rm{e}}} \right)}{\tau \chi_{1} }
   \\
   -\frac{v_{3} \omega }{2} &
   \phi^{1} v_{3}           &
   \phi^{2} v_{3}           &
   \CS^{2}+\phi^{3} v_{3}   &
   -\phi_{3}                &
   -p_{D_{\rm{e}}} v_{3}
   \\
   \frac{1}{4} (\omega -2 \CS\sqrt{\gamma_{11}}v^{1}) &
   \frac{1}{2} (\CS\sqrt{\gamma^{11}}-\phi^{1} )      &
   -\frac{1}{2} \phi^{2}                         &
   -\frac{1}{2} \phi^{3}                         &
   \frac{p_{\epsilon} \tau }{2}                  &
   \frac{p_{D_{\rm{e}}}}{2}
   \\
  \end{array}
  \right],
\end{align}

\begin{align}
  (\mathcal{R}^{2})^{-1} = \frac{1}{\CS^2}
  \left[
  \begin{array}{cccccc}
   \frac{1}{4} (\omega + 2 \CS\sqrt{\gamma_{22}}v^{2}) &
   -\frac{1}{2} \phi^{1}                               &
   -\frac{1}{2} (\CS\sqrt{\gamma^{22}}+\phi^{2})       &
   -\frac{1}{2} \phi^{3}                               &
   \frac{p_{\epsilon} \tau }{2}                        &
   \frac{p_{D_{\rm{e}}}}{2}
   \\
   -\frac{v_{1} \omega }{2} &
   \CS^2 + \phi^{1} v_{1}   &
   \phi^{2} v_{1}           &
   \phi^{3} v_{1}           &
   -\phi_{1}                &
   -p_{D_{\rm{e}}} v_{1}
   \\
   \frac{2 \chi_{2}  \CS^2+\alpha_{2}  \omega \tau^{-1} }{2 \chi_{2} } &
   -\frac{\phi^{1} \alpha_{2}  }{\chi_{2} \tau }                       &
   -\frac{\phi^{2} \alpha_{2}  }{\chi_{2} \tau }                       &
   -\frac{\phi^{3} \alpha_{2} }{\chi_{2} \tau }                        &
   \frac{p_{\epsilon} \alpha_{2} }{\chi_{2} }                          &
   \frac{p_{D_{\rm{e}}} \left(\alpha_{2} -2 \CS^2\right)}{\tau \chi_{2} }
   \\
   -\frac{\ye p_{D_{\rm{e}}} \omega }{\chi_{2} \tau }    &
   \frac{2 \ye p_{D_{\rm{e}}} \phi^{1} }{\chi_{2} \tau } &
   \frac{2 \ye p_{D_{\rm{e}}} \phi^{2} }{\chi_{2} \tau } &
   \frac{2 \ye p_{D_{\rm{e}}} \phi^{3} }{\chi_{2} \tau } &
   -\frac{2 \ye p_{D_{\rm{e}}} p_{\epsilon} }{\chi_{2} } &
   \frac{2 p_{D_{\rm{e}}} \left(\CS^{2}-\ye p_{D_{\rm{e}}} \right)}{\tau \chi_{2} }
   \\
   -\frac{v_{3} \omega }{2} &
   \phi^{1} v_{3}           &
   \phi^{2} v_{3}           &
   \CS^{2}+\phi^{3} v_{3}   &
   -\phi_{3}                &
   -p_{D_{\rm{e}}} v_{3}
   \\
   \frac{1}{4} (\omega -2 \CS\sqrt{\gamma_{22}}v^{2}) &
   -\frac{1}{2} \phi^{1}                              &
   \frac{1}{2} (\CS\sqrt{\gamma^{22}}-\phi^{2})       &
   -\frac{1}{2} \phi^{3}                              &
   \frac{p_{\epsilon} \tau }{2}                       &
   \frac{p_{D_{\rm{e}}}}{2}
   \\
  \end{array}
  \right],
\end{align}
and
\begin{align}
  (\mathcal{R}^{3})^{-1} = \frac{1}{\CS^{2}}
  \left[
  \begin{array}{cccccc}
   \frac{1}{4} (\omega + 2 \CS\sqrt{\gamma_{33}}v^{3} ) &
   -\frac{1}{2} \phi^{1}                                &
   -\frac{1}{2} \phi^{2}                                &
   -\frac{1}{2}(\CS\sqrt{\gamma^{33}}+\phi^{3})         &
   \frac{p_{\epsilon} \tau }{2}                         &
   \frac{p_{D_{\rm{e}}}}{2}
   \\
   -\frac{v_{1} \omega }{2} & 
   \CS^{2} + \phi^{1} v_{1} & 
   \phi^{2} v_{1}           &
   \phi^{3} v_{1}           & 
   -\phi_{1}                & 
   -p_{D_{\rm{e}}} v_{1}
   \\
   \frac{2 \chi_{3}  \CS^{2}+\alpha_{3}  \omega \tau^{-1} }{2 \chi_{3} } &
   -\frac{\phi^{1} \alpha_{3}  }{\chi_{3} \tau }                         &
   -\frac{\phi^{2} \alpha_{3}  }{\chi_{3} \tau }                         &
   -\frac{\phi^{3} \alpha_{3} }{\chi_{3} \tau }                          &
   \frac{p_{\epsilon} \alpha_{3} }{\chi_{3} }                            &
   \frac{p_{D_{\rm{e}}} \left(\alpha_{3} -2 \CS^{2}\right)}{\tau \chi_{3} }
   \\
   -\frac{\ye p_{D_{\rm{e}}} \omega }{\chi_{3} \tau }    &
   \frac{2 \ye p_{D_{\rm{e}}} \phi^{1} }{\chi_{3} \tau } &
   \frac{2 \ye p_{D_{\rm{e}}} \phi^{2} }{\chi_{3} \tau}  &
   \frac{2 \ye p_{D_{\rm{e}}} \phi^{3} }{\chi_{3} \tau}  &
   -\frac{2 \ye p_{D_{\rm{e}}} p_{\epsilon} }{\chi_{3} } &
   \frac{2 p_{D_{\rm{e}}} \left(\CS^{2}-\ye p_{D_{\rm{e}}} \right)}{\tau \chi_{3} }
   \\
   -\frac{v_{2} \omega }{2} &
   \phi^{1} v_{2}           &
   \CS^{2}+\phi^{2} v_{2}   &
   \phi^{3} v_{2}           &
   -\phi_{2}                &
   -p_{D_{\rm{e}}} v_{2}
   \\
   \frac{1}{4} (\omega -2 \CS\sqrt{\gamma_{33}}v^{3}) &
   -\frac{1}{2} \phi^{1}                              &
   -\frac{1}{2} \phi^{2}                              &
   \frac{1}{2} (\CS\sqrt{\gamma^{33}}-\phi^{3})       &
   \frac{p_{\epsilon} \tau }{2}                       &
   \frac{p_{D_{\rm{e}}}}{2}
   \\
  \end{array}
  \right],
\end{align}
\noindent where $\phi_{i} = p_{\epsilon}\,\tau\, v_{i}$, 
$\phi^{i} = p_{\epsilon}\,\tau\, v^{i}$, 
$\omega = \tau\, (p_{\epsilon}\,(v^{i}v_{i} - 2\epsilon) - 2\,p_{\tau}\,\tau)$, and
$\alpha_{i} = 2 \ye p_{D_{\rm{e}}} - \tau \chi_{i}$.

%% file: derivatives.tex
In Appendix~\ref{appendix:characteristic}, to compute the flux Jacobian matrices, we expressed the pressure $p \,=\, p(\tau, \epsilon, D_{\rm{e}})$ in terms of the independent variables $\tau \,=\, \rho^{-1}$, $\epsilon \,=\, e/ \rho$, and $D_{\rm{e}} = \rho \, \ye$. 
On the other hand, the tabulated EoS constructs thermodynamic variables in terms of $\rho$, $T$, and $\ye$. 
Thus, we need to express the thermodynamic derivatives of pressure necessary for the characteristic decomposition in terms of the independent variables from the EoS table.
We start with the differential of pressure
\begin{equation}
 \label{eq:pressure_var1}
  dp = (\partial_{\tau}p)_{\epsilon, D_{\rm{e}}}d\tau + (\partial_{\epsilon}p)_{\tau, D_{\rm{e}}}d\epsilon + (\partial_{D_{\rm{e}}}p)_{\epsilon, \tau}dD_{\rm{e}}.
\end{equation}
Similarly, we may express the differentials of $\tau$, $\epsilon$, and $D_{\rm{e}}$ in terms of differentials of the table variables
\begin{equation}
\begin{split}
  d\tau &= - \rho^{-2} d\rho, \\
  d\epsilon &= (\partial_{\rho}\epsilon)_{T,\ye}d\rho + (\partial_{T}\epsilon)_{\rho,\ye}dT + (\partial_{\ye}\epsilon)_{T,\rho}d\ye, \\
  dD_{\rm{e}} &= (\partial_{\rho}D_{\rm{e}})_{T,\ye}d\rho + (\partial_{T}D_{\rm{e}})_{\epsilon,\ye}dT + (\partial_{\ye}D_{\rm{e}})_{T,\epsilon}d\ye \\
   & = \ye d\rho + \rho d\ye.
\end{split}
\end{equation}
Inserting these differentials into Equation~\eqref{eq:pressure_var1}, we find another expression for the pressure differential
\begin{equation} \label{eq:pressure_dif}
\begin{split}
  dp = &\left[ -\rho^{2} (\partial_{\tau}p)_{\epsilon, D_{\rm{e}}} + (\partial_{\rho}\epsilon)_{T,\ye}\,(\partial_{\epsilon}p)_{\tau, D_{\rm{e}}}
    + \ye (\partial_{D_{\rm{e}}}p)_{\epsilon, \tau} \right] d\rho \\
    & + (\partial_{T}\epsilon)_{\rho,\ye} (\partial_{\epsilon}p)_{\tau, D_{\rm{e}}} dT \\
	& + \left[ (\partial_{\ye}\epsilon)_{T,\rho} (\partial_{\epsilon}p)_{\tau, D_{\rm{e}}} + \rho (\partial_{D_{\rm{e}}}p)_{\epsilon, \tau}\right]d\ye.
\end{split}
\end{equation}
On the other hand, we have the differential of pressure in terms of the table variables
\begin{equation}
 \label{eq:pressure_table}
  dp = (\partial_{\rho}p)_{T, \ye}d\rho + (\partial_{T}p)_{\rho, \ye}dT + (\partial_{\ye}p)_{T, \rho}d\ye.
\end{equation}

Comparing Equation~\eqref{eq:pressure_dif} and Equation~\eqref{eq:pressure_table}, we have the system of equations
\begin{gather}
  \begin{bmatrix} 
 	-\rho^{2} & (\partial_{\rho}\epsilon)_{T,\ye} & \ye \\ 
 	0 & (\partial_{T}\epsilon)_{\rho,\ye} & 0 \\
	 0 & (\partial_{\ye}\epsilon)_{T,\rho} & \rho
  \end{bmatrix}
  \begin{bmatrix}
	(\partial_{\tau}p)_{\epsilon, D_{\rm{e}}} \\ (\partial_{\epsilon}p)_{\tau, D_{\rm{e}}} \\ (\partial_{D_{\rm{e}}}p)_{\epsilon, \tau}
  \end{bmatrix}
 =
  \begin{bmatrix}
	(\partial_{\rho}p)_{T, \ye} \\ (\partial_{T}p)_{\rho, \ye} \\ (\partial_{\ye}p)_{T, \rho}
  \end{bmatrix}.
\end{gather}
Solving, with some simplifications, we find the derivatives of the pressure with respect to $\tau$, $\epsilon$, and $D_{\rm{e}}$ in terms of the table variables $\rho$, $T$, $\ye$
\begin{eqnarray}
	\left(\frac{\partial{p}}{\partial{\epsilon}}\right)_{\tau,D_{\rm{e}}} &=& \left(\frac{\partial{\epsilon}}{\partial{T}}\right)^{-1}_{\rho, \ye}\left(\frac{\partial{p}}{\partial{T}}\right)_{\rho, \ye}, \\
	\left(\frac{\partial{p}}{\partial{D_{\rm{e}}}}\right)_{\tau, \epsilon} &=& \tau \left[
					\left(\frac{\partial{p}}{\partial{\ye}}\right)_{\rho, T} -
          \left(\frac{\partial{\epsilon}}{\partial{\ye}}\right)_{\rho, T}
					\left(\frac{\partial{p}}{\partial{\epsilon}}\right)_{\tau, D_{\rm{e}}}\right], \\
	\left(\frac{\partial{p}}{\partial{\tau}}\right)_{\epsilon, D_{\rm{e}}} &=& \tau^{-2}
					\left[ \ye\left(\frac{\partial{p}}{\partial{D_{\rm{e}}}}\right)_{\tau, \epsilon}
					 + \left( \frac{\partial{\epsilon}}{\partial{\rho}} \right)_{\ye, T}
          \left(\frac{\partial{p}}{\partial{\epsilon}}\right)_{\tau, D_{\rm{e}}} - 
		  \left(\frac{\partial{p}}{\partial{\rho}}\right)_{T, \ye} \right].
\end{eqnarray}
We use these relations to relate derivatives needed for the characteristic decomposition in Appendix~\ref{appendix:characteristic} to derivatives obtained from table interpolations.